\documentclass{emulateapj}

\usepackage{graphicx}
\usepackage{amssymb}
\usepackage{longtable}
\usepackage{epsf}
\usepackage{apjfonts}

\slugcomment{Accepted by the Astrophysical Journal}

\shorttitle{The SMFs at $1.3 \leq z<4.0$ and their uncertainties}
\shortauthors{Marchesini et al.}

\begin{document}


\title{The Evolution of the Stellar Mass Function of Galaxies from $z=4.0$ and the First Comprehensive Analysis of its Uncertainties: Evidence for Mass-dependent Evolution \altaffilmark{1}}


\author{
Danilo~Marchesini\altaffilmark{2},
Pieter~G.~van Dokkum\altaffilmark{2},
Natascha~M.~F\"orster Schreiber\altaffilmark{3},
Marijn~Franx\altaffilmark{4},
Ivo~Labb\'e\altaffilmark{5,6},
Stijn~Wuyts\altaffilmark{7,8}}

\altaffiltext{1}{Based on observations with the 
\textit{Spitzer Space Telescope}, which is operated by the Jet 
Propulsion Laboratory (JPL), California Institute of Technology 
under NASA contract 1407. Based on observations with the NASA/ESA 
\textit{Hubble Space Telescope}, obtained at the Space Telescope 
Science Institute, which is operated by AURA, Inc., under NASA 
contract NAS5-26555. Based on observations collected at the European 
Southern Observatories, Chile (ESO Programme LP164.O-0612, 168.A-0485, 
170.A-0788, 074.A-0709, 275.A-5060, and 171.A-3045). Based on 
observations obtained at the Cerro Tololo Inter-American Observatory, 
a division of the National Optical Astronomy Observatories, which is 
operated by the Association of Universities for Research in Astronomy, 
Inc. under cooperative agreement with the National Science Foundation.}

\altaffiltext{2}{Department of Astronomy, Yale University, New Haven, 
CT, PO Box 208101, New Haven, CT  06520, USA}
\email{danilom@astro.yale.edu}
\altaffiltext{3}{MPE, Giessenbackstrasse, D-85748, Garching, Germany}
\altaffiltext{4}{Leiden Observatory, Leiden University, P.O. Box 9513, NL-2300 RA Leiden, Netherlands}
\altaffiltext{5}{Carnegie Observatories, 813 Santa Barbara Street, Pasadena, CA 91101}
\altaffiltext{6}{Hubble Fellow}
\altaffiltext{7}{Harvard-Smithsonian Center for Astrophysics, 60 Garden Street, Cambridge, MA 02138}
\altaffiltext{8}{W.~M.~Keck postdoctoral fellow}



\begin{abstract}
We present the evolution of the stellar mass function (SMF) of galaxies from 
$z=4.0$ to $z=1.3$ measured from a sample constructed from the deep NIR 
MUSYC, the FIRES, and the GOODS-CDFS surveys, all having very high-quality 
optical to mid-infrared data. This sample, unique in that it combines data 
from surveys with a large range of depths and areas in a self-consistent 
way, allowed us to 1) minimize the uncertainty due to cosmic variance and 
empirically quantify its contribution to the total error budget; 
2) simultaneously probe the high-mass end and the low-mass end (down to 
$\sim0.05$ times the characteristic stellar mass) of the SMF with good 
statistics; and 3) empirically derive the redshift-dependent completeness 
limits in stellar mass. We provide, for the first time, a comprehensive 
analysis of random and systematic uncertainties affecting the derived SMFs, 
including the effect of metallicity, extinction law, stellar population 
synthesis model, and initial mass function. We find that the mass density 
evolves by a factor of $\sim17^{+7}_{-10}$ since $z=4.0$, mostly driven by a 
change in the normalization $\Phi^{\star}$. If only random errors are taken 
into account, we find evidence for mass-dependent evolution, with the 
low-mass end evolving more rapidly than the high-mass end. However, we show 
that this result is no longer robust when systematic uncertainties due to 
the SED-modeling assumptions are taken into account. Another significant 
uncertainty is the contribution to the overall stellar mass density of 
galaxies below our mass limit; future studies with WFC3 will provide better 
constraints on the SMF at masses below $10^{10}$~M$_{\sun}$ at $z>2$. Taking 
our results at face value, we find that they are in conflict with 
semi-analytic models of galaxy formation. The models predict SMFs that are 
in general too steep, with too many low-mass galaxies and too few high-mass 
galaxies. The discrepancy at the high-mass end is susceptible to 
uncertainties in the models and the data, but the discrepancy at the 
low-mass end may be more difficult to explain.
\end{abstract}





%============================================================================

\section{INTRODUCTION}\label{sec-in}

Understanding the formation mechanisms and evolution with cosmic time 
of galaxies is one of the major goals of observational cosmology. In 
the current picture of structure formation, dark matter halos build up 
in a hierarchical fashion through the dissipationless mechanism of 
gravitational instability controlled by the nature of the dark matter, 
the power spectrum of density fluctuations, and the parameters of the 
cosmological model. The assembly of the stellar content of galaxies is 
governed by much more complicated physics, such as the mechanisms of star 
formation, gaseous dissipation, the feedback of stellar and central 
super-massive black hole energetic output on the baryonic material of the 
galaxies, and mergers (see \citealt{baugh06}, and references therein for 
a primer on hierarchical galaxy formation).

A powerful approach to understand these physical processes (including 
their relative importance as function of cosmic time) is to directly 
witness the growth of the stellar content in galaxies. Galaxies can 
grow their stellar mass both from in-situ star formation and/or merger 
events. Determining the growth of their stellar content as a function 
of both redshift and stellar mass provides insights into the physical 
processes governing the assembly and the evolution of galaxies. The 
stellar mass function (SMF) of galaxies and its evolution with cosmic 
time represents therefore a powerful tool to directly measure the 
build-up of the stellar mass content of galaxies.

In the past decade, significant observational progress has been made 
in the measurement of the SMF of galaxies and its evolution with 
redshift. Locally, the SMF has been measured from the 2dF Galaxy 
Redshift Survey \citep{cole01} and the Sloan Digital Sky Survey (SDSS; 
\citealt{bell03}), providing the $z\sim0$ benchmark. At intermediate 
redshifts ($z \lesssim 1.4$), the SMF has also been measured to a 
satisfactory degree from the VIMOS VLT Deep Survey (VVDS; 
\citealt{pozzetti07}; \citealt{vergani08}), the DEEP-2 Galaxy Redshift 
Survey \citep{bundy06}, and the COMBO-17 survey \citep{borch06}. The SMF 
appears to evolve slowly at $z \lesssim 1$, with about half of the 
total stellar mass density at $z\sim0$ already in place at $z\sim1$. %These 
%works have also studied the evolution with cosmic time of the SMFs of 
%different galaxy types, showing that out to $z\sim1$, non-starfoming red 
%sequence galaxies (as well as galaxies with an early-type morphology) 
%dominate the stellar mass budget at the high-mass end 
%($M_{\rm star}>10^{11}$~M$_{\sun}$).

The SMF has also been measured at higher redshifts, up to $z\sim5$, using 
photometric redshifts derived from multi-waveband imaging surveys (e.g., 
\citealt{dickinson03}; \citealt{fontana03}; \citealt{drory04}; 
\citealt{fontana04}; \citealt{conselice05} \citealt{drory05}; 
\citealt{fontana06}; \citealt{elsner08}; \citealt{perez08}). The general 
consensus is that at $z>1$ the stellar mass assembly proceeds much more 
quickly than at lower redshifts. The density of massive galaxies appears 
to have strongly evolved in the redshift range $1.5<z<3$ (e.g., 
\citealt{fontana06}), with roughly 40\% of the local stellar mass density 
assembled between $z=4$ and $z=1$ (e.g., \citealt{drory05}; \citealt{perez08}).

However, at redshift $z\gtrsim 2$, the derived SMFs are not in very good 
agreement (e.g., \citealt{elsner08}), possibly caused by differences in 
modeling techniques, field-to-field variations, and/or systematic differences 
in photometric redshift estimate errors. The cause of these differences is 
difficult to isolate, partly because most existing studies are characterized 
by either wide but very shallow data (e.g., \citealt{drory04}), or by deep 
data over relatively small areas (e.g., \citealt{dickinson03}; 
\citealt{fontana03}; \citealt{conselice05}; \citealt{drory05}; 
\citealt{fontana06}; \citealt{elsner08}). Therefore, they either only probe 
the high-mass end of the high-z SMF, or they survey volumes of the universe 
small enough that cosmic variance constitutes a significant, if not dominant, 
source of uncertainties. Only the IRAC-selected sample constructed by 
\citet{perez08} provides relatively deep data over three fields for a total 
of $\sim$660~arcmin$^2$. Furthermore, a comprehensive analysis of the errors 
affecting the derived SMFs at high redshift is still missing in the 
literature. Often, only Poisson errors are considered. Uncertainties from 
photometric redshift errors are seldomly included, and cosmic variance is 
never included (which, as we will show, is a very large, if not dominant, 
component of random errors even with relatively large surveyed area of 
$\sim$600~arcmin$^2$). Perhaps even more important is the missing analysis 
of systematic uncertainties in the derived SMF caused by the assumptions in 
the SED modeling to estimate the stellar masses. For example, a satisfactory 
agreement between different stellar population synthesis models has yet to 
be achieved, and there is evidence for evolution with redshift of some of 
the relevant parameters, such as the IMF (e.g., \citealt{dave08}; 
\citealt{vandokkum08}; \citealt{wilkins08}).
%Last, but not 
%least, disagreements among the measured SMFs can arise from the different 
%assumptions in the modeling of the observed spectral energy distributions 
%(SEDs) to estimate the stellar masses. In fact, assumptions have to be made 
%regarding the initial stellar mass function (IMF), the metallicity, the 
%extinction law, and the stellar population synthesis model, since none of 
%these can be constrained with broad-band photometry alone (e.g., 
%\citealt{muzzin08}).

In this paper we take advantage of the very high-quality data from the 
optical to the mid-infrared (MIR) available over the deep NIR MUSYC, the 
GOODS-CDFS, and the FIRES surveys to measure the evolution of the SMF of 
galaxies from $z=4.0$ to $z=1.3$ and to provide the first comprehensive 
analysis of its random and systematics uncertainties. The composite 
$K$-selected sample, constructed from a total surveyed area of 
$\sim$600~arcmin$^2$ and probing as deep as 
$K^{\rm tot}_{\rm S}(5~\sigma)=25.6$ , is unique in that it combines data from 
surveys with a large range of depths and areas in a self-consistent way. 
Crucially, the data handling, creation of catalogs, and determination of 
photometric redshifts and masses were all done with the same methods and 
techniques. The composite sample therefore self-consistently combines the 
advantages of deep, pencil beam surveys with those of shallow, wide surveys, 
allowing us to: 1) minimize the uncertainties due to cosmic variance, and 
empirically quantify its contribution to the total error budget by exploiting 
the large number of independent field of views; 2) simultaneously probe the 
high-mass end and the low-mass end of the SMF with good statistics; and 3) 
empirically derive the redshift-dependent completeness limits by 
exploiting the different depths of the three surveys.

This paper is structured as follows. In \S~\ref{sec-kss} we present the 
composite $K$-selected sample used to measure the SMF of galaxies at 
$1.3 \leq z < 4.0$; in \S~\ref{sec-sed} we describe the approach adopted 
to estimate stellar masses and the default set of SED-modeling assumptions.
The methods used to derive the SMF (the $1/V_{\rm max}$ and the 
maximum-likelihood methods) and the approach used to estimate the 
completeness limits in stellar mass are presented in \S~\ref{sec-mf}, 
as well as the SMFs of galaxies at $1.3 \leq z < 2.0$, $2.0 \leq z < 3.0$, 
and $3.0 \leq z < 4.0$. A detailed and comprehensive analysis of the 
random and systematic uncertainties of the derived SMFs is presented 
in  \S~\ref{sec-errors}, while the evolution of the stellar mass densities 
is presented in \S~\ref{sec-densities}. A comparison with the predictions 
from the latest generation of galaxy formation models is presented in 
\S~\ref{sec-comp_models}. Our results are summarized in \S~\ref{sec-concl}.
We assume $\Omega_{\rm M}=0.3$, $\Omega_{\rm \Lambda}=0.7$, and 
$H_{\rm 0}=70$~km~s$^{-1}$~Mpc$^{-1}$ throughout the paper. All magnitudes 
are on the AB system.

%============================================================================

\section{THE $K$-SELECTED SAMPLE}\label{sec-kss}

\subsection{Data}

The data set we have used to estimate the SMF consists of a 
composite $K$-selected sample of galaxies built from three 
deep multi-wavelength surveys, all having very high-quality 
optical to mid-IR photometry: the ``ultra-deep'' Faint InfraRed 
Extragalactic Survey (FIRES; \citealt{franx03}), the Great 
Observatories Origins Deep Survey (GOODS; \citealt{giavalisco04}; 
Chandra Deep Field South [CDFS]), and the MUlti-wavelength Survey 
by Yale-Chile (MUSYC; \citealt{gawiser06}). Photometric catalogs 
were created for all fields in the same way, following the 
procedures of \citet{labbe03}. 

%----------------------------------------------------------------------------

\subsubsection{MUSYC}

The deep NIR MUSYC survey consists of four 
$\sim 10^{\prime}\times10^{\prime}$ fields, namely, Hubble Deep 
Field-South 1 and 2 (HDFS-1, HDFS-2, hereafter), the SDSS-1030 
field, and the CW-1255 field, observed with the ISPI 
camera at the Cerro Tololo Inter-American Observatory (CTIO) 
Blanco 4~m telescope, for a total surveyed area of 
$\sim430$~arcmin$^{2}$ ($\sim416$~arcmin$^{2}$ not overlapping). 
A complete description of the deep NIR MUSYC observations, 
reduction procedures, and the construction of the $K$-selected 
catalog with $U$-to-$K$ photometry is presented in \citet{quadri07}.

We added deep {\it Spitzer}-IRAC 3.6-8.0~$\mu$m imaging to the NIR 
MUSYC survey. The IRAC data over the HDFS-1 are part of the GTO-214 
program (P.I.: Fazio), while the IRAC data over the other three fields 
come from the GO-30873 program (P.I.: Labb\'e). The average total 
limiting magnitudes of the IRAC images are $\sim$24.5, 24.2, 22.4, and 
22.3 (3~$\sigma$, AB~magnitude) in the 3.6, 4.5, 5.8, and 8.0~$\mu$m 
bands, respectively.\footnote{For the MUSYC survey, less than 4\% of the 
$K$-selected sources have IRAC $S/N<3$ in the 3.6~$\mu$m band; for the 
FIRES and FIREWORKS samples, which are significantly deeper than the 
MUSYC sample, $\sim$8\% of the galaxies have IRAC $S/N<3$.} The IRAC data, 
reduction, and photometry are described in detail in 
Appendix~\ref{app-iracdata}. 

The $K$-selected catalogs with IRAC photometry included is publicly 
available at \url{http://www.astro.yale.edu/musyc}. The SDSS-1030, 
CW-1255, HDFS-1, and HDFS-2 catalogs are $K_{\rm S}$ band-limited multicolor 
source catalogs down to $K^{\rm tot}_{\rm S}=$23.6, 23.4, 23.7, and 23.2, for 
a total of 3273, 2445, 2996, and 2118 sources, over fields of $\sim109$, 
$\sim105$, $\sim109$, $\sim106$~arcmin$^{2}$, respectively. All four fields 
were exposed in 14 different bands, $U$, $B$, $V$, $R$, $I$, $z$, $J$, $H$, 
$K$, and the four IRAC channels. The SDSS-1030, CW-1255, HDFS-1, and HDFS-2 
$K$-selected catalogs have 90\% completeness levels at 
$K^{\rm tot}_{\rm S}=$23.2, 22.8, 23.0 and 22.7, respectively. The final 
catalogs used in the construction of the composite sample have 2825, 2197, 
2266, and 1749 objects brighter than the 90\% completeness in the $K_{\rm S}$ 
band, over an effective area of 98.2, 91.0, 97.6, and 85.9 arcmin$^{2}$, 
respectively, for a total of 9037 sources over 372.7~arcmin$^{2}$.

%----------------------------------------------------------------------------

\subsubsection{FIRES}

FIRES consists of two fields, namely, the Hubble Deep Field-South 
proper (HDF-S) and the field around MS~1054-03, a foreground 
cluster at $z=0.83$. A complete description of the FIRES 
observations, reduction procedures, and the construction of 
photometric catalogs is presented in detail in \citet{labbe03} 
and \citet{forster06} for HDF-S and MS~1054-03 (hereafter HDFS 
and MS-1054, respectively). Both $K_{\rm S}$-selected catalogs 
were later augmented with {\it Spitzer}-IRAC data (\citealt{wuyts07}; 
\citealt{toft07}). The  HDFS catalog has 833 sources down to 
$K^{\rm tot}_{\rm S}=$26.0 over an area of $2.5^{\prime}\times2.5^{\prime}$. 
The MS-1054 catalog has 1858 sources down to $K^{\rm tot}_{\rm S}=$25.0 over 
an area of $5.5^{\prime}\times5.3^{\prime}$. The HDFS field was 
exposed in the WFPC2 $U_{\rm 300}$, $B_{\rm 450}$, $V_{\rm 606}$, 
$I_{\rm 814}$ pass-bands, the ISAAC $J_{\rm S}$, $H$, and $K_{\rm S}$ 
bands, and the four IRAC channels. The MS-1054 $K_{\rm S}$-selected 
catalog comprises FORS1 $U$, $B$, $V$, WFPC2 $V_{\rm 606}$ and 
$I_{\rm 814}$, ISAAC $J$, $H$, and $K_{\rm S}$, and IRAC 
3.6~$\mu$m - 8.0~$\mu$m photometry. The HDFS and MS-1054 catalogs have 
90\% completeness levels at $K^{\rm tot}_{\rm S}=$25.5 and 24.1, 
respectively. The final HDFS and MS-1054 catalogs used in the construction 
of the composite sample have 715 and 1547 objects brighter than the 
90\% completeness in the $K_{\rm S}$ band, over an effective area of 
4.5 and 21.0 arcmin$^{2}$, respectively.

%----------------------------------------------------------------------------

\subsubsection{FIREWORKS-CDFS}

In this work, we adopted the $K_{\rm S}$-selected catalog (dubbed 
FIREWORKS) of the CDFS field constructed based on the publicly available 
GOODS-CDFS data by \citet{wuyts08}. The photometry was performed in an 
identical way to that of the FIRES fields, and the included passbands 
are the ACS $B_{\rm 435}$, $V_{\rm 606}$, $i_{\rm 775}$, and 
$z_{\rm 850}$ bands, the WFI $U_{\rm 38}$, $B$, $V$, $R$, and $I$ bands, 
the ISAAC $J$, $H$, and $K_{\rm S}$ bands, and the 4 IRAC channels. 
The $K_{\rm S}$-selected catalog comprises 6308 objects down to 
$K^{\rm tot}_{\rm S}=$24.6 over a total surveyed area of 138~arcmin$^{2}$; 
the variation in exposure time and observing conditions between the 
different ISAAC pointings lead to an in-homogeneous depth over the whole 
GOODS-CDFS field (hereafter, CDFS). The final CDFS catalog used in the 
construction of the composite sample comprises 3559 
objects brighter than the 90\% completeness level 
($K^{\rm tot}_{\rm S}=23.7$), over an effective area of 113~arcmin$^{2}$ 
with coverage in all bands.

%----------------------------------------------------------------------------

\subsection{Photometric redshifts} \label{sec-zphot}

The fraction of $K$-selected galaxies with spectroscopic redshifts is 
$\sim$10\% (6\%) with $z_{\rm spec}>0$ ($z_{\rm spec} \geqslant 1.3$) 
in the composite $K$-selected sample. Consequently, we must rely primarily 
on photometric redshift estimates. Photometric redshifts $z_{\rm phot}$ 
for all galaxies were derived using the EAZY photometric redshift code 
\citep{brammer08}. EAZY fits the observed SED of each galaxy with a 
non-negative linear combination of galaxy templates. The template set 
used in this work is the default EAZY template set {\it eazy\_v1.0}.
\footnote{The template set needs to be large enough that it spans the 
broad range of multi-band galaxy colors and small enough that the color 
and redshift degeneracies are kept to a minimum (e.g., \citealt{benitez00}). 
The default template set used in this work to estimate photometric redshifts 
was carefully constructed and tested in \citet{brammer08}. It has been 
shown to satisfy the requirements for a satisfactory template sets, 
providing significantly reduced systematic effects and smaller scatter in 
the $z_{\rm phot}$ vs $z_{\rm spec}$ at all redshifts.} This template set was 
constructed from a large number of P\'EGASE models 
\citep{fioc97}. It consists of 5 principal component templates that span 
the colors of galaxies in the semi-analytic model by \citet{delucia07}, 
plus an additional template representing a young (50~Myr) and heavily 
obscured ($A_{\rm V}$=2.75) stellar population to account for the existence 
of dustier galaxies than present in the semi-analytic model. The default 
template error function ({\it TEMPLATE\_ERROR.eazy\_v1.0}) was applied to 
down-weight the rest-frame UV and rest-frame NIR during the fitting 
procedure. The $K$-band magnitude was used as a prior in constructing 
the redshift probability distribution for each galaxy, and the value 
$z_{\rm pm}$ (the redshift marginalized over the total probability 
distribution) was adopted as the best estimate of the galaxy redshift.

\begin{figure}
\epsscale{1}
\plotone{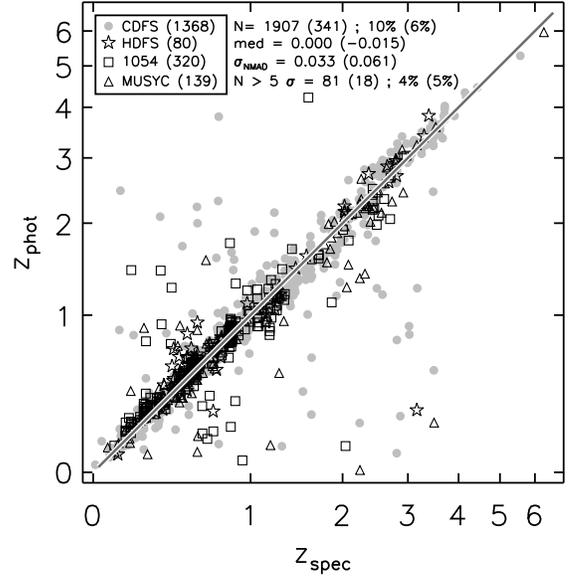}
\caption{Spectroscopic vs. photometric redshifts for the deep NIR 
MUSYC ({\it open triangles}), the CDFS FIREWORKS ({\it gray filled 
circles}), the MS-1054 ({\it open squares}) and the HDFS ({\it open 
stars}) FIRES surveys, shown on a ``pseudo-log'' scale. The total 
number N of $z_{\rm spec}$, the median in $\Delta z/(1+z_{\rm spec})$, 
the $\sigma_{\rm NMAD}$, and the number of catastrophic outliers is 
specified (values in parenthesis refer to galaxies with 
$z_{\rm spec}\geqslant1.3$). The comparison between $z_{\rm spec}$ 
and $z_{\rm phot}$ is extremely good, both at low and at high 
redshifts. \label{fig-zphot_zspec}}
\end{figure}

Figure~\ref{fig-zphot_zspec} shows the comparison of $z_{\rm phot}$ 
versus $z_{\rm spec}$ for the $K$-selected samples. For the full sample, 
the median in $\Delta z/(1+z_{\rm spec})$, with 
$\Delta z = z_{\rm phot} - z_{\rm spec}$ is 0.000, with the 
normalized median absolute deviation $\sigma_{\rm NMAD}\footnote{The 
normalized median absolute deviation $\sigma_{\rm NMAD}$, defined as 
$1.48 \times median[|(\Delta z - median(\Delta z))/(1+z_{\rm spec})|]$, 
is equal to the standard deviation for a Gaussian distribution, and 
it is less sensitive to outliers than the usual definition of the 
standard deviation (e.g. \citealt{ilbert06})}=$0.033. The fraction of 
catastrophic outliers, here defined as galaxies with 
$\Delta z / (1+z_{\rm spec}) > 5 \sigma_{\rm NMAD}$, is 4\%.
For galaxies with $z_{\rm spec}>1.3$, the median in 
$\Delta z/(1+z_{\rm spec})$ is -0.015, with $\sigma_{\rm NMAD}=0.061$, 
and the fraction of catastrophic outliers is 5\% (9\% if 
$\sigma_{\rm NMAD}=0.033$ corresponding to the entire sample is used in 
the definition of catastrophic outliers).

Restricting the comparison between $z_{\rm phot}$ and $z_{\rm spec}$ 
for the MUSYC sample alone gives similarly good results. We note that 
the estimate of the photometric redshifts in MUSYC has improved with 
respect to previously published works, due to the use of EAZY and the 
inclusion of the IRAC data.

The effects of photometric redshift errors (both random and systematic) 
on the derived SMFs and stellar mass densities are quantified and 
discussed in \S~\ref{sec-photozerrs}. Specifically, the sensitivity of 
the derived SMFs to the choice of template set is tested in 
\S~\ref{sec-prse}, and its contribution included in the total error budget.

%----------------------------------------------------------------------------

\subsection{The composite $K$-selected sample} \label{sec-compKsel}

Stars in all $K$-selected catalogs were identified by spectroscopy, 
by fitting the object SEDs with stellar templates from 
\citet{hauschildt99} and/or inspecting their morphologies, as in 
\citet{rudnick03}. On average, approximately 12\% of all objects 
were classified as stars.

We constructed a composite $K$-selected sample of high-redshift 
($1.3 \le z < 4.0$) galaxies to be used in the derivation of 
SMFs of galaxies in \S~\ref{sec-mf}. The large surveyed area of 
the deep NIR MUSYC and FIREWORKS-CDFS surveys allows us to probe 
the high-mass end of the SMF with good statistics, as well as 
empirically quantify uncertainties due to field-to-field variations. 
The very deep FIRES survey is critical in allowing us to constrain 
the low-mass end of the SMF. Finally, the FIREWORKS-CDFS survey 
bridges the two slightly overlapping regimes probed by the MUSYC 
and the FIRES surveys. The final composite sample includes 2014, 
830, and 213 $K$-selected galaxies in the three targeted redshift 
intervals $1.3\leq z<2.0$, $2.0\leq z<3.0$, and $3.0\leq z<4.0$, 
respectively, for a total of 3057 galaxies over an effective area 
of 511.2~arcmin$^{2}$ with $K_{\rm S}^{\rm tot}<25.5$ at $1.3\leq z<4.0$. 
Of these, $\sim$6\% have spectroscopic redshifts. 

%============================================================================

\section{SED MODELING: DEFAULT ASSUMPTIONS} \label{sec-sed}

Stellar masses for the HDFS, the MS-1054, and the CDFS fields 
were derived in F\"orster Schreiber et al. (in prep.), following 
the procedure described in \citet{wuyts07}. We adopted the same 
procedure to derive the stellar masses of the galaxies in the MUSYC 
HDFS-1, HDFS-2, SDSS-1030, and CW-1255 fields. In the following we 
describe our default assumptions, 
{(\it BC03,Z$_{\sun}$,Kroupa,Calzetti)}, to perform SED modeling.

We have generated stellar population synthesis models with the 
evolutionary synthesis code developed by \citet{bruzual03} (BC03). 
We selected the ``Padova~1994'' evolutionary tracks \citep{fagotto94}, 
which are preferred by Bruzual \& Charlot over the more recent 
``Padova 2000'' tracks because the latter may be less reliable and 
predict a hotter red giant branch leading to worse agreement with 
observed galaxy colors. The solar metallicity set of tracks was used. 
We fitted the BC03 templates to the observed optical-to-8~$\mu$m SED with 
the HYPERZ stellar population fitting code, version~1.1 \citep{bolzonella00}. 
We allowed the following star formation histories: a single stellar 
population (SSP) without dust, a constant star formation history (CSF) 
with dust, and an exponentially declining star formation history with 
an e-folding timescale of 300~Myr ($\tau_{\rm 300}$) with dust. 
The allowed $A_{\rm V}$ values ranged from 0 to 4 in step of 0.2~mag, 
and we used the attenuation law of \citet{calzetti00}, derived 
empirically from observations of local UV-bright starburst galaxies 
under the formalism of a foreground screen of obscuring dust. We 
constrained the time since the onset of star formation to lie between 
50~Myr and the age of the universe at the respective redshift. Finally, 
we scaled from a \citet{salpeter55} IMF with lower and upper mass 
cutoffs of 0.1~M$_{\sun}$ and 100~M$_{\sun}$ to a pseudo-\citet{kroupa01} 
IMF by dividing the stellar masses by a factor of 1.6. Therefore, the 
adopted default SED-modeling assumptions can be summarized with the 
combination ({\it BC03,Z$_{\sun}$,Kroupa,Calzetti}), where the first, 
second, third, and fourth elements are the adopted stellar population 
synthesis model, the metallicity, the IMF, and the extinction law, 
respectively. Note that we have used the mass of living stars plus stellar 
remnants, $M_{\rm star}$, instead of the total mass of stars formed, 
$M_{\rm tot}$. $M_{\rm tot}$ is the integral of the SFR and corresponds to 
the total gas mass consumed since the onset of star formation, whereas 
$M_{\rm star}$ is computed by subtracting from $M_{\rm tot}$ the mass 
returned to the ISM by evolved stars via stellar winds and supernova 
explosions (for details see \citealt{bruzual03}).

\begin{figure}
\epsscale{1}
\plotone{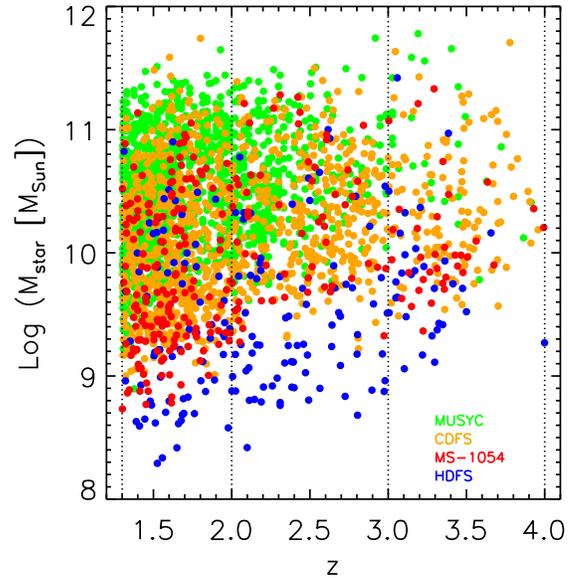}
\caption{Stellar mass versus redshift for FIRES HDFS (red filled 
circles), FIRES MS-1054 (blue open squares), FIREWORKS CDFS (orange 
open triangles), and MUSYC (green filled circles). 
\label{fig-mstar_z}}
\end{figure}

In Figure~\ref{fig-mstar_z} we show the stellar mass $M_{\rm star}$ 
versus the redshift for the $K$-selected composite sample in the 
targeted redshift range $1.3\leq z<4.0$.

Whereas ({\it BC03,Z$_{\sun}$,Kroupa,Calzetti}) represents our 
default SED-modeling assumptions, we have considered different 
stellar population synthesis models, metallicities, IMFs, and 
extinction curves to derive stellar masses. These perturbations 
on the default set of SED-modeling assumptions are described 
in detail in \S~\ref{sec-sedmodsys}, together with the resulting 
systematics effects on the derived SMFs. This has generally not 
been addressed in previous studies.

One possible limitation of our approach to derive stellar masses 
in our sample is contamination by obscured AGNs, as they could 
contribute to the emission in the rest-frame NIR (rest-frame 
wavelengths longer than 2~$\mu$m). Several recent studies (e.g., 
\citealt{papovich06}; \citealt{kriek07}; \citealt{daddi07}) have 
suggested that the fraction of obscured AGNs increases with redshift 
and stellar mass. According to \citet{kriek07}, the fraction is about 
$\sim$20\% for massive galaxies at $2<z<2.7$. To estimate the robustness 
of our estimated stellar masses in potential obscured AGNs, we have 
re-estimated the stellar masses of the galaxies with 8~$\mu$m excess 
(with respect to the best fit stellar population model) without including 
the two reddest IRAC channels in the SED-modeling (which probe the 
rest-frame NIR at $z>1.3$). The median difference of the stellar masses 
estimated with and without the 5.8 and 8.0~$\mu$m IRAC bands for 8~$\mu$m 
excess sources is $\sim$0\%, while the mean difference is $\sim6$\%, 
negligible with respect to the typical random error ($\gtrsim$30\%; e.g., 
\citealt{wuyts07}; \citealt{muzzin08}). We therefore conclude that our 
stellar mass estimates are robust even for galaxies which likely harbor an 
AGNs.

%============================================================================

\section{THE STELLAR MASS FUNCTION} \label{sec-mf}

In this section we describe a new technique used to properly 
quantify the completeness in stellar mass of the $K$-selected 
sample used in the derivation of the SMFs (\S~\ref{sec-compl}), 
and we discuss the two independent methods adopted to derived 
the SMFs of galaxies, namely an extended version of the 
$1/V_{\rm max}$ method (\S~\ref{sec-1vmaxmeth}) and the 
maximum-likelihood method (\S~\ref{sec-stymeth}). 

The uncertainties on the derived SMFs due to photometric 
redshifts errors, cosmic variance, and different SED-modeling 
assumptions are quantified and discussed in \S~\ref{sec-photozerrs}, 
\S~\ref{sec-sampvar}, and \S~\ref{sec-sedmodsys}, respectively.

%----------------------------------------------------------------------------

\subsection{Completeness in stellar mass} \label{sec-compl}

One of the most critical steps in properly deriving the SMF of 
galaxies is to understand the completeness in stellar mass as 
function of redshift of the sample. Determining the completeness 
for a flux-limited sample is particularly challenging, since there 
is not a sharp limit in stellar mass corresponding to the sharp 
limit in flux. This is a direct consequence of the fact that, 
at any given luminosity, galaxies exhibit a range in mass-to-light 
ratios ($M/L$). Because of this, galaxies that are not selected 
because their observed flux is below the detection limit can still 
have stellar masses well within the range of interest if their $M/L$ 
ratios are large enough. For the same reason, galaxies with small 
stellar masses are observed in even relatively shallow surveys if 
they have small $M/L$ ratios. 

\begin{figure*}
\epsscale{1.}
\plottwo{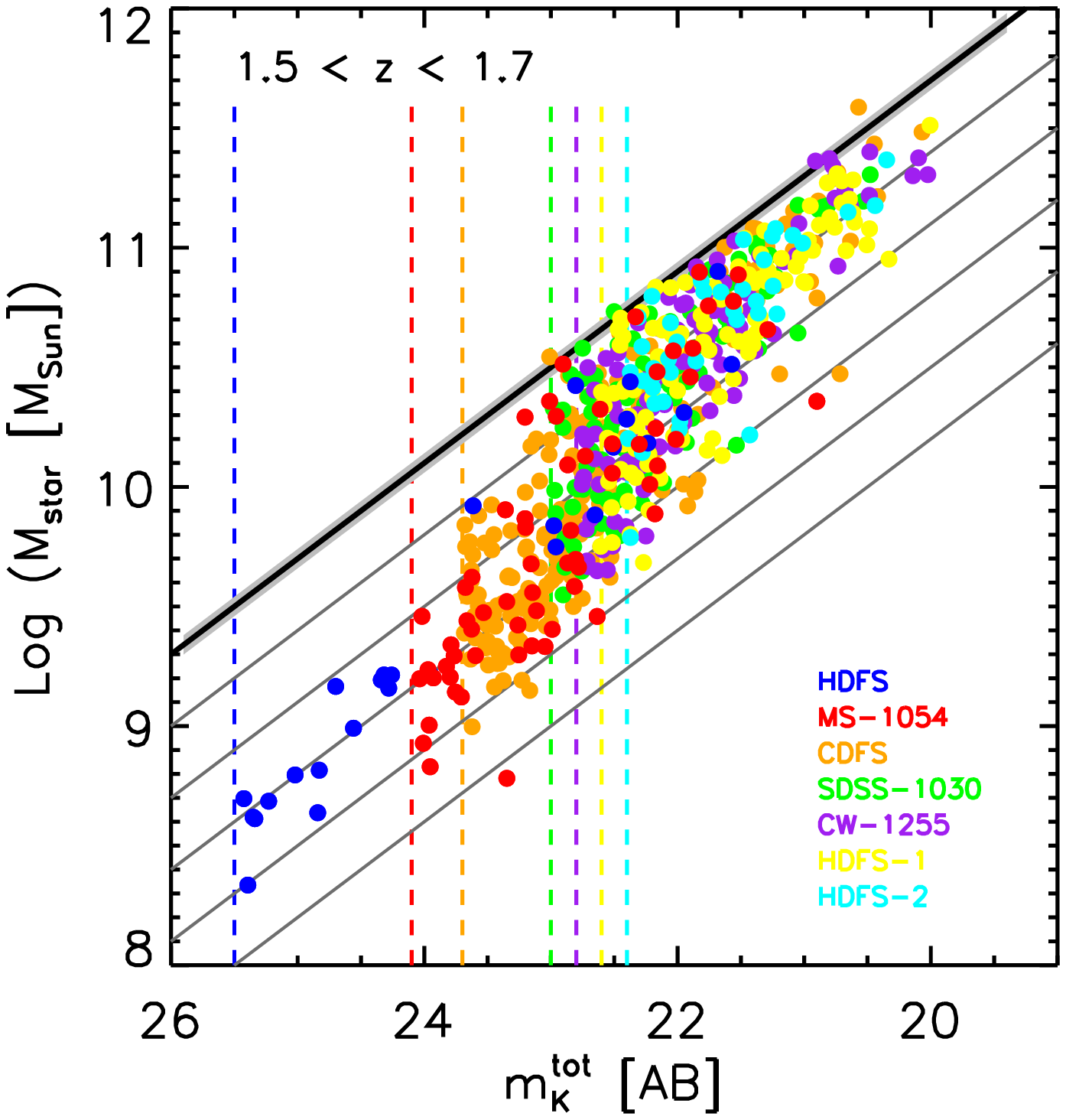}{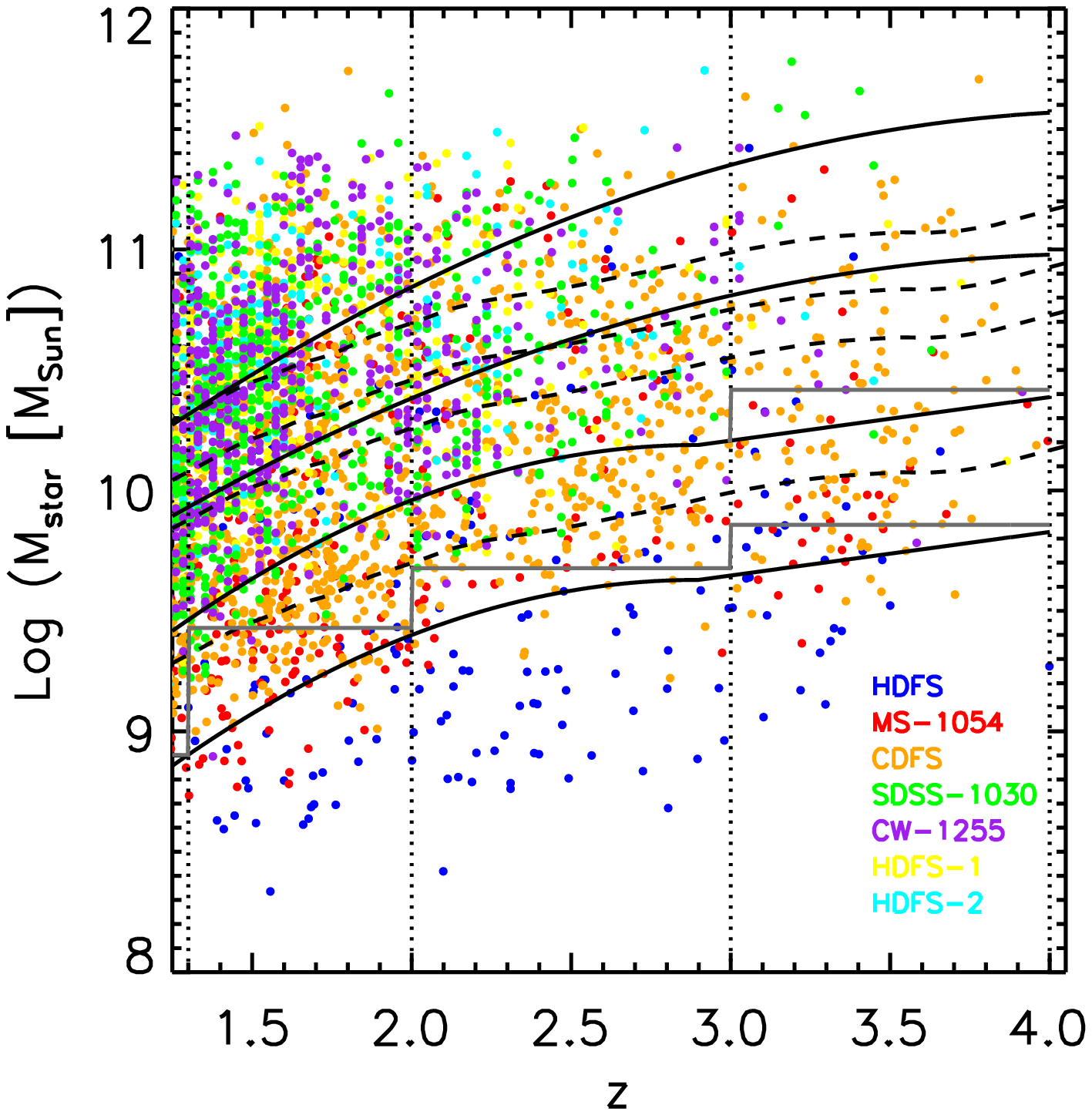}
\caption{{\it Left panel:} Stellar mass $M_{\rm star}$ versus 
observed total $K$-band magnitude for the galaxies at redshift 
$1.5<z<1.7$ in the $K$-selected sample. The thick black line 
represents the stellar mass of a single stellar population 
formed at $z\sim10$ and with no dust as a function of the 
observed $K$-band magnitude; the gray shaded region around it 
represents the scatter due to the width of the considered 
redshift interval. The gray curves represent $M/L$ ratios $\sim$2, 
4, 8, 16, and 32 times smaller than that of the single stellar 
population formed at $z\sim10$. The vertical dashed lines 
represent the $K$-band completeness limits, from left to right, 
of the HDFS, MS-1054, CDFS, SDSS-1030, CW-1255, HDFS-1, and HDFS-2 
samples. 
The largest range in $M/L$ ratios is observed at intermediate 
luminosities, while the range in $M/L$ is much narrower both at 
bright and faint luminosities. {\it Right panel:} Empirically-derived 
completeness limits in stellar mass as a function of redshift for 
our $K$-selected samples (the HDFS-1, HDFS-2, and CW-1255 samples are 
omitted for clarity), plotted as solid black curves (from top to 
bottom, SDSS-1030, CDFS, MS-1054, and HDFS). For comparison, the dashed 
curves represent the SSP-derived completeness limits. The gray solid lines 
represent the adopted conservative completeness limit at $3.0 \leq z<4.0$ 
for the MS-1054 sample, and at all redshifts for the HDFS sample to ensure 
robust derivation of the SMFs. Note the differences, as function of 
redshift and stellar mass, between the empirically- and SSP-derived 
completeness limits.
\label{fig-seccompl1}}
\end{figure*}

This is obvious from the left panel of Figure~\ref{fig-seccompl1}, where 
the stellar 
mass $M_{\rm star}$ is plotted versus the observed $K$-band magnitude 
for the $K$-selected galaxies in the narrow redshift range $1.5<z<1.7$ 
(because of the narrow redshift range, the observed $K$-band magnitude 
is approximately equivalent to the rest-frame luminosity). At a given 
luminosity, there is a range of stellar masses. The observed range in 
$M/L$ ratios at $1.5<z<1.7$ is largest for galaxies with intermediate 
luminosities and narrower for bright and for faint galaxies. 
For bright galaxies, the range in $M/L$ ratio is smaller due to the 
fact that the bright end of the luminosity function is dominated by 
red galaxies with generally high $M/L$ ratios. At the faint end, most 
galaxies are blue (e.g., \citealt{giallongo05}; \citealt{zucca06}; 
\citealt{marchesini07a}) and have low $M/L$ ratios, resulting in a narrow 
range in $M/L$ ratios. The average $M/L$ ratio for galaxies in the 
redshift range $1.5<z<1.7$ at the faint end is a factor of $\sim5$ 
smaller than those at the bright end.

The distribution of galaxies in the left panel of Figure~\ref{fig-seccompl1} 
is a function 
of redshift. First, as higher redshifts are probed, faint galaxies are 
progressively missed due to the limiting flux of the sample. 
Consequently, at higher redshifts the sample becomes progressively 
dominated by galaxies with larger $M/L$ ratios. Second, the galaxy 
properties will evolve with redshift, as well as their relative 
contributions in different luminosity ranges. For example, the bright 
end of the luminosity function is shown to be dominated by red galaxies 
at $z\sim2.2$, whereas at $z\sim3$ the contribution of red and blue 
galaxies to the observed number densities appears to be similar 
\citep{marchesini07a}. Consequently, the observed range of $M/L$ ratios 
at the bright end would increase going to higher redshifts.

Most of the previous works deriving the SMFs of galaxies (e.g., 
\citealt{drory05}; \citealt{fontana06};  \citealt{perez08}; see also 
Appendix~\ref{app-comp} for a detailed discussion of previously 
published works and their assumed completeness limits in stellar mass) 
have estimated their completeness in stellar mass based on the 
maximal stellar mass allowed for a galaxy at the flux limit of the sample. 
This maximal mass is typically taken to be the stellar mass of a passively 
evolving population with no dust extinction formed in a single burst at 
very high ($z\sim10$) redshift, scaled to match the limiting flux 
(SSP-derived completeness limit, hereafter). This approach, although easy 
to implement, is conceptually wrong, and potentially affected by several 
caveats. 

For example, the faint-end of the galaxy luminosity function is dominated 
by intrinsically blue galaxies (e.g., \citealt{zucca06}, 
\citealt{marchesini07a}), characterized by small $M/L$ ratios (see also 
left panel of Figure~\ref{fig-seccompl1}). If an SSP-derived completeness 
limit (shown by the dashed curves in the right panel of 
Figure~\ref{fig-seccompl1}) were to be used in deriving the completeness 
in stellar mass of a deep survey, the sample would be cut at conservatively 
too high stellar masses. More importantly, the derived number density at the 
low-mass end would be significantly affected due to miscalculation of 
$V_{\rm max}$ in the $1/V_{\rm max}$ method. Another complication is caused 
by the ubiquitous presence of dust in real galaxies. Dust extinction can 
reduce the completeness derived from no-extinction SSPs since, to first 
order, it moves the dashed curves in the right panel of 
Figure~\ref{fig-seccompl1} upwards. While at low redshift massive galaxies 
are usually characterized by passively evolving populations with no or little 
dust extinction, this does not seem to be the case at high redshifts, where 
significant amount of extinction ($A_{\rm V}\sim 1-2$) can be present even 
in massive galaxies (e.g., \citealt{blain99b}; \citealt{muzzin08}). 

In order to avoid these problems, we have used a different 
approach to estimate the redshift-dependent completeness limit in 
stellar mass. Our approach exploits the availability of several samples 
with different depths, and the completeness of a sample is estimated 
empirically from the available deeper samples. To estimate the 
redshift-dependent stellar mass completeness limit of one of the considered 
samples, we first selected galaxies belonging to the available deeper 
samples. Then, we scaled their fluxes and stellar masses to match the 
$K$-band completeness limit of the sample we want to derive the completeness 
limit in stellar mass for. The upper envelope of points in the 
$(M_{\rm star,scaled}-z)$ space, encompassing 95\% of the points, represents 
the most massive galaxies at the considered flux limit, and so provides a 
redshift-dependent stellar mass completeness limit for the considered sample. 
This method is illustrated in detail for the SDSS-1030 sample in 
Appendix~\ref{app-compl}. By repeating this procedure, we derived the 
redshift-dependent completeness limits in stellar mass for all samples. 
These limits are shown in the right panel of Figure~\ref{fig-seccompl1}, 
along with SSP-derived completeness limits. 

Interestingly, the empirically-derived completeness limit is similar to the 
SSP-derived completeness only for the CDFS sample. For shallower samples 
(e.g., SDSS-1030), the sample is actually less complete than what would be 
estimated with the SSP-derived completeness over the redshift range 
$2.5 \lesssim z \lesssim 4$. Conversely, the SSP-derived completeness is 
too conservative (by $\sim$0.3~dex) for deeper samples, such as the FIRES 
MS-1054 sample.

For the MS-1054 sample, the empirically-derived completeness at 
$3.0 \leq z<4.0$ is poorly derived due to the small number of galaxies in 
the HDFS sample in this redshift range. Therefore, we have conservatively 
assumed its largest value throughout the entire redshift range 
$3.0 \leq z<4.0$. For the HDFS sample, there is no deeper survey that can 
be used to empirically derive its completeness in stellar mass. The 
completeness of HDFS was therefore estimated by scaling the 
empirically-derived completeness limit of the MS-1054 sample to match the 
$K$-band 90\% completeness flux limit of the HDFS sample, and by taking 
the largest value of the scaled completeness within each individual redshift 
interval (shown as gray solid lines in the right panel of 
Figure~\ref{fig-seccompl1}). Although this is a very conservative approach, 
it ensures correct determination of the SMF at the low-mass end.

%----------------------------------------------------------------------------

\subsection{The $1/V_{\rm max}$ method} \label{sec-1vmaxmeth}

To estimate the observed SMF for our composite sample, 
we have applied an extended version of the $1/V_{\rm max}$ 
algorithm \citep{schmidt68} as defined in \citet{avni80} so that 
several samples with different depths can be combined in one 
calculation. This method is described in detail in 
\citet{marchesini07a}, where it was used to derive the rest-frame 
optical luminosity functions of galaxies at redshift $2.0<z<3.5$ 
from a similar $K$-selected sample. The empirically-derived 
redshift-dependent completeness limits in stellar mass as derived 
in \S~\ref{sec-compl} for the individual $K$-selected samples was 
used in the calculation of $V_{\rm max}$. The Poisson error in each 
stellar mass bin was computed adopting the recipe of \citet{gehrels86}.

The $1/V_{\rm max}$ estimator has the advantages of simplicity 
and no a priori assumption of a functional form for the stellar mass 
distribution; it also yields a fully normalized solution. However, 
it can be affected by the presence of clustering in the sample, 
leading to a poor estimate of the faint-end slope of the SMF. 
Field-to-field variation represents a significant source of 
uncertainty in deep surveys, since they are characterized by small 
areas and hence small probed volumes. The contribution due to sample 
variance to the total error budget is quantified in \S~\ref{sec-sampvar}.

%----------------------------------------------------------------------------

\subsection{The maximum likelihood method} \label{sec-stymeth}

We also measured the observed SMF using the STY method \citep{sandage79}, 
which is a parametric maximum-likelihood estimator. The STY method has 
been shown to be unbiased with respect to density inhomogeneities (e.g., 
\citealt{efstathiou88}), it has well-defined asymptotic error properties 
(e.g. \citealt{kendall61}) and does not require to select bin widths. 

We have assumed that the number density $\Phi(M)$ of galaxies is described 
by a \citet{schechter76} function,
\begin{eqnarray}
\Phi (M) = (\ln{10}) \Phi^{\star} 
\big[ 10^{(M-M^{\star})(1+\alpha)} \big] \times \exp{\big[ -10^{(M-M^{\star})} \big]},
\end{eqnarray}
where $M=\log{(M_{\rm star}/M_{\sun})}$, $\alpha$ is the low mass-end 
slope, $M^{\star}=\log{(M_{\rm star}^{\star}/M_{\sun})}$ is the 
characteristic stellar mass at which the SMF exhibits a rapid change 
in the slope, and $\Phi^{\star}$ is the normalization. 

The implementation of the STY method and the method of estimating 
errors is described in detail in \citet{marchesini07a}. The best-fit 
solution is obtained by maximizing the likelihood $\Lambda$ 
with respect to the parameters $\alpha$ and $M^{\star}$. The value of 
$\Phi^{\star}$ is then obtained by imposing a normalization on the 
best-fit SMF such that the total number of observed galaxies in the 
composite sample is reproduced. The 1 and 2~$\sigma$ errors on 
$\Phi^{\star}$ are estimated from the minimum and maximum values of 
$\Phi^{\star}$ allowed by the 1 and 2~$\sigma$ confidence contours in 
the $(\alpha-M^{\star}_{\rm star})$ parameter space, respectively.

%----------------------------------------------------------------------------

\subsection{Stellar Mass Functions} \label{subsec-mf}

\begin{figure}
\epsscale{1}
\plotone{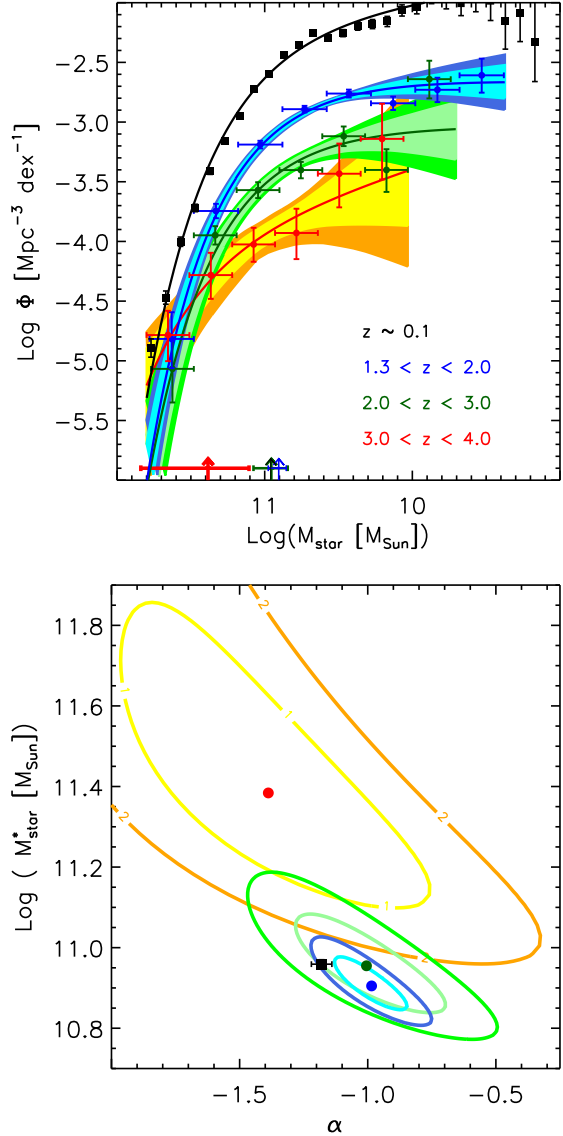}
\caption{{\it Top panel:} SMFs of galaxies at redshift 
$1.3\leq z<2.0$ (blue), $2.0\leq z<3.0$ (green), and 
$3.0\leq z<4.0$ (red). The filled symbols represent the 
SMFs derived with the $1/V_{\rm max}$ method, with error 
bars including only Poisson errors. The solid curves 
represent the SMFs derived with the maximum-likelihood 
analysis, with shaded regions representing the 1 and 
2~$\sigma$ uncertainties. The arrows show the best 
estimates of $M^{\star}_{\rm star}$ and the corresponding 
1~$\sigma$ errors derived with the maximum-likelihood 
analysis. The black solid curve and points represent 
the local ($z\sim0.1$) SMF from \citet{cole01}. 
{\it Bottom panel:} $(\alpha-M^{\star}_{\rm star})$ 
parameter space derived from the maximum-likelihood analysis. 
Filled circles are the best-fit values of $\alpha$ and 
$M^{\star}_{\rm star}$, while the curves represent their 1 
and 2~$\sigma$ contour levels; the colors are the same as 
in the top panel. The black filled square represent the 
redshift $z\sim0.1$ value from \citet{cole01}. Very little 
evolution of the shape of the SMF is observed from $z=3.0$ 
to $z=1.3$, and most of the evolution is in the characteristic 
density $\Phi^{\star}$. The shape of the SMF at $z=3.5$ is 
different, characterized by a much steeper low-mass end slope. 
The characteristic stellar mass $M^{\star}_{\rm star}$ seems 
to have evolved little, if any, from $z=2.5$ to $z\sim0.1$.
\label{fig-mfevol_onlyPoi}}
\end{figure}

Figure~\ref{fig-mfevol_onlyPoi} (top panel) shows the SMFs of galaxies 
at redshift $1.3\leq z<2.0$, $2.0 \leq z<3.0$, and $3.0\leq z<4.0$. 
Points with error bars show the SMFs derived using the $1/V_{\rm max}$ 
method. The solid curves show the SMFs derived with the maximum-likelihood 
analysis, while the shaded regions represent their 1 and 2~$\sigma$ 
uncertainties. The bottom panel of Figure~\ref{fig-mfevol_onlyPoi} shows the 
best-fit value and the 1 and 2~$\sigma$ confidence contour levels of the 
two Schechter function parameters $\alpha$ and $M^{\star}_{\rm star}$ in 
the three targeted redshift intervals. The local SMF derived by \citet{cole01} 
is also shown in Figure~\ref{fig-mfevol_onlyPoi}. The plotted uncertainties 
include Poisson errors only. The uncertainties on the derived SMFs due to 
cosmic variance, photometric redshift errors, and different SED-modeling 
assumptions are quantified and discussed in \S~\ref{sec-errors}. 

The large surveyed area allows for the determination of the high-mass 
end with unprecedented accuracy, while the depth of the FIRES survey 
allows us to constrain also the low-mass end. This is particularly 
important because of the well-known correlation between the two 
parameters $\alpha$ and $M^{\star}_{\rm star}$. 

\begin{deluxetable*}{cccccccc}
\centering
\tablecaption{SMFs derived with the $1/V_{\rm max}$ method
\label{tab-mf1}}
\tablehead{\colhead{$\log{M_{\rm star}}$} & \colhead{$\log{\Phi}$} & 
           \colhead{$\sigma$} & 
           \colhead{$\sigma_{\rm Poi}$} &
           \colhead{$\sigma_{\rm z,ran}$} & 
           \colhead{$\sigma_{\rm cv}$} &
           \colhead{$\sigma_{\rm z,sys}$} & 
           \colhead{$\sigma_{\rm sed,sys}$} \\
                    ($M_{\sun}$) & (Mpc$^{-3}$~dex$^{-1}$) &
                     & & & & &}
\startdata
$1.3 \leq z < 2.0$: &                 &                      &       &       &                       & \\
11.63 & -4.817 & $^{+0.283}_{-0.300}$ & $^{+0.224}_{-0.245}$ & 0.111 & 0.132 & $^{+0.000}_{-0.398}$  & $^{+0.342}_{-0.699}$ \\
11.33 & -3.745 & 0.168                & 0.060                & 0.029 & 0.155 & $^{+0.068}_{-0.142}$  & $^{+0.113}_{-0.470}$ \\
11.03 & -3.189 & 0.067                & 0.032                & 0.020 & 0.056 & $^{+0.000}_{-0.083}$  & $^{+0.000}_{-0.428}$ \\
10.73 & -2.892 & 0.110                & 0.028                & 0.020 & 0.105 & $^{+0.022}_{-0.041}$  & $^{+0.001}_{-0.417}$ \\
10.43 & -2.761 & 0.076                & 0.033                & 0.024 & 0.065 & $^{+0.000}_{-0.108}$  & $^{+0.052}_{-0.484}$ \\
10.13 & -2.843 & 0.106                & 0.056                & 0.039 & 0.082 & $^{+0.045}_{-0.031}$  & $^{+0.166}_{-0.272}$ \\
 9.83 & -2.730 & 0.245                & 0.102                & 0.061 & 0.214 & $^{+0.068}_{-0.014}$  & $^{+0.070}_{-0.153}$ \\
 9.53 & -2.608 & $^{+0.289}_{-0.292}$ & $^{+0.140}_{-0.145}$ & 0.135 & 0.214 & $^{+0.156}_{-0.050}$  & $^{+0.089}_{-0.164}$ \\
$2.0 \leq z < 3.0$: &                 &                      &       &       &                       & \\
11.63 & -5.067 & $^{+0.333}_{-0.356}$ & $^{+0.253}_{-0.282}$ & 0.152 & 0.156 & $^{+0.139}_{-0.000}$  & $^{+0.080}_{-0.523}$ \\
11.34 & -3.949 & 0.113                & 0.078                & 0.059 & 0.056 & $^{+0.001}_{-0.287}$  & $^{+0.041}_{-1.149}$ \\
11.05 & -3.570 & 0.103                & 0.067                & 0.045 & 0.064 & $^{+0.131}_{-0.068}$  & $^{+0.000}_{-0.493}$ \\
10.76 & -3.402 & 0.118                & 0.069                & 0.049 & 0.083 & $^{+0.107}_{-0.000}$  & $^{+0.093}_{-0.385}$ \\
10.47 & -3.119 & 0.215                & 0.082                & 0.070 & 0.186 & $^{+0.039}_{-0.100}$  & $^{+0.000}_{-0.482}$ \\
10.18 & -3.401 & $^{+0.291}_{-0.298}$ & $^{+0.174}_{-0.184}$ & 0.115 & 0.204 & $^{+0.136}_{-0.125}$  & $^{+0.271}_{-0.133}$ \\
 9.89 & -2.640 & $^{+0.298}_{-0.302}$ & $^{+0.154}_{-0.161}$ & 0.154 & 0.204 & $^{+0.142}_{-0.062}$  & $^{+0.000}_{-0.670}$ \\
$3.0 \leq z < 4.0$: &                 &                      &       &       &                       & \\
11.66 & -4.784 & $^{+0.446}_{-0.453}$ & $^{+0.203}_{-0.219}$ & 0.239 & 0.316 & $^{+0.000}_{-0.556}$  & $^{+0.096}_{-0.684}$ \\
11.37 & -4.282 & $^{+0.322}_{-0.329}$ & $^{+0.187}_{-0.199}$ & 0.126 & 0.229 & $^{+0.064}_{-0.428}$  & $^{+0.000}_{-0.527}$ \\
11.08 & -4.025 & $^{+0.217}_{-0.221}$ & $^{+0.140}_{-0.145}$ & 0.095 & 0.136 & $^{+0.146}_{-0.107}$  & $^{+0.062}_{-0.483}$ \\
10.79 & -3.929 & $^{+0.424}_{-0.431}$ & $^{+0.203}_{-0.219}$ & 0.195 & 0.316 & $^{+0.037}_{-0.000}$  & $^{+0.151}_{-0.281}$ \\
10.50 & -3.433 & $^{+0.453}_{-0.470}$ & $^{+0.253}_{-0.282}$ & 0.203 & 0.316 & $^{+0.000}_{-0.158}$  & $^{+0.000}_{-0.737}$ \\
10.21 & -3.141 & $^{+0.496}_{-0.524}$ & $^{+0.294}_{-0.340}$ & 0.244 & 0.316 & $^{+0.125}_{-0.000}$  & $^{+0.144}_{-0.406}$ \\
\enddata
\tablenotetext{}{$\sigma=(\sigma^{2}_{\rm Poi}+\sigma^{2}_{\rm cv}+\sigma^{2}_{\rm z,ran})^{1/2}$ 
is the total 1~$\sigma$ random error, including the Poisson errors 
($\sigma_{\rm Poi}$), the errors due to photometric redshift random 
uncertainties ($\sigma_{\rm z,ran}$; see \S~\ref{sec-photozerrs}), and 
the error due to cosmic variance ($\sigma_{\rm cv}$; see 
\S~\ref{sec-sampvar}); $\sigma_{\rm z,sys}$ is the systematic 
uncertainty due to different template sets in the photometric 
redshift estimate, while $\sigma_{\rm sed,sys}$ is the systematic 
uncertainty due to different SED-modeling assumptions (see 
\S~\ref{sec-photozerrs} and \ref{sec-sedmodsys}).}
\end{deluxetable*}

\begin{deluxetable*}{cccccccccc}
\centering
\tablecaption{Best-fit Schechter function parameters for the SMFs
\label{tab-mf2}}
\tablehead{\colhead{Redshift} & \colhead{$\alpha$}     & 
           \colhead{$\log{M^{\star}_{\rm star}}$}      & 
           \colhead{$\Phi^{\star}$}                    &
           \colhead{$\alpha_{\rm z,sys}$} & \colhead{$\alpha_{\rm sed,sys}$} & 
           \colhead{$(\log{M^{\star}_{\rm star}})_{\rm z,sys}$} & 
           \colhead{$(\log{M^{\star}_{\rm star}})_{\rm sed,sys}$} &
           \colhead{$\Phi^{\star}_{\rm z,sys}$} & 
	   \colhead{$\Phi^{\star}_{\rm sed,sys}$} \\
                    Range        &                     &
                    ($M_{\sun}$) & (10$^{-4}$~Mpc$^{-3}$~dex$^{-1}$) &
		                 & &  &  &  & }
\startdata
$z \sim 0.1$       & $-1.18\pm0.03$ & 
                     $10.96\pm0.01$ & 
                     $30.87\pm4.80$ & & & & & & \\
$1.3 \leq z < 2.0$ & $-0.99^{+0.13,0.23}_{-0.14,0.23}$ & 
                     $10.91^{+0.07,0.11}_{-0.05,0.09}$ & 
                     $10.17^{+1.71,3.02}_{-1.99,3.19}$ &
                     $^{+0.08}_{-0.10}$ & $^{+0.16}_{-0.31}$ & 
                     $^{+0.05}_{-0.06}$ & $^{+0.41}_{-0.17}$ &
                     $^{+0.60}_{-1.71}$ & $^{+3.61}_{-8.16}$ \\
$2.0 \leq z < 3.0$ & $-1.01^{+0.30,0.50}_{-0.27,0.46}$ & 
                     $10.96^{+0.12,0.23}_{-0.11,0.18}$ & 
                     $3.95^{+1.44,2.43}_{-1.34,2.15}$ &
                     $^{+0.03}_{-0.12}$ & $^{+0.16}_{-0.35}$ & 
                     $^{+0.03}_{-0.01}$ & $^{+0.22}_{-0.33}$ &
                     $^{+0.85}_{-1.03}$ & $^{+1.07}_{-3.20}$ \\
$3.0 \leq z < 4.0$ & $-1.39^{+0.63,1.06}_{-0.55,0.98}$ & 
                     $11.38^{+0.46,1.36}_{-0.28,0.40}$ & 
                     $0.53^{+0.81,1.32}_{-0.45,0.52}$ &
                     $^{+0.30}_{-0.20}$ & $^{+0.42}_{-0.53}$ & 
                     $^{+0.08}_{-0.25}$ & $^{+0.25}_{-0.30}$ & 
                     $^{+0.37}_{-0.16}$ & $^{+0.12}_{-0.42}$ \\
\enddata
\tablenotetext{}{The quoted errors correspond to the 1 and 
2~$\sigma$ errors estimated from the maximum-likelihood 
analysis as described in \S~\ref{sec-stymeth}. Also listed 
are the systematic uncertainties on the Schechter function 
parameters due to different SED-modeling assumptions and 
different template sets in the photometric redshift estimate 
(see \S~\ref{sec-photozerrs} and \ref{sec-sedmodsys}). The 
local ($z\sim0.1$) values are taken from \citet{cole01}.}
\end{deluxetable*}

Figure~\ref{fig-mfevol_onlyPoi} clearly shows dramatic evolution of 
the SMF, qualitatively consistent with other studies (e.g., 
\citealt{fontana06}; \citealt{perez08}). The main trend is a gradual 
decrease with redshift of the characteristic density $\Phi^{\star}$, 
rather than a change in the slope $\alpha$ or the characteristic 
stellar mass $M^{\star}_{\rm star}$. 
The density at $M_{\rm star}\sim10^{11}$ has evolved by a factor of 
$\sim20$ since $z=3.5$, a factor of $\sim8$ since $z=2.5$, and a factor 
of $\sim3.5$ since $z=1.65$. The data points shown in 
Figure~\ref{fig-mfevol_onlyPoi}, along with the best-fitting Schechter 
function parameters, are listed in Table~\ref{tab-mf1} and Table~\ref{tab-mf2}.

\begin{figure}
\epsscale{1}
\plotone{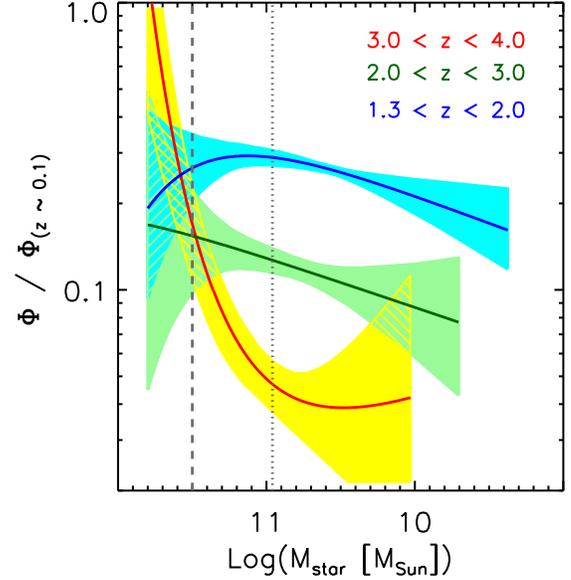}
\caption{Ratio of the high-$z$ SMFs ($\Phi$) and the local 
SMF ($\Phi_{\rm z\sim0.1}$) plotted as function of the stellar 
mass as measured from the maximum-likelihood analysis. The 
shaded regions represent the 1~$\sigma$ uncertainties. Colors 
are as in Fig.~\ref{fig-mfevol_onlyPoi}. The vertical dashed 
and dotted lines represent the value of 
$3 \times 10^{11}$~M$_{\sun}$ and the $z=0.1$ characteristic 
stellar mass, $M_{\rm star}^{\star}\sim10^{11}$~M$_{\sun}$, 
respectively. Evidence for mass-dependent evolution is present, 
with the evolution to $z\sim0.1$ being larger at the low-mass 
end and smallest for the most massive galaxies.
\label{fig-massdepevol}}
\end{figure}

We also find evidence for mass-dependent evolution. In particular, the 
data suggest a remarkable lack of evolution for the most massive galaxies, 
with $M_{\rm star} > 3 \times 10^{11}$, over the redshift range 
$1.3 \leq z < 4.0$. The average density of these galaxies is 
$1.3 \times 10^{-5}$~Mpc$^{-3}$. The differential evolution of galaxies 
with different masses is shown more clearly in Figure~\ref{fig-massdepevol}, 
where the high-redshift SMFs divided by the local SMF have been plotted 
as function of stellar mass. If the form of the mass function does not 
evolve with redshift, the curves in Figure~\ref{fig-massdepevol} would be 
constant lines as function of stellar mass. On the contrary, the observed 
evolution of the number densities is larger for less massive galaxies and 
smallest for the most massive galaxies.

%============================================================================

\section{UNCERTAINTIES} \label{sec-errors}

The results found in the previous sections are very intriguing. However, 
only Poisson errors have been considered, and, as previously noted, 
uncertainties due to photometric redshift errors (both random and 
systematic), cosmic variance, and different SED-modeling assumptions 
also affect the measurement of the high-redshift SMF.\footnote{We note 
that the results from the maximum-likelihood analysis are unbiased with 
respect to density inhomogeneities, hence not affected by cosmic variance.}
In this section we quantify the uncertainties on the 
measured SMFs due to these sources of errors, providing the first 
comprehensive analysis of random and systematic uncertainties affecting 
the high-$z$ SMFs.

\subsection{Uncertainties due to photometric redshift errors} \label{sec-photozerrs}

Studies of high-redshift galaxies largely rely on photometric redshift 
estimates. It is therefore important to understand how the photometric 
redshift uncertainties affect the derived SMFs and densities. 

\subsubsection{Photometric redshift random errors}

To quantify the uncertainties on the SMFs due to photometric redshift 
random errors we have proceeded as follows. First, for each galaxy in 
the $K$-selected sample, a set of 100 mock SEDs was created by 
perturbing each flux point according to its formal error bar. Second, 
we estimated photometric redshift $z_{\rm phot}$ in the same way as 
described in \S~\ref{sec-zphot}. Third, we fitted the mock SEDs to 
estimate stellar masses as described in \S~\ref{sec-sed}, using the 
default set of SED-modeling assumptions. Finally, we have derived 
completeness limits in stellar mass and SMFs of galaxies with the 
$1/V_{\rm max}$ and the maximum-likelihood analysis for each of the 
100 Monte Carlo realizations of the composite $K$-selected sample.
This approach naturally addresses the fact that fainter sources tend to 
be characterized by less accurate $z_{\rm phot}$ estimates due to the 
larger errors in their photometry, as well as sources characterized by 
power-law SEDs and consequently by very poorly-constrained $z_{\rm phot}$ 
estimates and very broad $z_{\rm phot}$ distributions derived from the 
Monte Carlo realizations. Moreover, the adopted Monte Carlo approach to 
estimate the uncertainties on the SMF due to the photometric redshift 
random error is to be preferred to the approach using the comparison of 
$z_{\rm phot}$ with $z_{\rm spec}$, as this comparison is strongly affected 
by the very biased and incomplete sub-sample of galaxies at $z \gtrsim1.5$ 
with available spectroscopic redshifts (see, e.g., \citealt{brammer08}).

The contribution to the total error budget of the SMFs derived using 
the $1/V_{\rm max}$ method due to photometric redshift random errors 
($\sigma_{\rm z,ran}$) was estimated by taking, for each stellar mass 
bin, the lower and upper errors on $\Phi(M)$ comprising the central 
68\% of the Monte Carlo distribution. The values of $\sigma_{\rm z,ran}$ 
for each stellar mass bin in the three targeted redshift intervals are 
listed in Table~\ref{tab-mf1}. The contribution of photometric redshift 
random uncertainties to the total error budget of the SMFs derived with 
the $1/V_{\rm max}$ method is generally smaller (although non-negligible) 
than $\sigma_{\rm Poi}$ and $\sigma_{\rm cv}$ (the error due to cosmic 
variance; see \S~\ref{sec-sampvar}), the latter dominating the random 
error budget. This is true at all redshifts. The contribution of 
$\sigma_{\rm z,ran}$ is largest (although still relatively small) for the 
largest stellar mass bin ($M\sim11.65$), which is usually populated only 
by a handful of sources, and for the SMF at redshift $3.0 \leq z < 4.0$.

The uncertainties on the SMFs derived using the maximum-likelihood 
analysis due to photometric redshift random errors appears to be 
negligible. This is due to the fact that, when the maximum-likelihood 
analysis is used to derive the SMFs, the whole stellar mass range 
contributes to the determination of the Schechter function parameters, 
significantly reducing the impact of photometric redshift random errors. 
The derived 1~$\sigma$ contour level from the maximum-likelihood analysis 
contains $\sim$95\% of the Monte Carlo realizations.\footnote{These results 
are fully consistent with the Monte Carlo simulations performed in 
\citet{marchesini07a} to address the systematic uncertainties on the 
Schechter function parameters of the high-redshift rest-frame optical 
luminosity functions due to photometric redshift random uncertainties.} 
Therefore, the errors on the Schechter function parameters due to 
photometric redshift random errors can be neglected. This is true for all 
three targeted redshift intervals.

\subsubsection{Photometric redshift systematic errors}\label{sec-prse}

In addition to random errors, systematic errors can be caused by the 
specific choice of the templates or the template error function 
used in the estimate of the photometric redshifts. To quantify 
these systematic errors, we have repeated the $z_{\rm phot}$ 
estimates using the following different combination of template 
set and template error function: 1) {\it eazy\_v1.0\_nodust} and 
{\it TE.eazy\_v1.0\_nodust}, with {\it eazy\_v1.0\_nodust} equal 
to the default set {\it eazy\_v1.0} without the dusty template, 
and {\it TE.eazy\_v1.0\_nodust} the template error function 
specifically constructed for the {\it eazy\_v1.0\_nodust} template 
set; 2) {\it eazy\_v1.0} and {\it TE.eazy\_v1.0\_nodust}; 
3) {\it br07\_default} and {\it TE.eazy\_v1.0}, with {\it br07\_default} 
the default template set of \citet{blanton07}. These three 
combinations were chosen because they resulted in 
$z_{\rm phot}-z_{\rm spec}$ comparisons of similar quality (or only 
slightly worse) as that derived in \S~\ref{sec-zphot} using the 
default EAZY template set and template error function. We decided 
not to use the {\it cww+kin}\footnote{The {\it cww+kin} template 
set comprises the empirical templates from \citet{coleman80} plus 
the ``SB1'' starburst spectrum from \citet{kinney96}; these templates 
have been extended in the UV and IR by \citet{arnouts99} 
(\url{http://www.oamp.fr/people/arnouts/LE\_PHARE.html})} and the 
{\it pegase13}\footnote{The {\it pegase13} template set contains the 
set of constant star formation rate models with additional dust 
reddening applied using the extinction curve of \citet{calzetti00} 
(see \citealt{brammer08} for details).} template sets (also distributed 
with the EAZY code), due to the significantly worse resulting 
$z_{\rm phot}-z_{\rm spec}$ comparisons.
Modeling of the observed SEDs was then performed using the new sets 
of $z_{\rm phot}$ to derived stellar masses, and the completeness limits 
in stellar mass were then re-estimated and the SMFs re-derived with both 
the $1/V_{\rm max}$ and the maximum-likelihood methods. The last three 
columns of Table~\ref{tab-appsedmod1} list the SMFs of galaxies at 
$1.3 \leq z<2.0$, $2.0 \leq z<3.0$, and $3.0 \leq z<4.0$ derived using 
the $1/V_{\rm max}$ method for each combination of template set and template 
error function. Table~\ref{tab-appsedmod2} lists the best-fit Schechter 
function parameters $\alpha$, $M^{\star}_{\rm star}$, and $\Phi^{\star}$ of 
the SMFs of galaxies at $1.3 \leq z<2.0$, $2.0 \leq z<3.0$, and 
$3.0 \leq z<4.0$ derived using the maximum-likelihood analysis for each 
combination of template set and template error function.

Systematic errors were then quantified by comparing the resulting SMFs 
with the SMFs derived using the preferred default EAZY template set and 
template error function, following the same approach (described in 
\S\ref{sec-sedmodsys}) used to quantify the systematic uncertainties due 
to different SED modeling assumptions. These systematic uncertainties, 
$\sigma_{\rm z,sys}$ for the SMFs derived with the $1/V_{\rm max}$ method, 
and $\alpha_{\rm z,sys}$, $(\log{M_{\rm star}^{\star}})_{\rm z,sys}$, 
and $\Phi^{\star}_{\rm z,sys}$ for the Schechter function parameters 
derived with the maximum-likelihood analysis, are listed in 
Table\ref{tab-mf1} and \ref{tab-mf2}, respectively. The values of 
$\sigma_{\rm z,sys}$ are generally asymmetric and larger than 
$\sigma_{\rm z,ran}$, and larger than the total 1~$\sigma$ random error 
$\sigma$ for $\sim$1/3 of the considered stellar mass bins. On the contrary, 
the systematic uncertainties on the Schechter function parameters due to 
different template sets or template error functions are always smaller than 
the 1~$\sigma$ error estimated from the maximum-likelihood analysis. We 
note that our results may be influenced by unknown systematic effects in 
the redshifts, particularly at the high-mass end. Spectroscopic redshifts, 
or photometric redshifts with very small errors and systematics (e.g., 
\citealt{vandokkum09}), are needed to confirm the shape of the mass function 
in the highest mass bins.

%----------------------------------------------------------------------------

\subsection{Uncertainties due to cosmic variance} \label{sec-sampvar}

As already pointed out, cosmic variance represents a significant 
source of uncertainty in deep surveys, since they are characterized 
by small areas and hence small probed volumes. Our composite sample 
is made of several independent fields with a large total effective 
area of $\sim511$~arcmin$^{2}$, which significantly reduces the 
uncertainties due to cosmic variance. Also, the large number of fields 
considered in this work with their large individual areas allows us 
to empirically quantify the field-to-field variations from one field 
to the other in the estimate of the SMF with the $1/V_{\rm max}$ method, 
especially at the high-mass end, and to properly account for it in the 
error budget. 

In order to quantify the uncertainties due to field-to-field variations 
in the determination of the SMF, we proceeded as in \citet{marchesini07a}. 
Briefly, using the $1/V_{\rm max}$ method, we measured $\Phi^{\rm j}$, 
where $\Phi^{\rm j}$ is the galaxy number density in the stellar mass 
bin $\Delta M$ for the $j$th field. For each stellar mass bin with 
$n \geqslant3$, we estimated the contribution to the error budget of 
$\Phi$ from cosmic variance using:
\begin{equation} \label{eq-cv1}
\sigma_{\rm cv} = \frac{rms(\Phi^{\rm j})}{\sqrt{n}},
\end{equation}
with $n$ the number of individual fields used. For the stellar mass 
bins with $n \leqslant2$, we adopted the mean of the $rms(\Phi^{\rm j})$ 
with $n \geqslant3$. The final 1~$\sigma$ random error associated to 
$\Phi(M)$ is then 
$\sigma=(\sigma_{\rm Poi}^{2}+\sigma_{\rm cv}^{2}+\sigma_{\rm z,ran}^{2})^{1/2}$, 
with $\sigma_{\rm Poi}$ the Poisson error in each magnitude bin, and 
$\sigma_{\rm z,ran}$ the error due to photometric redshift random 
uncertainties as derived in \S~\ref{sec-photozerrs}.\footnote{We note 
that uncertainties related to field-to-field variations could be 
potentially correlated with photometric redshift uncertainties. However, 
spectroscopic complete high-mass samples at high redshift ($z>1$) do not 
yet exist, resulting in very little knowledge about an existing 
covariance between density estimates and photometric redshift systematics.} 

The values of $\sigma_{\rm cv}$ for each stellar mass bin in the three 
targeted redshift intervals are listed in Table~\ref{tab-mf1}. In the 
redshift range $1.3 \leq z < 2.0$, cosmic variance is the dominant source 
of random errors over almost the entire probed stellar mass range, with 
the exception of the most massive bin which is populated only by a handful 
of sources. At $z \geq 2.0$, cosmic variance is generally comparable, or 
slightly smaller than, Poisson errors, due to the larger probed volumes 
and to the smaller number of galaxies with respect to the $1.3 \leq z < 2.0$ 
redshift interval. We stress that the results from the maximum-likelihood 
analysis are not affected by cosmic variance, since the adopted STY method 
is unbiased with respect to density inhomogeneities 
(e.g. \citealt{efstathiou88}).

%----------------------------------------------------------------------------

\subsection{Systematic effects due to different SED modeling assumptions} \label{sec-sedmodsys}

As described in \S~\ref{sec-sed}, the default set of SED-modeling 
assumptions is represented by {\it (BC03,$Z_{\sun}$,Kroupa,Calzetti)}, 
i.e. BC03 stellar population synthesis models with a 
pseudo-\citet{kroupa01} IMF and solar metallicity have been used in 
combination with the \citet{calzetti00} extinction law to derived 
stellar masses. With broad-band photometry alone, it is not possible 
to constrain the metallicity, the IMF, the extinction law, or the 
stellar population synthesis model. Even with high-quality 
optical-to-MIR photometry and NIR spectroscopy, it is not possible 
to statistically constrain any of the above, as shown by 
\citet{muzzin08} for a sample of $z~\sim2$ galaxies. Therefore, we 
have chosen {\it (BC03,$Z_{\sun}$,Kroupa,Calzetti)} as our default 
set of SED-modeling assumptions, instead of having the metallicity, 
the IMF, the extinction curve, and the stellar models as free 
parameters. 

In the following we describe the adopted approach to quantify the 
systematic effects on the derived SMFs of galaxies due to the different 
choices of SED-modeling settings. 

\subsubsection{Variations on the default SED-modeling assumptions}

We have derived stellar masses by fitting the observed SEDs with 
different sets of SED-modeling assumptions, by changing the stellar 
population synthesis models, the IMF, the metallicity, and the extinction 
law. 

For the additional metallicities, we have used super-solar 
($Z=2.5 \times Z_{\sun}$) and sub-solar ($Z=0.2 \times Z_{\sun}$) 
metallicities. 

To explore the systematic effects due to different attenuation laws, 
the Milky Way (MW) extinction curve by \citet{allen76} and the Small 
Magellanic Cloud (SMC) extinction curve (\citealt{prevot84}; 
\citealt{bouchet85}) were also used. The main differences between the 
\citet{calzetti00} and the MW extinction laws lie in the ratio of 
total-to-selective absorption $R_{\rm V}=A_{\rm V}/E(B-V)$ (4.05 versus 
3.1, respectively) and in the \citet{calzetti00} law lacking the 
2175~\AA~bump characteristic of MW dust mixtures. Otherwise, their 
wavelength dependence are fairly similar. The SMC law with 
$R_{\rm V}=2.72$ also lacks the 2175~\AA~bump. In addition, it rises 
more steeply with decreasing wavelengths in the near-UV than the 
other two laws; in other words, the \citet{calzetti00} and the MW 
laws are much ``grayer'' at near-UV wavelengths. For (self-)consistency, 
the MW extinction law was used in combination with solar and super-solar 
metallicities, whereas the SMC curve was used with the sub-solar metallicity.

In addition to the pseudo-\citet{kroupa01} IMF (discussed in 
\S~\ref{sec-sed}), we have used three additional IMFs, namely the 
\citet{chabrier03} and two bottom-light IMFs.\footnote{We note that (because 
of our fitting procedure) the shape of the SMF is identical for a 
\citet{salpeter55} IMF, with a simple systematic shift of a factor of 
$\sim$1.6 to larger stellar masses.} Theoretical arguments and 
indirect observational evidence suggest that the stellar IMF may evolve 
with cosmic time, such that it is more weighted toward high-mass stars at 
higher redshift (see, e.g., \citealt{dave08}; \citealt{vandokkum08}; 
\citealt{wilkins08}). Recently, \citet{vandokkum08} provided new 
constraints on the IMF at high 
redshift by comparing the evolution of the $M/L$ ratios of early-type 
galaxies to their color evolution, finding a logarithmic slope of the 
IMF around 1~$M_{\sun}$ ($x=-0.3$) significantly flatter than the 
present-day value ($x\sim1.3$). Moreover, assuming a 
\citet{chabrier03}-like parameterization of the IMF with an evolving 
characteristic mass $m_{\rm c}$, the analysis in \citet{vandokkum08} 
implies a characteristic mass $m_{\rm c}=1.9~M_{\sun}$ at $z=3-6$ 
(for solar metallicity). This IMF is best described as ``bottom-light''
rather than top-heavy, since it does not have a larger number of 
massive stars than a standard \citet{chabrier03} IMF, but has a deficit 
of low-mass stars. For the bottom-light IMF, we adopted the 
parameterization defined in eq.~18 of \citet{vandokkum08}, with 
$m_c=1.9~M_{\sun}$. We also used a second bottom-light IMF by adopting 
a smaller value for the characteristic mass, $m_{\rm c}=0.3$~M$_{\sun}$. 
This is the characteristic mass required to reproduced the top-heavy IMF 
with a simple cutoff at 1~M$_{\sun}$ invoked by \citet{blain99a} for 
submillimeter galaxies. The \citet{chabrier03} IMF is recovered by using 
$m_c=0.079~M_{\sun}$. Note that the bottom-light IMFs have been used in 
combination with the \citet{maraston05} stellar population synthesis 
models.

Different stellar population synthesis models do not paint a consistent 
picture of evolution in the rest-frame NIR (probed by the IRAC bands).
Therefore, we have explored systematic effects due to different stellar 
population synthesis models by performing SED modeling with the 
\citet{maraston05} (MA05) and the \citet{charlot08} (CB08) stellar 
population models. The BC03 and the MA05 models differ in several 
aspects: the stellar evolutionary tracks adopted to construct the 
isochrones, the synthesis technique and the treatment of the thermally 
pulsating asymptotic giant branch (TP-AGB) phase. The Padova stellar 
tracks used in BC03 include a certain amount of convective-core 
overshooting, whereas the Frascati tracks \citep{cassisi97} used in 
MA05 do not. The two stellar evolutionary models also differ for the 
temperature distribution of the red giant branch phase. The differences 
in the rest-frame NIR originates mainly from a different implementation 
of the TP-AGB phase \citep{maraston06}. Following the fuel consumption 
approach, \citet{maraston05} finds that this phase in stellar evolution 
has a substantial impact on the NIR luminosity for ages between 0.2 and 
2~Gyr. \citet{bruzual03} follow the isochrone synthesis approach, 
characterizing properties of the stellar population per mass bin. The 
latter method leads to smaller luminosity contributions by TP-AGB stars. 
The CB08 stellar population synthesis models are generated with a recent 
version of the \citet{bruzual03} stellar population synthesis code which 
incorporates a new prescription by \citet{marigo07} for the TP-AGB 
evolution of low- and intermediate-mass stars. Whereas the 
\citet{marigo07} tracks used in CB08 account for 9 evolutionary stages 
in the TP-AGB (three in the O-rich phase, three in the C-rich phase, 
and three in the superwind phase), the BC03 models include only 1 
evolutionary stage on each of these phases. The main effect of this 
added prescription is to improve the predicted NIR colors of 
intermediate-age stellar populations (\citealt{bruzual07}; see also 
\citealt{charlot08}).

We note that the star formation history is also a significant source of 
uncertainty. We have treated this implicitly in our Monte Carlo simulations, 
as we chose the best-fitting star formation history (out of three models) 
for each realization (see \S~\ref{sec-sed}). However, it is well known that 
masses can be altered significantly by adding "maximally old" components in 
the fits and generally by allowing more complex forms of the star formation 
history than simple exponentially declining models (e.g., 
\citealt{papovich01}; \citealt{wuyts07}). Fitting such complex star 
formation history models is beyond the scope of the present paper, but we 
note that multiple component fits tend to increase the masses, particularly 
for galaxies whose light is dominated by star bursts (see \citealt{wuyts07}; 
\citealt{pozzetti07}).

The considered sets of SED-modeling assumptions are summarized in 
Table~\ref{tab-sedmod-sets}. 

\begin{deluxetable}{c}
\tablecaption{Considered sets of SED-modeling assumptions
\label{tab-sedmod-sets}}
\tablehead{\colhead{{\it (model,Z,IMF,dust)}}}
\startdata
{\it (BC03,$Z_{\sun}$,Kroupa,Calzetti)} \\
{\it (BC03,$2.5~Z_{\sun}$,Kroupa,Calzetti)} \\
{\it (BC03,$0.2~Z_{\sun}$,Kroupa,Calzetti)} \\
{\it (BC03,$Z_{\sun}$,Kroupa,MW)} \\
{\it (BC03,$2.5~Z_{\sun}$,Kroupa,MW)} \\
{\it (BC03,$0.2~Z_{\sun}$,Kroupa,SMC)} \\
{\it (BC03,$Z_{\sun}$,Chabrier,Calzetti)} \\
{\it (CB08,$Z_{\sun}$,Kroupa,Calzetti)} \\
{\it (MA05,$Z_{\sun}$,Kroupa,Calzetti)} \\
{\it (MA05,$Z_{\sun}$,Bottom-light $m_{\rm c}=0.3$,Calzetti)} \\
{\it (MA05,$Z_{\sun}$,Bottom-light $m_{\rm c}=1.9$,Calzetti)} \\
\enddata
\tablenotetext{}{The first element of the table 
is the default set of SED-modeling assumptions.}
\end{deluxetable}

\subsubsection{Derivation of the SMFs}

\begin{figure*}
\epsscale{0.95}
\plotone{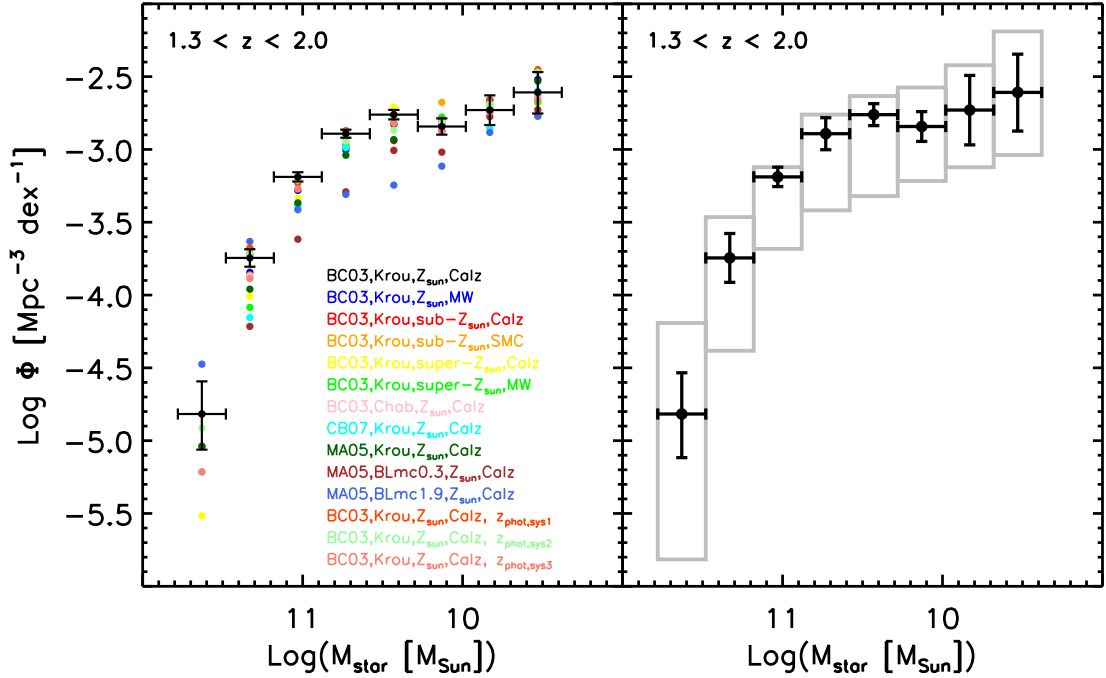}
\caption{{\it Left panel:} SMF of galaxies at $1.3 \leq z<2.0$ 
derived with the $1/V_{\rm max}$ method. The SMF corresponding 
to the default set of SED-modeling assumptions is plotted with 
black filled circles and 1~$\sigma$ Poisson errors; the SMFs 
corresponding to different sets of SED-modeling settings and 
different combinations of template sets and template error 
functions are plotted with different colors (no errors plotted 
for clarity). {\it Right panel:} SMF of galaxies at $1.3 \leq z<2.0$ 
derived with the $1/V_{\rm max}$ method and assuming the default set 
of SED-modeling settings; the black error bars now include the 
Poisson error, the error due to field-to-field variations, and 
the error due to photometric redshift random uncertainties. 
The gray boxes represent the total 1~$\sigma$ errors, with the 
systematic uncertainties added linearly to the 1~$\sigma$ random 
errors 
$\sigma = (\sigma^{2}_{\rm Poi}+\sigma^{2}_{\rm cv}+\sigma^{2}_{\rm z,ran})^{1/2}$.
\label{fig-appsedmod1}}
\end{figure*}

\begin{figure*}
\epsscale{0.95}
\plotone{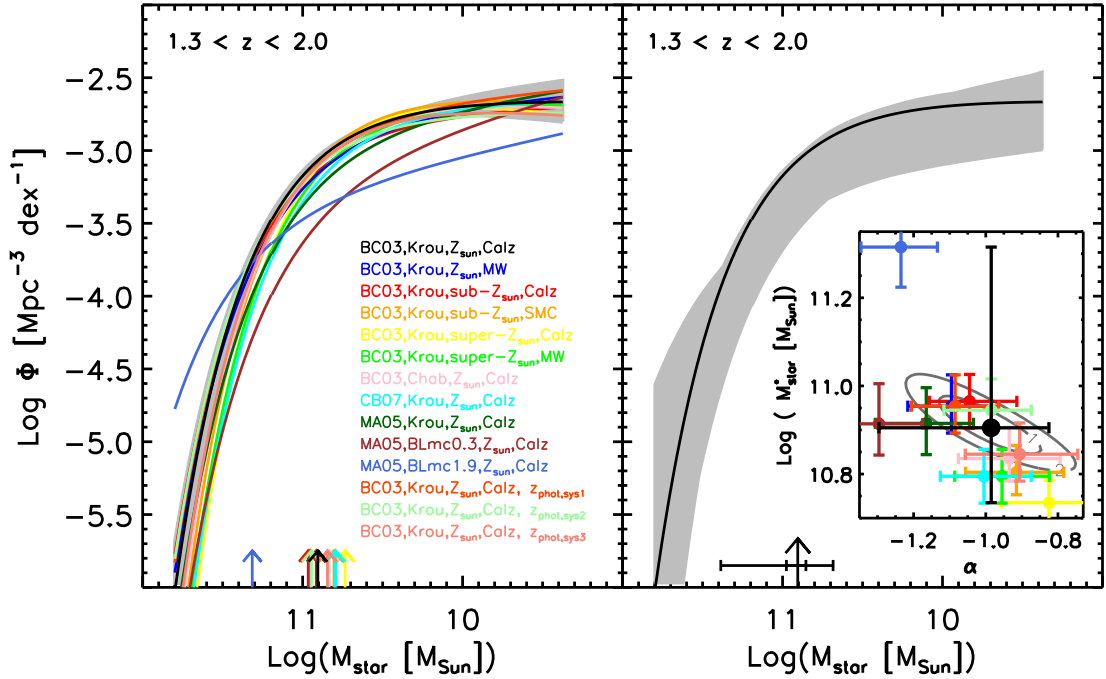}
\caption{{\it Left panel:} SMF of galaxies at $1.3 \leq z<2.0$ 
derived with the maximum-likelihood analysis. The SMF and its 
1~$\sigma$ error corresponding to the default set of SED-modeling 
assumptions is plotted with black line and gray shaded region; 
the SMFs corresponding to different sets of SED-modeling settings 
and different combinations of template sets and template error 
functions are plotted with different colors (no errors plotted for 
clarity); the arrows represent the characteristic stellar masses 
$M^{\star}_{\rm star}$. {\it Right panel:} SMF of galaxies at 
$1.3 \leq z<2.0$ derived with the maximum-likelihood analysis and 
assuming the default set of SED-modeling settings (black solid 
curve); the gray shaded region represent the total 1~$\sigma$ 
uncertainty, included the systematic uncertainties. The arrow 
represents $M^{\star}_{\rm star}$; the smaller error bars 
represent the 1~$\sigma$ error derived from the maximum-likelihood 
analysis; the larger error bars represent the total 1~$\sigma$ 
error,with the systematic uncertainties added linearly. The 
insert shows the parameter space $(\alpha-M^{\star}_{\rm star})$, 
with the best-fit values corresponding to the default set of 
SED-modeling settings (black filled circle) and its corresponding 
1~ and 2~$\sigma$ contour levels (solid gray ellipsoids), and the 
best-fit values corresponding to the other SED-modeling assumption 
sets and different combinations of template sets and template error 
functions (colored filled circles). If the bottom-light IMFs are 
not considered, the largest systematic effects on the derived SMFs 
are caused by the changes in the stellar population synthesis models 
and the combination of super-solar metallicity with the MW extinction 
law. Much larger systematic effects are found when the bottom-light 
IMFs are adopted (brown and light-blue symbols), both at the high- 
and low-mass ends.
\label{fig-appsedmod2}}
\end{figure*}

For each new combination of SED-modeling assumptions, we have derived 
the completeness limits in stellar mass and the SMFs with both the 
$1/V_{\rm max}$ method and the maximum-likelihood analysis. 
Table~\ref{tab-appsedmod1} lists the SMFs of galaxies at 
$1.3 \leq z<2.0$, $2.0 \leq z<3.0$, and $3.0 \leq z<4.0$ derived using 
the $1/V_{\rm max}$ method for each combination of SED-modeling settings. 
Table~\ref{tab-appsedmod2} lists the best-fit Schechter function 
parameters $\alpha$, $M^{\star}_{\rm star}$, and $\Phi^{\star}$ of the SMFs 
of galaxies at $1.3 \leq z<2.0$, $2.0 \leq z<3.0$, and $3.0 \leq z<4.0$ 
derived using the maximum-likelihood analysis for each combination of 
SED-modeling settings. In the left panel of Figures~\ref{fig-appsedmod1}, 
the SMF of galaxies at $1.3 \leq z<2.0$ derived with the $1/V_{\rm max}$ 
method for the default set {\it (BC03,$Z_{\sun}$,Kroupa,Calzetti)} is 
compared to the SMFs derived for the other considered sets of 
SED-modeling assumptions. Similarly, the left panel of 
Figure~\ref{fig-appsedmod2} shows the different SMFs at $1.3 \leq z<2.0$ 
corresponding to the various sets of SED-modeling assumptions used in 
the maximum-likelihood analysis.

\begin{deluxetable*}{ccccccccccccccc}
\tablecaption{SMFs from the $1/V_{\rm max}$ method for the different SED-modeling assumptions
\label{tab-appsedmod1}}
\tablehead{\colhead{ } & \colhead{Set~1} & \colhead{Set~2} & \colhead{Set~3} & 
 \colhead{Set~4} & \colhead{Set~5} & \colhead{Set~6} & \colhead{Set~7} & \colhead{Set~8} & \colhead{Set~9} &
 \colhead{Set~10} & \colhead{Set~11} & \colhead{Set~12} & \colhead{Set~13} & \colhead{Set~14} \\
 $\log{(M_{\rm star}/M_{\sun})}$ & \multicolumn{14}{c}{$\log (\Phi$ [Mpc$^{-3}$ dex$^{-1}$]$)$}
}
\startdata
$1.3 \leq z<2.0$: & \multicolumn{14}{c}{} \\
11.63 & -4.817 & -5.516 & -4.817 & -5.039 & -5.039 & -5.039 & -5.039 & -5.215 & -5.039 & $<$-5.3 & -4.475 & -4.817 & -4.914 & -5.215\\
11.33 & -3.745 & -4.011 & -3.717 & -3.844 & -4.085 & -3.863 & -3.873 & -4.154 & -3.960 & -4.215  & -3.632 & -3.677 & -3.718 & -3.887\\
11.03 & -3.189 & -3.334 & -3.194 & -3.281 & -3.378 & -3.217 & -3.211 & -3.390 & -3.367 & -3.617  & -3.414 & -3.231 & -3.211 & -3.272\\
10.73 & -2.892 & -2.891 & -2.985 & -3.003 & -2.968 & -2.891 & -2.942 & -2.988 & -3.040 & -3.291  & -3.308 & -2.870 & -2.933 & -2.874\\
10.43 & -2.761 & -2.709 & -2.939 & -2.828 & -2.762 & -2.757 & -2.768 & -2.782 & -2.933 & -3.006  & -3.245 & -2.824 & -2.870 & -2.818\\
10.13 & -2.843 & -2.795 & -2.872 & -2.785 & -2.778 & -2.677 & -2.817 & -2.839 & -2.823 & -3.019  & -3.115 & -2.798 & -2.813 & -2.873\\
 9.83 & -2.730 & -2.719 & -2.730 & -2.716 & -2.688 & -2.722 & -2.749 & -2.855 & -2.660 & -2.775  & -2.883 & -2.662 & -2.690 & -2.743\\
 9.53 & -2.608 & -2.686 & -2.631 & -2.519 & -2.673 & -2.768 & -2.630 & -2.599 & -2.530 & -2.729  & -2.772 & -2.452 & -2.469 & -2.658\\
$2.0 \leq z<3.0$: & \multicolumn{14}{c}{} \\
11.63 & -5.067 & -5.590 & -4.988 & -5.389 & -5.590 & -5.385 & -5.383 & -5.389 & -5.389 & $<$-5.6 & -5.213 & -5.066 & -4.991 & -4.929\\
11.34 & -3.949 & -4.133 & -3.908 & -4.455 & -4.331 & -4.078 & -3.981 & -4.500 & -4.201 & -5.098  & -4.439 & -3.955 & -3.947 & -4.235\\
11.05 & -3.570 & -3.641 & -3.685 & -3.701 & -3.800 & -3.619 & -3.700 & -3.711 & -3.770 & -3.999  & -4.063 & -3.638 & -3.439 & -3.630\\
10.76 & -3.402 & -3.309 & -3.424 & -3.451 & -3.331 & -3.354 & -3.357 & -3.417 & -3.462 & -3.741  & -3.787 & -3.368 & -3.295 & -3.309\\
10.47 & -3.119 & -3.292 & -3.360 & -3.353 & -3.346 & -3.148 & -3.168 & -3.259 & -3.334 & -3.479  & -3.600 & -3.218 & -3.080 & -3.203\\
10.18 & -3.401 & -3.208 & -3.188 & -3.175 & -3.210 & -3.235 & -3.229 & -3.281 & -3.131 & -3.258  & -3.534 & -3.265 & -3.435 & -3.526\\
 9.89 & -2.640 & -2.911 & -3.078 & -2.813 & -2.934 & -2.938 & -2.794 & -2.963 & -3.024 & -3.311  & -3.089 & -2.703 & -2.698 & -2.499\\
$3.0 \leq z<4.0$: & \multicolumn{14}{c}{} \\
11.66 & -4.784 & -5.191 & -4.831 & -4.978 & -5.337 & -4.688 & -4.924 & -5.468 & -5.342 & $<$-5.5 & -5.179 & -4.820 & -4.969 & -5.340\\
11.37 & -4.282 & -4.767 & -4.438 & -4.540 & -4.676 & -4.426 & -4.417 & -4.522 & -4.577 & -4.794  & -4.809 & -4.218 & -4.273 & -4.710\\
11.08 & -4.025 & -3.964 & -4.131 & -4.109 & -4.078 & -4.068 & -3.996 & -4.160 & -4.068 & -4.509  & -4.508 & -4.043 & -3.879 & -4.132\\
10.79 & -3.929 & -3.886 & -4.099 & -3.778 & -3.973 & -3.814 & -3.964 & -4.012 & -3.983 & -4.073  & -4.210 & -3.903 & -3.920 & -3.892\\
10.50 & -3.433 & -3.892 & -3.788 & -3.892 & -3.892 & -3.670 & -4.068 & -4.069 & -4.068 & -4.169  & -3.968 & -3.486 & -3.486 & -3.591\\
10.21 & -3.141 & -3.280 & -3.092 & -2.997 & -3.070 & -3.317 & -3.214 & -3.280 & -3.070 & -4.019  & -3.547 & -3.141 & -3.141 & -3.016\\
\enddata
\tablenotetext{}{Set~1 = {\it (BC03,$Z_{\sun}$,Kroupa,Calzetti)}, 
default set of SED-modeling assumptions; 
Set~2 = {\it (BC03,$2.5~Z_{\sun}$,Kroupa,Calzetti)}; 
Set~3 = {\it (BC03,$0.2~Z_{\sun}$,Kroupa,Calzetti)};
Set~4 = {\it (BC03,$Z_{\sun}$,Kroupa,MW)}; 
Set~5 = {\it (BC03,$2.5~Z_{\sun}$,Kroupa,MW)};
Set~6 = {\it (BC03,$0.2~Z_{\sun}$,Kroupa,SMC)}; 
Set~7 = {\it (BC03,$Z_{\sun}$,Chabrier,Calzetti)};
Set~8 = {\it (CB08,$Z_{\sun}$,Kroupa,Calzetti)};
Set~9 = {\it (MA05,$Z_{\sun}$,Kroupa,Calzetti)};
Set~10 = {\it (MA05,$Z_{\sun}$,Bottom-light $m_{\rm c}=0.3$,Calzetti)};
Set~11 = {\it (MA05,$Z_{\sun}$,Bottom-light $m_{\rm c}=1.9$,Calzetti)};
Set~12 = {\it (BC03,$Z_{\sun}$,$Z_{\sun}$,Kroupa,Calzetti)}, with 
{\it eazy\_v1.0\_nodust} and {\it TE.eazy\_v1.0\_nodust};
Set~13 = {\it (BC03,$Z_{\sun}$,$Z_{\sun}$,Kroupa,Calzetti)}, with 
{\it eazy\_v1.0} and {\it TE.eazy\_v1.0\_nodust};
Set~14 = {\it (BC03,$Z_{\sun}$,$Z_{\sun}$,Kroupa,Calzetti)}, with 
{\it br07\_default} and {\it TE.eazy\_v1.0}.}
\end{deluxetable*}

\begin{deluxetable*}{lcccccccccccccc}
\tablecaption{Best-fit Schechter parameters for the different SED-modeling assumptions
\label{tab-appsedmod2}}
\tablehead{\colhead{Parameter} & \colhead{Set~1} & \colhead{Set~2} & \colhead{Set~3} & 
 \colhead{Set~4} & \colhead{Set~5} & \colhead{Set~6} & \colhead{Set~7} & \colhead{Set~8} & \colhead{Set~9} &
 \colhead{Set~10} & \colhead{Set~11} & \colhead{Set~12} & \colhead{Set~13} & \colhead{Set~14}}
\startdata
$1.3 \leq z<2.0$: & \multicolumn{14}{c}{} \\
$\alpha$                                            & -0.99 & -0.83 & -1.05 & -1.10 & -0.96 & -0.92 & -0.94 & -1.01 & -1.17 & -1.30 & -1.24 & -1.09 & -0.99 & -0.91\\
$\log{(M^{\star}_{\rm star}/M_{\sun})}$             & 10.91 & 10.73 & 10.97 & 10.95 & 10.80 & 10.80 & 10.84 & 10.80 & 10.92 & 10.91 & 11.32 & 10.95 & 10.95 & 10.85\\
$\Phi^{\star}$ [10$^{-4}$ Mpc$^{-3}$ dex$^{-1}$] & 10.17 & 13.78 &  7.49 &  7.35 & 10.65 & 12.72 & 11.01 &  9.62 &  6.38 &  3.64 &  2.01 &  8.46 &  8.54 & 10.77\\
\hline
$2.0 \leq z<3.0$: & \multicolumn{14}{c}{} \\
$\alpha$                                            & -1.01 & -0.85 & -1.03 & -1.24 & -0.89 & -1.03 & -1.03 & -0.97 & -1.21 & -0.94 & -1.36 & -1.09 & -0.98 & -1.13\\
$\log{(M^{\star}_{\rm star}/M_{\sun})}$             & 10.96 & 10.83 & 10.99 & 10.94 & 10.80 & 10.88 & 10.90 & 10.83 & 10.93 & 10.62 & 11.17 & 10.96 & 10.94 & 10.98\\
$\Phi^{\star}$ [10$^{-4}$ Mpc$^{-3}$ dex$^{-1}$] &  3.95 &  5.02 &  3.14 &  2.59 &  4.33 &  4.41 &  4.10 &  3.75 &  2.78 &  3.45 &  0.75 &  3.70 &  4.80 &  2.92\\
\hline
$3.0 \leq z<4.0$: & \multicolumn{14}{c}{} \\
$\alpha$                                            & -1.39 & -1.39 & -1.31 & -1.92 & -1.74 & -1.61 & -1.49 & -0.96 & -1.44 & -1.06 & -1.69 & -1.44 & -1.09 & -1.59\\
$\log{(M^{\star}_{\rm star}/M_{\sun})}$             & 11.38 & 11.36 & 11.36 & 11.64 & 11.44 & 11.43 & 11.41 & 11.13 & 11.26 & 11.09 & 11.41 & 11.46 & 11.24 & 11.13\\
$\Phi^{\star}$ [10$^{-4}$ Mpc$^{-3}$ dex$^{-1}$] &  0.53 &  0.42 &  0.44 &  0.11 &  0.22 &  0.34 &  0.40 &  0.65 &  0.49 &  0.42 &  0.13 &  0.37 &  0.90 &  0.65\\
\enddata
\tablenotetext{}{SED-modeling assumption sets as in 
Tab.~\ref{tab-appsedmod1}.}
\end{deluxetable*}

%----------------------------------------------------------------------------

\subsubsection{The effects of different SED-modeling assumptions}

In this section we discuss in details the effects on the 
derived SMFs when changing the SED-modeling assumptions. A 
detailed analysis of the effects of the different SED-modeling 
assumptions on the estimated stellar masses is presented in 
\citet{muzzin08} for a sample of 34 $K$-selected galaxies at 
$z\sim2$.

\paragraph{Stellar population synthesis models}

With respect to the default SED-modeling assumptions, using the 
\citet{maraston05} models results in derived SMFs with generally 
steeper low-mass end slopes $\alpha$, slightly smaller characteristic 
stellar masses $M^{\star}_{\rm star}$ (by $<0.1$~dex), and smaller 
normalizations $\Phi^{\star}$ (by $\sim$40\%-50\%). If the 
\citet{charlot08} models are instead used, the derived SMFs have 
similar $\alpha$, significantly smaller $M^{\star}_{\rm star}$ 
(by $\sim$0.1-0.2~dex), but similar $\Phi^{\star}$. However, due to the 
correlation between the Schechter function parameters $\alpha$ and 
 $M^{\star}_{\rm star}$, the SMFs derived using the \citet{maraston05} 
and the \citet{charlot08} models are in general very similar, resulting 
in a general decrease of the number densities of galaxies. This decrease 
is larger at the high-mass end, and smaller at the low-mass end.

\paragraph{Metallicities}

Changing the metallicity from solar to sub-solar results in smaller 
characteristic densities $\Phi^{\star}$ by $\sim$20\%-30\%, but no 
significant effect on $\alpha$ and $M^{\star}_{\rm star}$. Conversely, 
using super-solar metallicity results in shallower $\alpha$, smaller 
$M^{\star}_{\rm star}$ (by $\sim$0.1-0.2~dex), and larger $\Phi^{\star}$ 
(by $\sim$30\%-40\%). The SMFs derived with sub-solar metallicities are 
similar to those derived using the default SED-modeling assumptions at 
the high-mass end, but with generally smaller number densities at the 
low-mass end. The SMFs derived with super-solar metallicities are instead 
characterized by a smaller number densities with respect to the SMFs 
derived with the default SED-modeling assumptions. This decrease is larger 
at the high-mass end, and much smaller at the low-mass end.

\paragraph{Extinction laws}

Changing the adopted extinction law from \citet{calzetti00} to the 
MW law results in steeper $\alpha$, similar or slightly larger 
$M^{\star}_{\rm star}$, and significantly smaller $\Phi^{\star}$ 
(by $\sim$20\%-50\%). The net results on the derived SMFs is a decrease 
in the number densities with respect to the SMFs derived with the 
\citet{calzetti00} extinction law. This decrease is small at the high-mass 
end and much larger at the low-mass end.

Using the SMC extinction curve in combination with sub-solar metallicity 
results in slightly shallower $\alpha$, smaller $M^{\star}_{\rm star}$ 
(by $\sim$0.1-0.15~dex), and larger $\Phi^{\star}$ (by $\sim$40\%-60\%) 
compared to the SMFs derived with \citet{calzetti00} extinction law and 
sub-solar metallicity. The net result is a decrease of the number 
densities at the high-mass end, and an increase of the number densities 
at the low-mass end. With respect to the default SED-modeling assumptions, 
using a SMC extinction curve in combination with sub-solar metallicity 
results in smaller number densities at the high-mass end and similar number 
densities at the low-mass end. The latter is due to the fact that the 
effects at the low-mass end of changing the extinction curve and the 
metallicity are broadly similar, but opposite in sign.

\paragraph{IMFs}

Using the \citet{chabrier03} IMF in place of the pseudo-\citet{kroupa01} 
IMF does not have a significant effect on the derived shape of the SMFs, 
with only a small decrease in the characteristic stellar mass 
$M^{\star}_{\rm star}$ by $\sim$0.05~dex.

A more complex behavior is however found when the two bottom-light 
IMFs are considered. As shown in the left panel of 
Figures~\ref{fig-appsedmod2}, the shapes of the SMFs derived using the 
bottom-light IMFs are significantly different from that of the SMF 
derived with the default SED-modeling assumptions. This is especially 
true for the bottom-light IMF with $m_{\rm c}=1.9$~M$_{\sun}$, 
characterized by a steeper low-mass end and a characteristic stellar 
mass larger by a factor of $\sim2.5$. This results is particularly 
important since it is commonly assumed that changing IMF in the 
SED-modeling results in a systematic shift of the derived SMF, leaving 
the shape of the SMF unchanged. This is clearly not the case for 
bottom-light IMFs: the more the IMF is skewed toward high-mass stars 
(i.e., the more deficient in low-mass stars the IMF is), the larger 
is the effect on the shape of the derived SMF. 

\begin{figure}
\epsscale{1.0}
\plotone{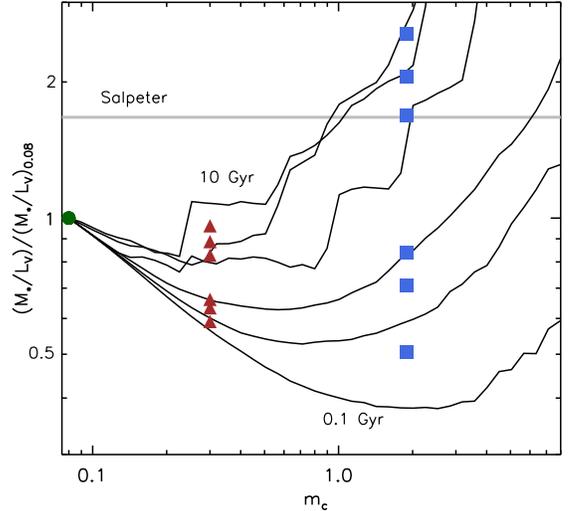}
\caption{Effect of changing characteristic mass on the 
$M/L_{\rm V}$ ratio for stellar populations of ages 0.1, 
0.3, 0.5, 3, 5, and 10~Gyr (from bottom to top) using 
stellar population synthesis models (filled symbols), 
following \citet{vandokkum08}. The solid curves, derived 
using simple stellar evolutionary tracks, were taken from 
\citet{vandokkum08}. Three different characteristic 
masses have been considered: $m_{\rm c}=0.08$ (i.e., 
\citealt{chabrier03} IMF; dark green circle), 
$m_{\rm c}=0.3$ (brown triangles), and 
$m_{\rm c}=1.9$~M$_{\sun}$ (blue squares). For 
$m_{\rm c}=1.9$~M$_{\sun}$ and old ages, the mass 
function becomes remnant dominated, and the $M/L$ ratios 
approach, and even exceed, those implied by a 
\citet{salpeter55} IMF (gray solid line).
\label{fig-mlratio}}
\end{figure}

Another very interesting result is the resulting higher number density 
of massive galaxies when using the bottom-light IMF with 
$m_{\rm c}=1.9$~M$_{\sun}$ with respect to the SMFs derived with the 
other IMFs. This result might come unexpectedly at first. Naively, one 
would expect that, by making the IMF more deficient in low-mass stars, 
which dominate the stellar mass of a galaxy but contribute little to 
the integrated light, the derived stellar masses would be smaller 
compared to those derived from the other considered IMFs, as a 
consequence of a lowering of the $M/L$ ratio. However, as 
already pointed out by \citet{vandokkum08}, the number of turn-off stars 
is also reduced for $m_{\rm c} \ga 0.4$~M$_{\sun}$, and these stars 
dominate the light at rest-frame optical wavelengths. Moreover, the 
turn-off mass can be similar to $m_{\rm c}$, meaning that the effect 
on the $M/L$ ratio is not a constant, but depends on the age of the 
population. A final complication is the mass in stellar remnants, which 
is a larger fraction of the total stellar mass for more top-heavy IMFs. 
Using simple stellar evolutionary tracks but not full stellar population 
synthesis modeling, \citet{vandokkum08} calculated the effects of changing 
characteristic mass on the $M/L_{\rm V}$ ratio for stellar populations of 
different ages, from 0.1 to 10~Gyr. They found that for young ages the 
$M/L$ ratio steadily declines with increasing $m_{\rm c}$, but the behavior 
is more complex when $m_{\rm c}$ becomes similar to the turn-off mass. 
Specifically, they found that for $m_{\rm c}\sim1$~M$_{\sun}$ and old ages 
the mass function becomes remnant dominated, and the $M/L$ ratios approach, 
or even exceed, those implied by a \citet{salpeter55} IMF. We can directly 
test their conclusions by correctly treating the above issues with the 
available stellar population synthesis models constructed with the 
bottom-light IMFs. The effect of changing characteristic mass on the 
$M/L_{\rm V}$ ratio for different population ages is shown in 
Figure~\ref{fig-mlratio}.

These results are consistent with those obtained by \citet{vandokkum08}, 
confirming that the $M/L_{\rm V}$ ratios for old stellar populations 
and high characteristic mass can approach, and even exceed, the 
$M/L_{\rm V}$ ratios implied by \citet{chabrier03} and 
\citet{salpeter55} IMFs, due to the stellar population becoming 
remnant-dominated.

The SMFs derived with the bottom-light IMFs in the SED-modeling 
assumptions can be now easily explained with the shown behavior of 
the $M/L_{\rm V}$ ratio in mind. When an IMF with $m_{\rm c}=0.3$ 
is adopted, the $M/L$ ratios are always smaller, or at most 
comparable, to the $M/L$ ratios derived with a Chabrier-like IMFs. 
Therefore, the derived stellar masses are always smaller, and the 
derived SMF is, broadly speaking, shifted to smaller masses. Note, 
however, that the dependence of the $M/L$ ratio on age will affect 
the specific shape of the SMF. When an IMF with $m_{\rm c}=1.9$ is 
adopted, the $M/L$ ratios are larger than those derived with a 
\citet{chabrier03} IMF when the age of the population is larger than 
$\sim$0.9~Gyr. Consequently, those galaxies which are fitted by ages 
older than 0.9~Gyr will have much larger stellar masses than those 
derived assuming a \citet{chabrier03} IMF. On the contrary, those 
galaxies fitted by ages younger than $\sim$0.9~Gyr will have smaller 
stellar masses compared to those derived assuming a \citet{chabrier03} 
IMF. The net effect on the derived SMF is a significant increase of 
the number densities of massive galaxies, which are usually characterized 
by old stellar population, and a decrease of the number densities of 
low-mass galaxies, which are usually characterized by young stellar 
population. 

Note that the net effect on the derived SMF caused by assuming a 
bottom-light IMF is also a function of redshift, since the maximal age 
of the stellar population is limited by the age of the universe at that 
redshift. As the age of the stellar populations get younger going to 
higher redshifts, the effect on the SMF due to a a bottom-light IMF will 
be closer to a systematic shift to smaller stellar masses without changing 
the shape significantly.

\paragraph{Summary}

Broadly speaking, the different combinations of SED-modeling 
assumptions result in smaller estimates of the stellar masses 
with respect to the stellar masses derived using the default 
set. Consequently, the systematic effects on the SMFs are largest 
at the high-mass end of the SMFs, due to its steep slope and rapid 
changes in number density as function of stellar mass. The net 
effect on the derived SMFs is an average decrease of the number 
densities of galaxies at the high-mass end, while the systematic 
effects are generally smaller at the low-mass end. If the bottom-light 
IMFs are not considered, the largest systematic effects are caused 
by the changes in the stellar population synthesis models and the 
combination of super-solar metallicity with the MW extinction law. 
The largest systematic effects are caused by the use of the 
bottom-light IMFs.

%----------------------------------------------------------------------------

\subsubsection{The systematic uncertainties in the SMF due to different SED-modeling assumptions}

The systematic effects on the SMFs due to different SED-modeling 
assumptions have been quantified by comparing the resulting SMFs with 
those derived using the default set of settings 
{\it (BC03,$Z_{\sun}$,Kroupa,Calzetti)}. Note that we implicitly assume 
that changes in the derived SMFs are the result of the 
changes we made to the model parameters. We cannot exclude subtle 
second-order effects that may influence the fitting procedure, but given 
the excellent agreement between the maximum-likelihood and the $1/V_{\rm max}$ 
estimators these are likely much smaller than the effects that we are 
measuring here.

For the $1/V_{\rm max}$ method, the systematic uncertainties of 
$\Phi(M)$ have been estimated by taking, for each stellar mass bin, 
the difference between the maximum (and minimum) value of $\Phi(M)$ 
allowed by all the considered combinations of SED-modeling settings 
and the value of $\Phi(M)$ derived with the default set. These 
systematic uncertainties ($\sigma_{\rm sys}$) are listed in 
Table~\ref{tab-mf1} and were then added linearly to the 1~$\sigma$ 
errors 
$\sigma = (\sigma^{2}_{\rm Poi}+\sigma^{2}_{\rm cv}+\sigma^{2}_{\rm z,ran})^{1/2}$
(which include the Poisson error, the error due to field-to-field 
variations, and the error due to photometric redshift random 
uncertainties) to obtain the total 1~$\sigma$ errors. In the right 
panel of Figure~\ref{fig-appsedmod1}, we show the SMF of galaxies 
at $1.3 \leq z<2.0$, plotting the 1~$\sigma$ errors with and without the 
contribution of the systematic effects due to different SED-modeling 
assumptions and different combinations of template sets and template 
error functions. 

For the maximum-likelihood analysis, the systematic uncertainties 
on the Schechter function parameters have been estimated by taking the 
difference between the maximum and minimum values derived when using all 
the considered combinations of SED-modeling settings and the value 
corresponding to the default set. These systematic uncertainties 
($\alpha_{\rm sys}$, $M^{\star}_{\rm sys}$, and $\Phi^{\star}_{\rm sys}$), 
are listed in Table~\ref{tab-mf2}. The right panel of 
Figure~\ref{fig-appsedmod2} shows the SMF of galaxies 
at $1.3 \leq z<2.0$ derived with the default set of SED-modeling settings 
and the total 1~$\sigma$ uncertainties after including the systematic 
uncertainties due to different SED-modeling assumptions and different 
combinations of template sets and template error functions; also plotted 
is the parameter space $(\alpha-M^{\star}_{\rm star})$.

%----------------------------------------------------------------------------

\subsection{Stellar Mass Functions with all uncertainties} \label{subsec-mfsys}

\begin{figure*}
\centering
\includegraphics[width=17cm]{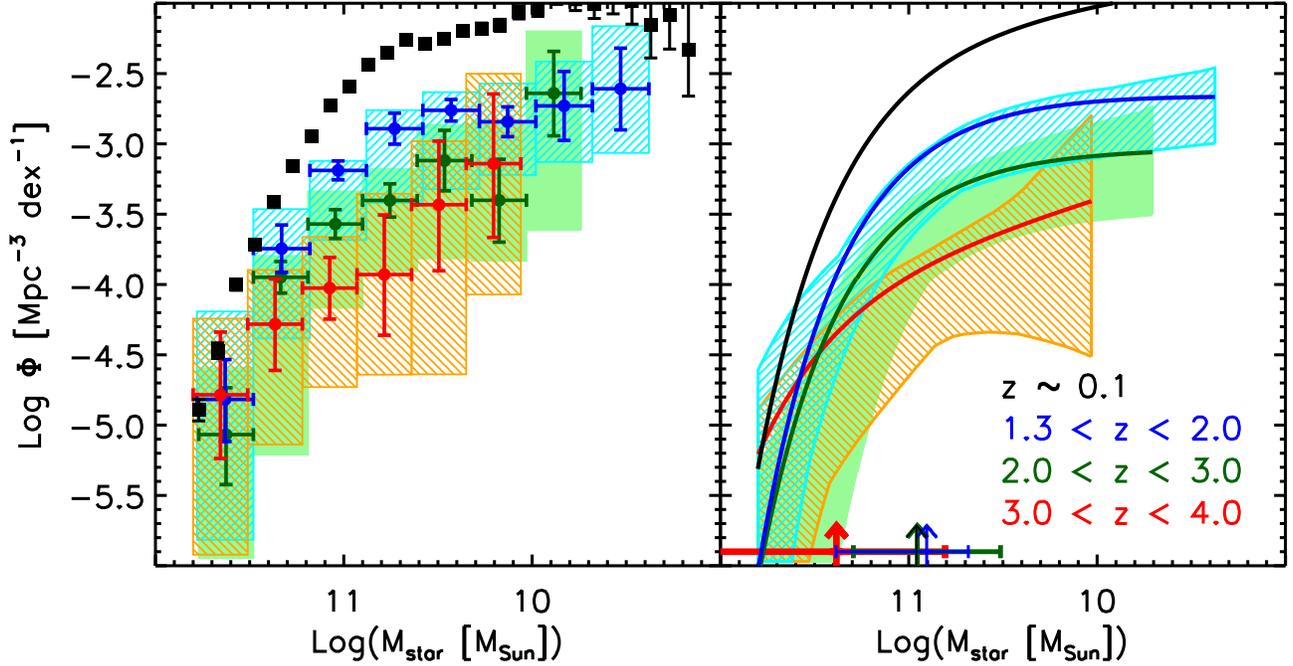}
\caption{SMFs of galaxies at redshift $1.3 \leq z < 2.0$ 
(blue), $2.0 \leq z < 3.0$ (green), and $3.0 \leq z < 4.0$ 
(red). {\it Left panel:} SMFs of galaxies derived using 
the $1/V_{\rm max}$ method (filled circles); the error bars 
include Poisson errors, photometric redshift uncertainties, 
and errors due to cosmic variance. The shaded boxes (orange, 
green, and cyan, corresponding to the redshift intervals 
$3.0 \leq z<4.0$, $2.0 \leq z<3.0$, and $1.3 \leq z<2.0$, 
respectively) represent the total 1~$\sigma$ uncertainties 
in the measurements of the SMFs as described in 
\S~\ref{sec-errors}, with systematic errors added linearly to the 
plotted error bars. {\it Right panel:} SMFs of galaxies derived 
using the maximum-likelihood analysis (solid curves); the shaded 
regions represent the total 1~$\sigma$ uncertainties as described 
in \S~\ref{sec-errors}, including the systematic uncertainties. 
The arrows show the best estimates of $M^{\star}_{\rm star}$, with 
their error bars including also systematic uncertainties.
\label{fig-mfevol_sys}}
\end{figure*}

\begin{figure}
\epsscale{1}
\plotone{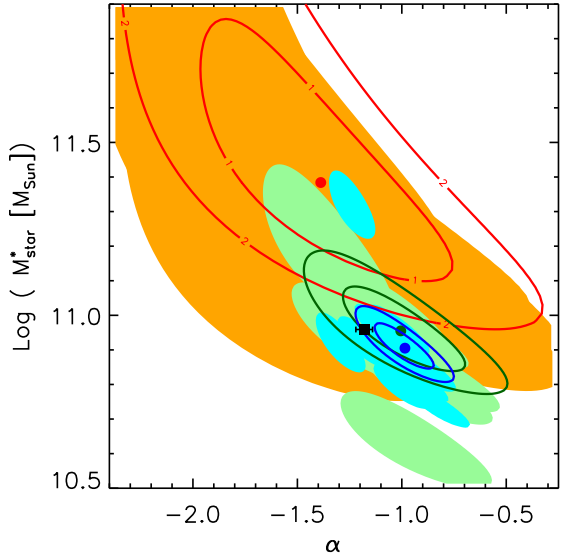}
\caption{Parameter space $(\alpha-M^{\star}_{\rm star})$ 
derived from the maximum-likelihood analysis. The red, 
dark green, and blue filled circles are the best-fit 
values of $\alpha$ and $M^{\star}_{\rm star}$ at 
redshift $3.0 \leq z < 4.0$, $2.0 \leq z < 3.0$, and 
$1.3 \leq z < 2.0$, respectively. The red, dark green, 
and blue curves represent their 1 and 2~$\sigma$ contour 
levels, respectively. The filled regions show the 1~$\sigma$ 
allowed values of $\alpha$ and $M^{\star}_{\rm star}$ 
after inclusion of the systematic uncertainties in the 
error analysis. The black filled square represents the 
redshift $z\sim0.1$ value from \citet{cole01}.
\label{fig-mfevol_sys_err}}
\end{figure}

Figure~\ref{fig-mfevol_sys} shows the evolution of the SMF of 
galaxies from $z=4.0$ to $z=1.3$ including the contribution of 
random and systematic uncertainties in the error budget, i.e., 
the Poisson errors, the uncertainties due to cosmic variance and 
photometric redshift random errors, and the systematic uncertainties 
due to different SED-modeling assumptions and different combinations  
of template sets and template error functions. These errors are also 
listed in Table~\ref{tab-mf1} and \ref{tab-mf2}. Most of the systematic 
effects are in the same direction, with a resulting net effect of 
decreasing the observed number densities, especially at the high-mass 
end and in the highest targeted redshift interval. The 
$(M^{\star}_{\rm star}-\alpha)$ plane is also plotted in 
Figure~\ref{fig-mfevol_sys_err} showing the effect of systematic 
uncertainties on the Schechter function parameters. 

The systematics uncertainties are a dominant contribution to the 
overall error budget. The largest contribution to the systematic 
uncertainties due to different SED-modeling assumptions are due to 
changes in the adopted IMF, specifically when a bottom-light IMF is 
used. The systematic uncertainties due to different combinations of 
template sets and template error functions in the estimate of the 
photometric redshifts are always smaller than the systematic 
uncertainties due to different SED-modeling assumptions, especially 
when the maximum-likelihood analysis is used. The maximum-likelihood 
analysis is indeed quite robust against photometric redshift errors, 
both random and systematic, and the dominant source of uncertainties 
are the systematic errors due to different SED-modeling assumptions. 
This is true at all redshifts but the highest redshift range, where 
Poisson errors represent a significant contribution to the error budget. 
As shown in the insert of Figure~\ref{fig-appsedmod2} for the redshift 
range $1.3 \leq z < 2.0$, the changes in the Schechter function parameters 
when using different SED-modeling assumptions, in comparison with the 
random errors, are very significant ($>2~\sigma$). At $2.0 \leq z < 3.0$, 
the changes are slightly less significant, but still at the $\sim2~\sigma$ 
level, while at $3.0 \leq z < 4.0$, where Poisson uncertainties are very 
large, the changes are mostly at the $1~\sigma$ level.
When the $1/V_{\rm max}$ method is used, cosmic variance is the dominant 
source of random errors at $1.3 \leq z < 2.0$ in all stellar mass bins, 
but it becomes comparable to the Poisson errors at $2.0 \leq z < 3.0$. The 
contribution of photometric redshift random uncertainties to the total 
error budget is generally smaller than Poisson errors, and increases 
going to higher redshifts. The relative contribution of systematic 
uncertainties is smallest at the highest targeted redshift interval, 
$3.0 \leq z < 4.0$, where random errors contribute significantly to the 
total error budget.

If the systematic uncertainties are included, the results highlighted 
in \S~\ref{subsec-mf} are no longer robust. In particular, we cannot 
exclude a strong evolution (by as much as a factor of $\sim50$) in the 
number density of the most massive ($M_{\rm star} > 10^{11.5}$) galaxies 
from $z=4.0$ to $z=1.3$. We note that the effects of systematic 
uncertainties due to different SED-modeling assumptions are likely 
smaller when the redshift evolution is considered, as some errors would 
cancel out when comparing the SMFs at two different epochs. However, as 
the metallicity, IMF, and appropriate extinction law may all evolve with 
redshift, it is unclear to what extent this cancellation of errors 
actually occurs.

%----------------------------------------------------------------------------

\subsection{Comparison with previous works} \label{subsec-comp}

\begin{figure*}
\centering
\epsscale{0.8}
%\plotone{f12.eps}
\includegraphics[width=16cm]{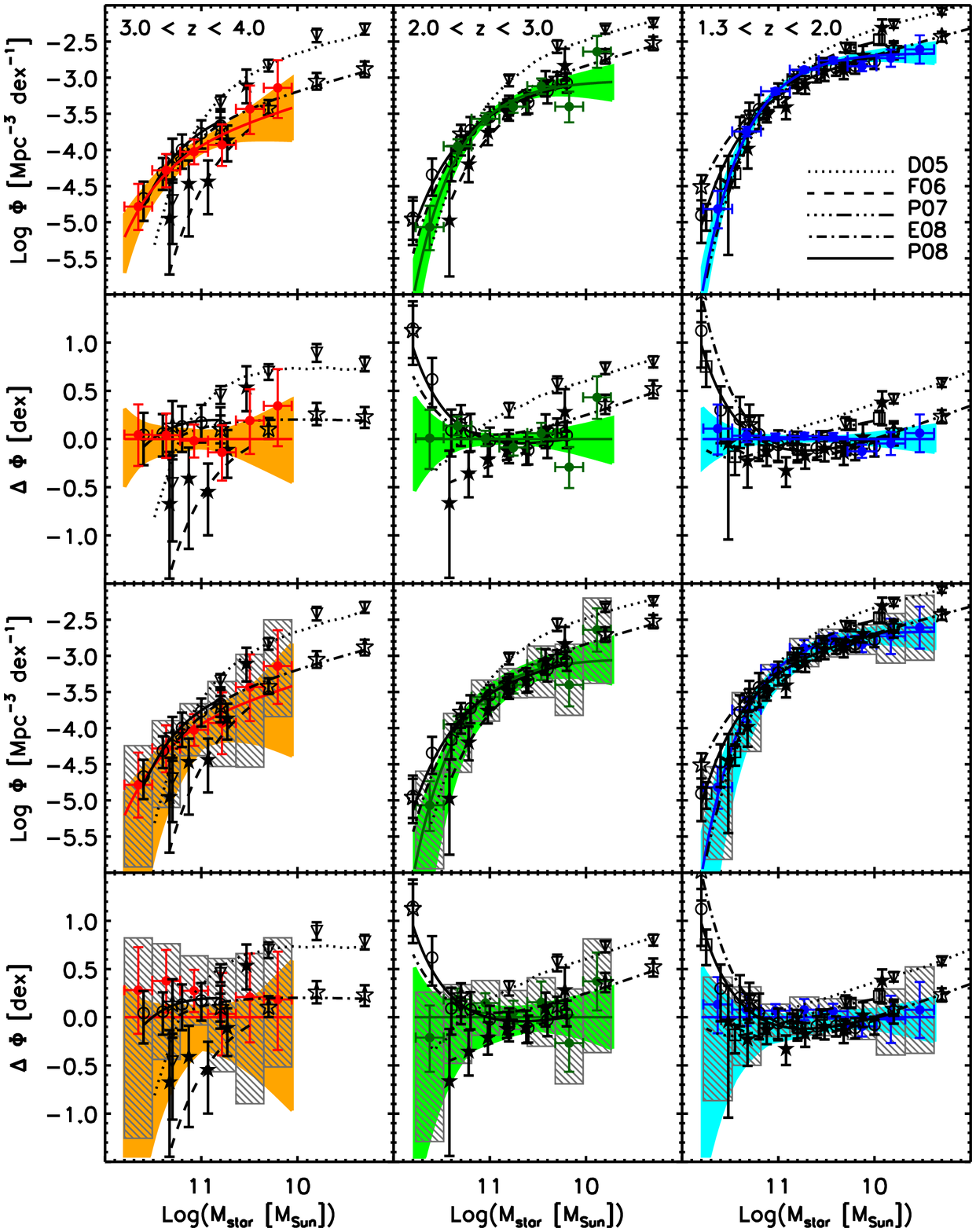}
\caption{Comparison between the SMFs derived from this work and 
previous measurements from the literature. {\it First row:} The 
SMFs derived from this work are shown as filled red, dark green, 
and blue circles ($1/V_{\rm max}$ method) and solid curves 
(maximum-likelihood analysis). The 1~$\sigma$ error bars of the 
$1/V_{\rm max}$ measurements include Poisson errors and uncertainties 
from photometric redshift random uncertainties, but not cosmic variance 
and systematic uncertainties. Similarly, the 1~$\sigma$ error of the 
maximum-likelihood measurements (orange, green, and cyan shaded 
regions) do not include systematic uncertainties. Previous works are 
plotted as filled stars and dashed curves (\citealt{fontana06}; F06); 
open circles and solid curves (\citealt{perez08}; P08); open stars and 
dot-dashed curves (\citealt{elsner08}; E08); open triangles and dotted 
curves (\citealt{drory05}; D05); open squares and long dot-dashed curves 
(\citealt{pozzetti07}; P07). {\it Second row:} Symbols as in the 
first row panels, but now the differences between the SMFs from the 
literature and those derived in this work,
$\Delta \Phi = \log{\Phi_{\rm others}} - \log{\Phi_{\rm ours}}$, are plotted 
as a function of stellar mass. {\it Third and forth row panels:} Symbols 
as in first and second row panels, respectively, with cosmic variance added 
in quadrature to the error bars, and systematic uncertainties now included 
in the total error budget represented by the shaded gray boxes (for the 
$1/V_{\rm max}$ points) and shaded orange, green, and cyan regions (for the 
maximum-likelihood measurements). The systematic effects due to the 
bottom-light IMFs are not included. Most of the disagreements between the 
different measurements of the SMFs stem from an incomplete analysis of the 
errors. A comprehensive analysis of random and systematic uncertainties is 
necessary to reconcile the different measurements of the high-$z$ SMFs.
\label{fig-all_comp}}
\end{figure*}

Figure~\ref{fig-all_comp} shows the comparison of the SMFs 
derived in this study with other works from the literature 
(see Appendix~\ref{app-comp} for a detailed discussion of the 
individual works). As discussed above, cosmic variance and 
systematic uncertainties dominate the errors. However, most 
literature studies do not give estimates for these errors. 
Therefore, we show the comparison twice in Figure~\ref{fig-all_comp}: 
the panels in the top two rows include only Poisson errors and 
uncertainties due to photometric redshift random errors, while the 
panels in the bottom two rows include all sources of error (with the 
exclusion of the systematic effects due to the bottom-light IMFs). 
To highlight similarities and differences between our SMFs and 
those derived in other works, we also plot 
$\Delta \Phi = \log{\Phi_{\rm others}} - \log{\Phi_{\rm ours}}$ 
as function of stellar mass in the second- and fourth-row panels.
We note that the error bars of the different surveys cannot be 
compared directly, as they were not derived in a uniform way.

From the top panels of Figure~\ref{fig-all_comp}, it is obvious 
that our SMFs agree with those from the literature for some 
redshift and stellar mass ranges, but disagree for others. Our 
SMFs are generally in good agreement with those from 
\citet{elsner08}, \citet{perez08}, and \citet{pozzetti07}. Broad 
agreement is also found with the SMFs at $z<3$ from \citet{fontana06}, 
and with the high-mass end of the SMFs at $z<3$ from \citet{drory05}.
However, there is also significant disagreement between our SMFs and 
those from the literature, and also between the works in the literature 
themselves, for some redshift and stellar mass ranges. The disagreements 
between the different SMFs increase with increasing redshift. Our SMFs 
fall somewhere in the middle of the SMFs from the literature. 
The largest disagreement is with the SMFs from \citet{drory05} at the 
low-mass end at all redshifts. At the high-mass end, the largest 
disagreement is with the SMFs from \citet{fontana06} at $z\sim3.5$, 
from \citet{perez08} at $z\sim2.5$, and from \citet{elsner08} at 
$z\sim1.6$. The large differences between the SMFs from \citet{fontana06} 
and \citet{elsner08} are interesting, since both were derived from the 
GOODS-MUSIC catalog. The former was derived from a $K$-selected catalog, 
while the latter from a $z$-selected catalog. We note, however, that 
\citet{fontana06} claim that their $z$-selected SMF is very similar to 
their $K$-selected one.

Once the systematic uncertainties are taken into account, as shown 
in the bottom panels of Figure~\ref{fig-all_comp}, the SMFs derived 
in this work become consistent with most of the SMFs from the 
literature. The low-mass end of the SMFs at $z\sim2.5$ and $z\sim1.6$ 
from \citet{drory05} is still significantly steeper with respect to 
both our SMFs and the other SMFs from the literature. A possible 
explanation is the different way the completeness limits in stellar 
mass has been derived by \citet{drory05} (SSP-derived completeness), 
as this potentially over-corrects densities at the low-mass end. 

We stress that most of the disagreements between the different 
measurements of the SMFs stem from an incomplete analysis of the 
errors. The errors on the SMFs from the literature include only 
Poisson errors (e.g., \citealt{drory05}), or Poisson errors and 
errors from photometric redshift uncertainties (but not cosmic 
variance, e.g. \citealt{fontana06}; \citealt{perez08}; 
\citealt{elsner08}). Field-to-field variations is a significant 
source of errors when the SMF is derived using the $1/V_{\rm max}$ 
method. This is true at all redshifts, especially at the high-mass 
end, but cosmic variance dominates the error budget at $z\sim1.6$. 
The maximum-likelihood estimator, unbiased with respect to density 
inhomogeneities, has been applied only by \citet{fontana06} and 
\citet{pozzetti07}, while the other works have simply fitted the 
SMFs derived with the $1/V_{\rm max}$ with a Schechter function. 
Finally, it is extremely important to include the systematic 
uncertainties due to different SED-modeling assumptions, which 
dominate the total error budget and are necessary to reconcile 
the different measurements of the high-$z$ SMFs. 

%============================================================================

\section{Densities} \label{sec-densities}

In this section we present estimates of the stellar mass density 
$\rho_{\rm star}$ derived by integrating the best-fit Schechter 
function obtained from the maximum-likelihood analysis (no use of 
the $1/V_{\rm max}$ results has been done in the estimate of the 
stellar mass densities). The stellar mass density (obtained by 
integrating the SMF derived from the maximum-likelihood analysis) 
is a more robust measurement than the Schechter function parameters 
$M_{\rm star}^{\star}$, $\alpha$, and $\Phi^{\star}$, because the errors 
in these parameters are highly correlated. Stellar mass densities have 
been estimated adopting three different integration intervals: 
$\rho_{\rm star}^{\rm 8<M<13}$, where the integration was 
performed for stellar masses $10^{8}<M_{\rm star}/M_{\sun}<10^{13}$; 
$\rho_{\rm star}^{\rm 10<M<11}$, where the integration was performed 
for stellar masses $10^{10}<M_{\rm star}/M_{\sun}<10^{11}$; and 
$\rho_{\rm star}^{\rm 11<M<12}$, where the integration was performed 
for stellar masses $10^{11}<M_{\rm star}/M_{\sun}<10^{12}$, i.e. 
massive galaxies. These estimates are listed in Table~\ref{tab-mdall}, 
along with the 1~$\sigma$ errors and the values of the stellar 
mass density at $z\sim0.1$ estimated from the SMF of \citet{cole01}.
Note that the contribution of galaxies less massive than 
$M_{\rm star}=10^{8}$~M$_{\sun}$ to the global stellar mass density 
is negligible if the Schechter parameterization of the SMF is a good 
approximation and valid also at stellar masses smaller than probed by 
our composite sample.

The 1~$\sigma$ errors on the stellar mass densities have been 
estimated by deriving the distribution of all of the values of 
$\rho_{\rm star}$ allowed within the 1~$\sigma$ solutions of the Schechter 
function parameters from the maximum-likelihood analysis. The 
contribution to the total error budget from photometric redshift random 
uncertainties (derived from the 100 Monte Carlo realizations described 
in \S~\ref{sec-photozerrs}) was added in quadrature. The 1~$\sigma$ 
errors including the systematic uncertainties were estimated in the same 
way by deriving the distribution of all of the values of $\rho_{\rm star}$ 
allowed within the 1~$\sigma$ solutions obtained using different 
SED-modeling assumptions and different combinations of template sets and 
template error functions to estimate the photometric redshifts.

\begin{deluxetable*}{cccc}
\centering
\tabletypesize{\small}
\tablecaption{Stellar mass densities \label{tab-mdall}}
\tablehead{\colhead{Redshift} & 
  \colhead{$\log{\rho_{\rm star}^{\rm  8<M<13}}$} & 
  \colhead{$\log{\rho_{\rm star}^{\rm 10<M<11}}$} &
  \colhead{$\log{\rho_{\rm star}^{\rm 11<M<12}}$} \\
                    Range          &  (M$_{\sun}$~Mpc$^{-3}$)  & 
          (M$_{\sun}$~Mpc$^{-3}$)  & 
          (M$_{\sun}$~Mpc$^{-3}$)  }
\startdata
$z \sim 0.1$       & $8.51\pm0.07$ 
                   & $8.27\pm0.07$
                   & $7.92\pm0.07$ \\
$1.3 \leq z < 2.0$ & $7.91^{+0.02(0.02,0.02)}_{-0.02(0.35,0.16)}$ 
                   & $7.68^{+0.02(0.04,0.04)}_{-0.02(0.40,0.16)}$ 
                   & $7.38^{+0.07(0.09,0.08)}_{-0.06(0.62,0.38)}$ \\
$2.0 \leq z < 3.0$ & $7.55^{+0.05(0.12,0.12)}_{-0.04(0.43,0.18)}$ 
                   & $7.31^{+0.05(0.11,0.11)}_{-0.05(0.48,0.14)}$ 
                   & $7.07^{+0.08(0.13,0.13)}_{-0.09(1.10,0.39)}$ \\
$3.0 \leq z < 4.0$ & $7.27^{+0.37(0.93,0.93)}_{-0.13(0.61,0.39)}$ 
                   & $6.88^{+0.19(0.19,0.19)}_{-0.22(0.67,0.50)}$ 
                   & $6.91^{+0.13(0.13,0.13)}_{-0.16(0.94,0.73)}$ \\
\enddata
\tablenotetext{}{Stellar mass density estimated by integrating 
the best-fit Schechter SMF over the specified stellar 
mass range. The quoted 1~$\sigma$ errors include Poisson 
errors and errors due to photometric redshift uncertainties; 
the numbers in parenthesis are the 1~$\sigma$ errors including 
the systematic uncertainties due to different SED-modeling 
assumptions and different combination of template set and template 
error function used in the estimate of the photometric redshifts; 
in the second number in parenthesis, the effect of the bottom-light 
IMFs is excluded.}
\end{deluxetable*}

\begin{figure}
\centering
\epsscale{1.1}
\plotone{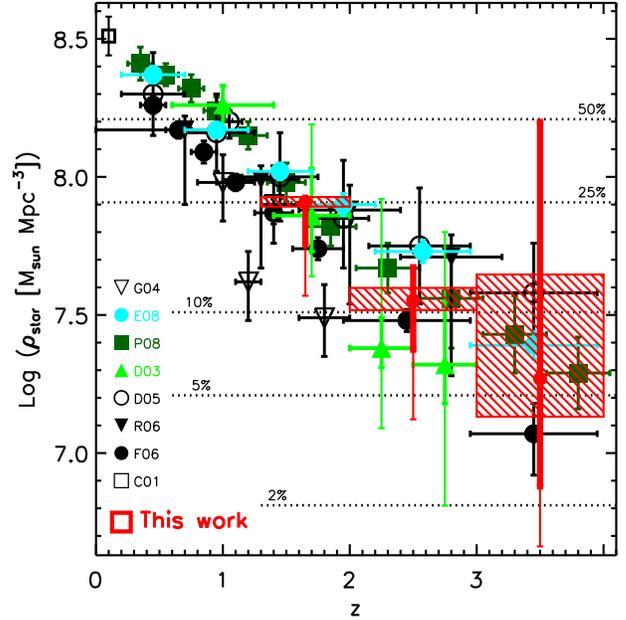}
\caption{Evolution as a function of redshift of the global 
stellar mass density estimated by integrating the SMFs over 
the stellar mass range $10^{8}<M_{\rm star}/M_{\sun}<10^{13}$
(the contribution to the total stellar mass density of galaxies 
with $M_{\rm star}<10^{8}$~M$_{\sun}$ is negligible if the Schechter 
parameterization of the SMF is valid also at stellar masses smaller 
than probed by our sample). Red symbols represent the total stellar 
mass densities estimated in this work (shaded boxes do not include 
the systematic uncertainties while error bars do; the thick error 
bars do not include the systematic effect of the two bottom-light 
IMFs, while the thin error bars do). The estimates of the total 
stellar mass densities from the literature were taken from 
\citet{cole01} (open square); 
\citet{dickinson03} (filled green triangles; thick error bars 
include random errors and cosmic variance alone, while the 
thin error bars include the systematic uncertainties due to 
different SED-modeling assumptions); \citet{glazebrook04} (open 
upside-down triangles; error bars including Poisson errors only); 
\citet{drory05} (open circles); \citet{elsner08} (filled cyan circles); 
\citet{fontana06} (filled circles); \citet{perez08} (filled dark green 
squares); and \citet{rudnick06} (filled  upside-down triangles; error 
bars including random errors and uncertainties due to cosmic variance). 
For the measurements from the literature, only Poisson errors and errors 
due to photometric redshift uncertainties (as derived in the corresponding 
work), are plotted without uncertainties due to cosmic variance and 
systematic errors, unless stated otherwise. The horizontal dotted lines 
represent 50\%, 25\%, 10\%, 5\%, and 2\% (from top to bottom, respectively) 
of the total stellar mass density at $z=0.1$. The stellar mass 
density has increased by a factor $\sim$17$^{+7}_{-10}$, $\sim$9$\pm1$, and 
$\sim$4$\pm0.2$ from $z=3.5$, $z=2.5$, and $z=1.65$, respectively, down to 
$z=0.1$. A much stronger evolution with redshift of the stellar 
mass density is however allowed once the systematic uncertainties 
are taken into account. \label{fig-massdens}}
\end{figure}

In Figure~\ref{fig-massdens} we show the evolution of the total 
stellar mass density as a function of redshift, together with 
a compilation of results from the literature. The values from the 
literature, derived assuming a \citet{salpeter55} IMF, have been scaled 
to a pseudo-\citet{kroupa01} IMF by dividing the stellar mass densities 
by a factor of 1.6. The values from the literature of the stellar mass 
density converted to our IMF are listed in Table~\ref{tab-othersmd}. The 
measured evolution of the total stellar mass 
density from $z=4.0$ to $z=1.3$ is broadly consistent with most previous 
measurements in the literature. Our measurements are among the currently 
most accurate measurements of the total stellar mass density at these 
redshifts. This is due to the large surveyed area, the large number of 
independent fields, and the high-quality of the optical-to-MIR data. Only 
the value at $z\sim3.5$ is characterized by a large random error, due to 
the large uncertainties on the low-mass end slope of the derived SMF. 
Moreover, our measurements are the first to include a comprehensive 
analysis of random and systematic uncertainties in the error budget.
Note that the mass density at $z>3$ is very poorly constrained when 
systematic uncertainties are included. We stress that this arises despite 
the fact that our sample is arguably the best suited for studying the total 
stellar mass density (as it samples both the low-mass and high-mass end in 
a homogeneous way). All previous studies in the literature suffer from 
similar, or larger, uncertainties.

\begin{deluxetable}{cccc}
\centering
\tabletypesize{\small}
\tablecaption{Stellar mass densities from the literature\label{tab-othersmd}}
\tablehead{\colhead{Redshift} & 
  \colhead{$\log{\rho_{\rm star}^{\rm  8<M<13}}$} & \colhead{Redshift} & 
  \colhead{$\log{\rho_{\rm star}^{\rm  8<M<13}}$} \\
                    Range          &  (M$_{\sun}$~Mpc$^{-3}$)  &     Range          &  (M$_{\sun}$~Mpc$^{-3}$)}
\startdata
%\multicolumn{2}{c}{\citet{dickinson03}}          & \multicolumn{2}{c}{\citet{fontana06}}\\
%$0.6<z<1.4$ & 8.26$\pm$0.08                      & $1.3<z<1.6$ & 7.87$\pm$0.05\\
%$1.4<z<2.0$ & 7.86$^{+0.17(0.33)}_{-0.13(0.22)}$ & $1.6<z<2.0$ & 7.74$\pm$0.04\\
%$2.0<z<2.5$ & 7.58$^{+0.11(0.54)}_{-0.07(0.29)}$ & $2.0<z<3.0$ & 7.48$\pm$0.04\\
%$2.5<z<3.0$ & 7.52$^{+0.23(0.48)}_{-0.14(0.51)}$ & $3.0<z<4.0$ & 7.17$^{+0.11}_{-0.15}$\\
%\multicolumn{2}{c}{\citet{rudnick06}}            & \multicolumn{2}{c}{\citet{elsner08}}\\
%$0.0<z<1.0$ & 8.17$^{+0.05}_{-0.27}$             & $1.25<z<1.75$ & 8.02$\pm$0.03\\
%$1.0<z<1.6$ & 7.99$^{+0.05}_{-0.32}$             & $1.75<z<2.25$ & 7.90$\pm$0.04\\
%$1.6<z<2.4$ & 7.88$^{+0.09}_{-0.34}$             & $2.25<z<3.00$ & 7.73$\pm$0.04\\
%$2.4<z<3.2$ & 7.71$^{+0.08}_{-0.43}$             & $3.00<z<4.00$ & 7.39$\pm$0.05\\
%\multicolumn{2}{c}{\citet{perez08}}              & \multicolumn{2}{c}{\citet{drory05}}\\
%$0.8<z<1.0$ & 8.24$\pm$0.05                      & $1.0<z<1.2$ & 8.20$\pm$0.06\\
%$1.0<z<1.3$ & 8.15$\pm$0.05                      & $1.2<z<1.8$ & 8.00$\pm$0.16\\
%$1.3<z<1.6$ & 7.95$\pm$0.07                      & $1.8<z<2.2$ & 7.85$\pm$0.20\\
%$1.6<z<2.0$ & 7.82$\pm$0.07                      & $2.2<z<3.0$ & 7.75$\pm$0.20\\
%$2.0<z<2.5$ & 7.67$\pm$0.08                      & $3.0<z<4.0$ & 7.58$\pm$0.20\\
%$2.5<z<3.0$ & 7.56$\pm$0.18                      & & \\
%$3.0<z<3.5$ & 7.43$\pm$0.14                      & & \\
%$3.5<z<4.0$ & 7.29$\pm$0.13                      & & \\
\multicolumn{2}{c}{\citet{dickinson03}}          & \multicolumn{2}{c}{\citet{glazebrook04}}\\
$0.6<z<1.4$ & 8.26$\pm$0.08                      & $0.8<z<1.1$ & 7.98$^{+0.10}_{-0.14}$\\
$1.4<z<2.0$ & 7.86$^{+0.17(0.33)}_{-0.13(0.22)}$ & $1.1<z<1.3$ & 7.62$^{+0.11}_{-0.14}$\\
$2.0<z<2.5$ & 7.58$^{+0.11(0.54)}_{-0.07(0.29)}$ & $1.3<z<1.6$ & 7.90$^{+0.14}_{-0.14}$\\
$2.5<z<3.0$ & 7.52$^{+0.23(0.48)}_{-0.14(0.51)}$ & $1.6<z<2.0$ & 7.49$^{+0.12}_{-0.14}$\\
\multicolumn{2}{c}{\citet{drory05}}          & \multicolumn{2}{c}{\citet{fontana06}}\\
$0.25<z<0.75$ & 8.30$\pm$0.15                & $0.4<z<0.6$ & 8.26$\pm$0.03\\
$0.75<z<1.25$ & 8.16$\pm$0.15                & $0.6<z<0.8$ & 8.17$\pm$0.02\\
$1.25<z<1.75$ & 8.00$\pm$0.16                & $0.8<z<1.0$ & 8.09$\pm$0.03\\
$1.75<z<2.25$ & 7.85$\pm$0.20                & $1.0<z<1.3$ & 7.98$\pm$0.02\\
$2.25<z<3.00$ & 7.75$\pm$0.20                & $1.3<z<1.6$ & 7.87$\pm$0.05\\
$3.00<z<4.00$ & 7.58$\pm$0.20                & $1.6<z<2.0$ & 7.74$\pm$0.04\\
\multicolumn{2}{c}{\citet{perez08}}          & $2.0<z<3.0$ & 7.48$\pm$0.04\\
$0.2<z<0.4$ & 8.41$\pm$0.06                  & $3.0<z<4.0$ & 7.07$^{+0.11}_{-0.15}$\\
$0.4<z<0.6$ & 8.37$\pm$0.04                  & \multicolumn{2}{c}{\citet{rudnick06}}\\
$0.6<z<0.8$ & 8.32$\pm$0.05                  & $0.0<z<1.0$ & 8.17$^{+0.05}_{-0.27}$\\
$0.8<z<1.0$ & 8.24$\pm$0.05                  & $1.0<z<1.6$ & 7.99$^{+0.05}_{-0.32}$\\
$1.0<z<1.3$ & 8.15$\pm$0.05                  & $1.6<z<2.4$ & 7.88$^{+0.09}_{-0.34}$\\
$1.3<z<1.6$ & 7.95$\pm$0.07                  & $2.4<z<3.2$ & 7.71$^{+0.08}_{-0.43}$\\
$1.6<z<2.0$ & 7.82$\pm$0.07                  & \multicolumn{2}{c}{\citet{elsner08}}\\
$2.0<z<2.5$ & 7.67$\pm$0.08                  & $0.25<z<0.75$ & 8.37$\pm$0.03\\
$2.5<z<3.0$ & 7.56$\pm$0.18                  & $0.75<z<1.25$ & 8.17$\pm$0.02\\
$3.0<z<3.5$ & 7.43$\pm$0.14                  & $1.25<z<1.75$ & 8.02$\pm$0.03\\
$3.5<z<4.0$ & 7.29$\pm$0.13                  & $1.75<z<2.25$ & 7.90$\pm$0.04\\
            &                                & $2.25<z<3.00$ & 7.73$\pm$0.04\\
            &                                & $3.00<z<4.00$ & 7.39$\pm$0.05\\
\enddata
\tablenotetext{}{All the values from the literature 
have been divided by 1.6 to convert to the 
pseudo-\citet{kroupa01} IMF used in this work. The 
quoted errors include only Poisson errors and errors 
due to photometric redshift uncertainties, without 
uncertainties due to cosmic variance and systematic 
errors, unless stated otherwise. The quoted errors 
of \citet{dickinson03} include random errors and 
cosmic variance, with the numbers in parenthesis 
including the systematic uncertainties due to 
different SED-modeling assumptions; The quoted 
errors of \citet{rudnick06} include random errors 
and uncertainties due to cosmic variance.}
\end{deluxetable}

The total stellar mass density has increased by a factor of 
$\sim17^{+7}_{-10}$, $\sim9\pm1$, and $\sim4\pm0.2$, from redshift 
$z=3.5$, $z=2.5$, and $z=1.65$, respectively, down to redshift 
$z\sim0.1$. Due to the systematic uncertainties on the derived stellar 
mass densities (represented with thin error bars in 
Figure~\ref{fig-massdens}), a much stronger evolution with redshift of 
the total stellar mass density than previously measured is actually allowed. 
Systematic uncertainties still allow for an increase as large as a factor 
of $\sim70$, $\sim25$, and $\sim9$ of the global stellar mass density 
($\sim44$, $\sim14$, and $\sim6$, if the effects of the bottom-light IMFs 
are excluded) from redshift $z=3.5$, $z=2.5$, and $z=1.65$, respectively, 
to $z\sim0.1$. However, the effects of systematic uncertainties due to 
different SED-modeling assumptions are likely smaller when the redshift 
evolution is considered, as some errors would cancel out when comparing 
the stellar mass densities at two different epochs. 

If only random errors are taken into account, there are a few measurements 
from the literature that are in significant disagreement with ours. The 
stellar mass density in \citet{elsner08} at $z\sim2.5$ is significantly larger 
than our measurement by about 60\%. We note however that the error bars of 
the measurements from \citet{elsner08} only include random errors, but not 
uncertainties from cosmic variance and systematic errors. Because the 
analysis of \citet{elsner08} is based on the single, relatively small, 
field of GOODS-CDFS, field-to-field variance is a particular important 
source of errors. We stress the importance of a comprehensive analysis of 
the errors, including systematic uncertainties. Excluded our works, the 
only measurements of the stellar mass density including a complete analysis 
of all the errors, including systematic uncertainties, come from the work 
of \citet{dickinson03}. Because \citet{dickinson03} include both cosmic 
variance and systematic uncertainties in their error budget, they are 
consistent with ours, despite their estimates of the stellar mass densities 
lying systematically below ours. Our measurements are characterized by much 
smaller errors due to our much larger surveyed area (by a factor of $\sim80$). 

We finally note that at $z>2$, we cannot constrain the low-mass end of 
the SMFs very well, becoming incomplete at $\sim10^{10}$~M$_{\sun}$. 
If the SMF is much steeper at the low-mass end than indicated by our 
maximum-likelihood analysis, we could be missing a significant fraction 
of the integrated stellar mass. Very recently, \citet{reddy08} suggested 
that up to $\sim$50\% of the total stellar mass at $1.9 \le z<3.4$ is in 
faint-UV galaxies with masses smaller than $\sim10^{10}$~M$_{\sun}$ 
(compared to $\sim$10\%-20\% from an extrapolation of our Schechter fits). 
However, \citet{reddy08} do not measure stellar masses directly, but convert 
UV luminosity to stellar mass. Significantly deeper NIR observations are 
required to directly probe and constrain the low-mass end of the high-$z$ 
SMF.

%============================================================================

\section{COMPARISON WITH MODEL PREDICTIONS} \label{sec-comp_models}

In this section we compared the derived SMFs of galaxies at 
$1.3 \leq z < 4.0$ with those predicted by the latest generation 
of galaxy formation models. Specifically, we have included the 
predictions from the semi-analytic models of \citet{monaco07}, 
\citet{somerville08}, and \citet{wang08}, which include active 
galactic nuclei (AGN) feedback. We refer to those papers for 
detailed descriptions of their models, and to \citet{fontanot09}
 for a detailed comparison of these models. 
A \citet{chabrier03} IMF has been assumed in all models. We have 
therefore scaled their predictions to match our pseudo-Kroupa IMF 
by multiplying their stellar masses by 1.12. The model-predicted 
SMFs have been convolved with a normal distribution of standard 
deviation 0.25~dex, intended to represent measurement errors in 
$\log{M_{\rm star}}$.

The semi-analytic model of \citet{monaco07}, MORGANA, attempts, 
through modeling of cooling, star formation, feedback, galactic 
winds and superwinds, AGN activity and AGN feedback, to move from 
a phenomenological description of galaxy formation to a fully physically 
motivated one. We refer to \citet{monaco07} for a detailed description 
of all physical processes included in MORGANA. The predicted SMFs and 
stellar mass densities from the MORGANA model adopted in this work were 
derived assuming a WMAP-3 cosmology \citep{spergel07} and including 
some minor improvements with respect to \citet{monaco07} (Lo Faro et al., 
in prep.).

The model predictions from \citet{wang08} were derived using the Garching 
semi-analytic model implemented on the Millennium dark matter simulation 
described in \citet{springel05}. Specifically, the semi-analytic model 
described in \citet{delucia07} was used, which built on previous works by 
the ``Munich'' galaxy formation group (see \citealt{kauffmann00}; 
\citealt{springel01}; and \citealt{delucia04} for detailed descriptions 
of the scheme for building the merger tree and the prescriptions 
adopted to model the baryonic physics, most notably those associated 
with the growth of and the feedback from black holes in galaxy nuclei 
and the cooling model). Here, we specifically consider the ``C'' model in 
\citet{wang08}, which assumes a WMAP-3 cosmology. This change in cosmology 
results in a significant delay of structure formation in comparison with 
WMAP-1 results. Therefore, to compensate for the delay in structure 
formation, model ``C'' has twice as much the star formation efficiency 
with respect to the efficiency assumed by \citet{delucia07}. This 
increase in efficiency has to be compensated by much higher feedback 
efficiencies (both from supernovae and from AGN) to prevent the 
overproduction of stars at late times.

The semi-analytic model of \citet{somerville08}, built on the 
previous models described in \citet{somerville99} and \citet{somerville01}, 
presents several improvements, including, but not limited to, tracking 
of a diffuse stellar halo component built up of tidally destroyed 
satellites and stars scattered in mergers, galaxy-scale AGN-driven 
winds, fueling of black holes with hot gas via Bondi accretion, and 
heating by radio jets. The prediction from the \citet{somerville08} 
semi-analytic model are taken from their fiducial model WMAP-3 
model, which adopts a fraction $f_{\rm scatter}=0.4$ of the stars 
in merged satellite galaxies added to a diffuse component distributed 
in a very extended halo or envelope.

\begin{figure*}
\centering
\epsscale{0.95}
\plotone{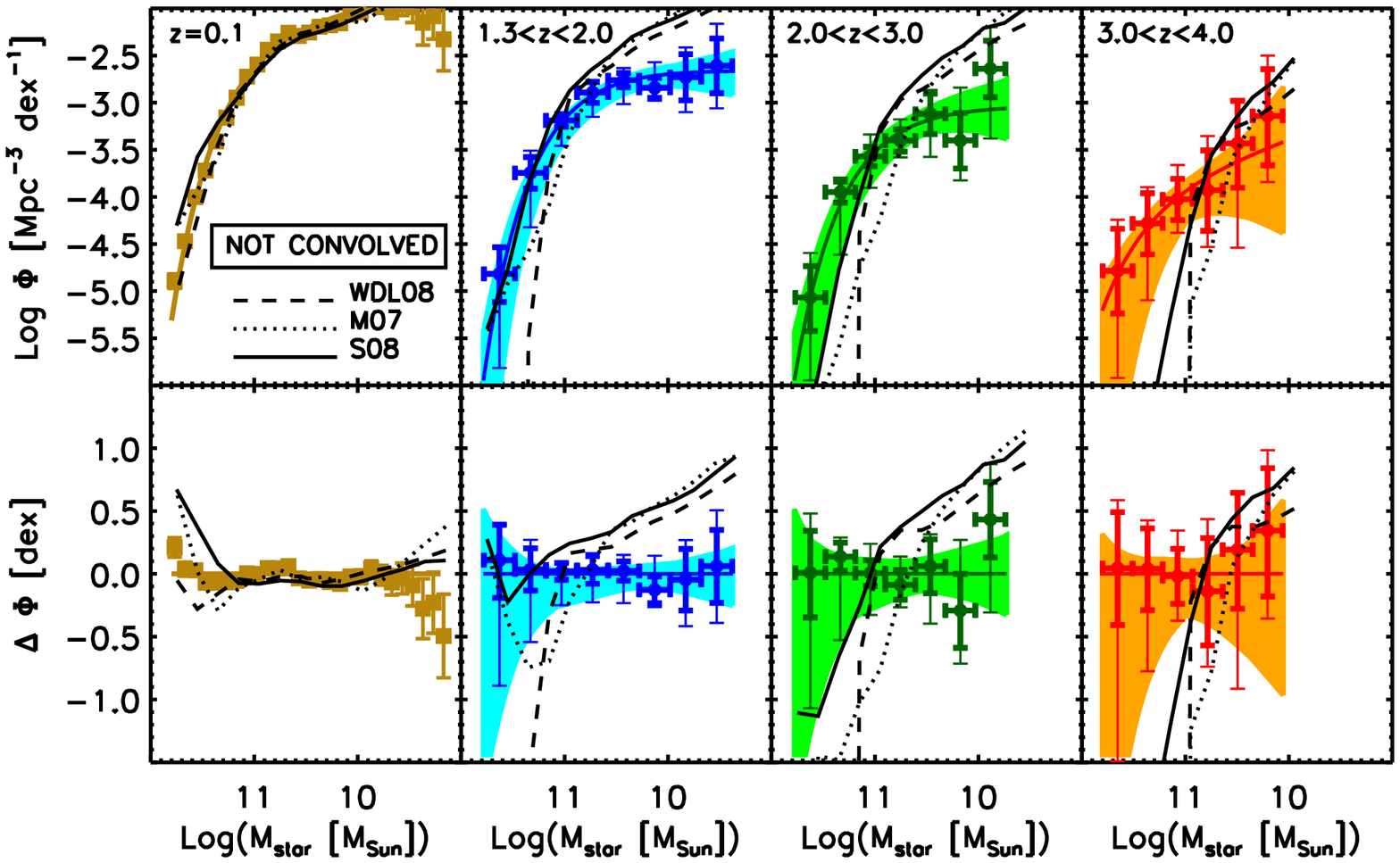}
\caption{Comparison between the observed SMFs and the semi-analytic 
model predicted SMFs not convolved with a normal distribution of 
standard deviation 0.25~dex. The predicted SMFs are represented 
by black curves: solid for the \citet{somerville08} model, dotted 
for the \citet{monaco07} model, and dashed for the \citet{wang08} 
model. Red, green, and blue filled circles represent the SMFs 
measured from the $1/V_{\rm max}$ method; the thick error bars include 
Poisson errors, cosmic variance, and the uncertainties from 
photometric redshift random errors; the thin error bars include the 
systematic uncertainties, with the exclusion of the effect due to the 
use of the bottom-light IMFs for consistency with the theoretical 
models. Red, green, and blue solid curves represent the SMFs measured 
from the maximum-likelihood analysis. The shaded regions represent the 
total 1~$\sigma$ errors, including the systematic uncertainties. The 
top panels show the comparisons between the observed and the 
model-predicted SMFs. The bottom panels show the differences between 
the model-predicted SMFs and those derived in this work, 
$\Delta \Phi = \log{\Phi_{\rm models}} - \log{\Phi_{\rm ours}}$, plotted 
as function of stellar mass. The SMFs predicted from the models are in 
general too steep, significantly over-predicting the number densities of 
galaxies below the characteristic stellar mass at all redshifts, and 
severely under-predicting the number density of the massive galaxies at 
$z>2.0$. The good agreement at $z=0.1$ is the direct result of the 
optimization of the free parameters in the models to match the $z=0$ 
universe.
\label{fig-SMFcomp_model}}
\end{figure*}

\begin{figure*}
\centering
\epsscale{0.95}
\plotone{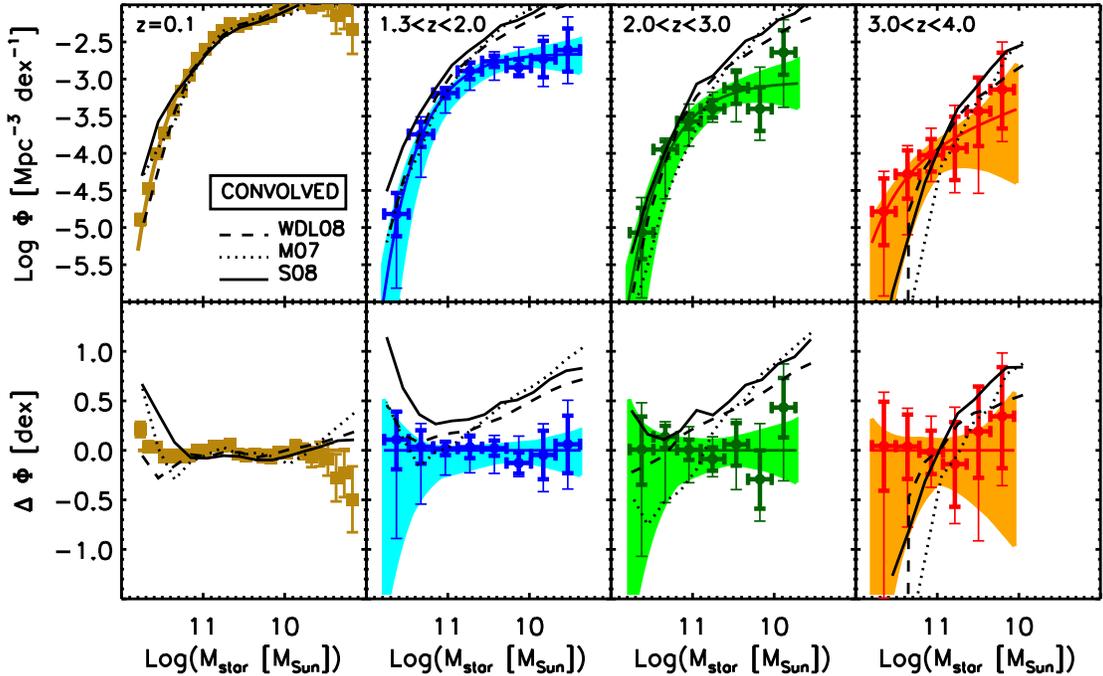}
\caption{Same as in Fig.~\ref{fig-SMFcomp_model}, except here the 
SMFs predicted by the semi-analytic models have been convolved 
with a normal distribution of standard deviation 0.25~dex; no 
convolution has been applied to the model-predict SMF at $z=0.1$. 
While the discrepancies at the low-mass end are still present, the 
discrepancies at the high-mass end are significantly reduced.
\label{fig-SMFcomp_modelB}}
\end{figure*}

Figures~\ref{fig-SMFcomp_model} and \ref{fig-SMFcomp_modelB} show the 
comparison between the SMFs measured in this work (with the exclusion 
of the systematic effects due to the bottom-light IMFs) and those 
predicted by the models. Figure~\ref{fig-SMFcomp_model} shows the 
comparison for the model-predicted SMFs without convolution with a 
normal distribution, while Figure~\ref{fig-SMFcomp_modelB} shows the 
comparison for the model-predicted SMFs convolved with a normal 
distribution of standard deviation 0.25~dex. To highlight similarities 
and differences between our SMFs and the model-predicted SMFs, we also 
plot $\Delta \Phi = \log{\Phi_{\rm models}} - \log{\Phi_{\rm ours}}$ as 
function of stellar mass in the bottom panels of 
Figures~\ref{fig-SMFcomp_model} and \ref{fig-SMFcomp_modelB}. Broadly 
speaking, the predicted SMFs are too steep with respect to the observed 
SMFs. If the comparison is done with the not-convolved model-predicted 
SMFs (see Figure~\ref{fig-SMFcomp_model}), in the redshift range 
$1.3 \leq z < 2.0$, where the high-mass end is reasonably well reproduced 
by all but the model of \citet{wang08}, all models significantly 
over-predict the number density of galaxies below the characteristic 
stellar mass. This is true also in the redshift range $2.0 \leq z < 3.0$, 
but now all models also significantly under-predicting the number densities 
of the massive galaxies. The disagreement at the high-mass end is even more 
pronounced in the redshift range $3.0 \leq z < 4.0$, where all models show 
a significant deficiency of massive galaxies with respect to observations. 
If the comparison between the observed and the model-predicted SMFs is 
instead performed using the convolved model-predicted SMFs, intended to 
represent measurement errors in $\log{M_{\rm star}}$ (see 
Figure~\ref{fig-SMFcomp_modelB}), the disagreements at the high-mass end are 
significantly reduced, although the model of \citet{somerville08} now tends 
to over-predict the number density of galaxies in the redshift range 
$1.3 \leq z < 2.0$ at all stellar masses, while the model of \citet{monaco07} 
still significantly under-predicts the number density of massive galaxies at 
$3.0 \leq z < 4.0$. Convolving the model-predicted SMFs does not instead 
help at all to solve the discrepancies below the characteristic stellar mass. 
These results are in qualitatively agreement with the comparison of observed 
and predicted rest-frame optical luminosity functions performed at 
$2.0 \leq z \leq 3.3$ by \citet{marchesini07b}, who found that all models 
significantly over-predict the observed number density of galaxies at the 
faint-end. 

As already pointed out in \citet{somerville08}, potentially serious 
discrepancies, common to all of the CDM-based semi-analytic models, are 
connected with low-mass galaxies. This is indeed what we find, although 
the presence of a significant population of very massive galaxies out to 
redshift $z\sim3-4$ and the little observed evolution in its number 
densities from $z=4.0$ to $z=1.3$ highlight other potential problems 
within the theoretical models. 

\begin{figure}
\centering
\epsscale{1.15}
\plotone{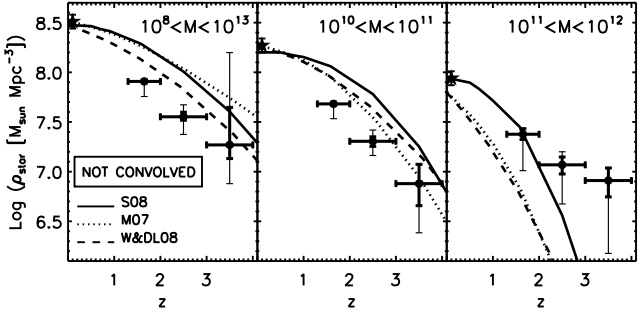}
\caption{Comparison of the observed stellar mass densities with 
those predicted by the semi-analytic models as function of redshift, 
with no convolution with a normal distribution of standard deviation 
0.25~dex applied to the model-predicted SMFs. Filled circles represent 
the observed stellar mass densities, while the curves are those 
predicted by the models. The left, middle, and right panels show the 
comparison between observations and model predictions for 
$\rho_{\rm star}^{\rm 8<M<13}$, $\rho_{\rm star}^{\rm 10<M<11}$, and 
$\rho_{\rm star}^{\rm 11<M<12}$, respectively. The thick error bars include 
Poisson errors, cosmic variance, and the uncertainties from photometric 
random errors; the thin error bars include the systematic uncertainties, 
with the exclusion of the effect due to the use of the bottom-light IMFs. 
The models fail in reproducing the observed evolution of the stellar mass 
density of low-mass galaxies ($10^{10} < M_{\rm star}/M_{\sun} < 10^{11}$), 
and severely under-predict the stellar mass density of massive galaxies 
($10^{11} < M_{\rm star}/M_{\sun} < 10^{12}$). 
\label{fig-MDcomp_model}}
\end{figure}

\begin{figure}
\centering
\epsscale{1.15}
\plotone{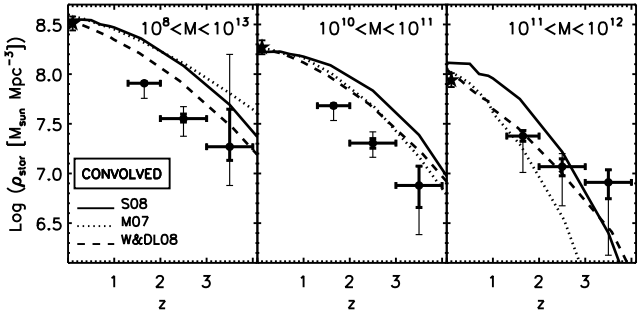}
\caption{Same as Fig.~\ref{fig-MDcomp_model}, except here the 
predictions of the semi-analytic models are convolved with a 
normal distribution of standard deviation 0.25~dex. While all 
models fail in reproducing the observed evolution of the stellar 
mass density of low-mass galaxies 
($10^{10} < M_{\rm star}/M_{\sun} < 10^{11}$), the disagreements 
between the observed and predicted stellar mass densities of 
massive galaxies are significantly reduced.
\label{fig-MDcomp_modelB}}
\end{figure}

The different comparison between observed and predicted SMFs in the two 
stellar mass regimes (below and above $M_{\rm star}\sim10^{11}$~M$_{\sun}$) 
is also shown in Figures~\ref{fig-MDcomp_model} and \ref{fig-MDcomp_modelB}, 
where the evolution of $\rho_{\rm star}^{\rm 10<M<11}$ and 
$\rho_{\rm star}^{\rm 11<M<12}$ as function of redshift are compared to the 
predictions from the semi-analytic models before and after convolution of 
the model-predicted SMFs with a normal distribution of standard deviation 
0.25~dex, respectively.

If no convolution with a normal distribution of standard deviation 0.25~dex 
is applied to the model-predicted SMFs (Figure~\ref{fig-MDcomp_model}), 
all models severely under-predict the stellar mass density of the massive 
galaxies at all redshifts. Only the model of \citet{somerville08} well match 
the stellar mass density of massive galaxies at $1.3 \leq z < 2.0$, although 
it suffers from the same deficit of massive galaxies at $z>2.0$. As in the 
comparison with the SMFs, if the convolved model-predicted SMFs are adopted 
(Figure~\ref{fig-MDcomp_modelB}), the disagreements for the stellar mass 
densities of massive galaxies is significantly reduced, especially for the 
model of \citet{wang08} which now reasonably well reproduces the evolution 
of the stellar mass density of massive galaxies over the entire redshift 
range $0.0 \leq z < 4.0$. However, the model of \citet{monaco07} still 
under-predicts significantly the stellar mass density of massive galaxies 
at $3.0 \leq z < 4.0$, while the model of \citet{somerville08} now 
over-predicts the stellar mass density of massive galaxies at 
$1.3 \leq z < 2.0$. The moderate success (after convolution) at the 
high-mass end is however counter-balanced by the failure in predicting the 
evolution of the stellar mass density of galaxies below the characteristic 
stellar mass ($10^{10} < M_{\rm star}/M_{\sun} < 10^{11}$). Over the entire 
redshift interval $1.3 \leq z < 4.0$, all models over-predict the observed 
stellar mass density of low-mass galaxies by a factor of a few, with the 
predictions from the \citet{somerville08} model showing the largest 
disagreements with the observations.

We conclude that the models succeed for some redshift and stellar mass 
ranges and fail for others, although this may partly be due to systematic 
errors in the observed SMFs and stellar mass densities (see 
\S~\ref{sec-errors} and \S~\ref{sec-densities}). In general, the 
model-predicted SMFs are too steep compared to the observed SMFs, 
resulting in a significant over-prediction of the number densities of 
low-mass galaxies. Models also fail in predicting the observed number 
density of massive galaxies observed at $3.0 \leq z < 4.0$.

%============================================================================

\section{SUMMARY AND CONCLUSIONS} \label{sec-concl}

In this paper we have measured the stellar mass functions of galaxies 
at redshifts $1.3 \leq z < 4.0$ from a composite $K$-selected sample 
constructed with the deep NIR MUSYC, the ultra-deep FIRES, and the 
GOODS-CDFS surveys, all having very high-quality optical-to-MIR data. 
This sample is unique in that it combines data from surveys with a large 
range of depths and areas in a self-consistent way.
%for its combination of depth and surveyed area. 
  
The total effective surveyed area of $\sim511$~arcmin$^{2}$ over several 
independent field of views allowed us to minimize the uncertainties 
due to cosmic variance, and empirically quantify its contribution 
to the total error budget. Moreover, we were able to probe the 
high-mass end of the SMFs with unprecedented good statistics. The 
three adopted surveys allowed us to empirically derive the 
redshift-dependent completeness limits of the $K$-selected samples by 
exploiting their different depths. This is a significant improvement 
with respect to previous studies, since it does not rely on stellar 
population synthesis models to assess the completeness in stellar 
mass of the sample. Finally, the ultra-deep FIRES survey allowed 
us to probe the low-mass end of the SMF down to stellar masses as 
small as $\sim0.05$ times the characteristic stellar mass.

We provide, for the first time, a comprehensive analysis of all 
random and systematic uncertainties affecting the derived SMFs. 
We have quantified the uncertainties on the SMF due to photometric 
redshift random errors by repeating the full analysis on a set of 
100 mock $K$-selected catalogs created by perturbing the observed 
photometry of each object according to its formal errors. Systematic 
uncertainties due to different sets of SED-modeling assumptions 
and different input parameters in the estimate of the photometric 
redshifts have also been quantified by changing IMF, metallicity, 
stellar population synthesis model, extinction curve in the SED-modeling, 
and template set and template error function in EAZY, the code used 
to estimate photometric redshifts.

We found that the stellar mass density evolves by a factor of 
$\sim17^{+7}_{-10}$ from $z=4.0$, and a factor of $\sim4\pm0.2$ since $z=1.3$. 
The observed evolution appears to be mostly driven by a change in the 
normalization $\Phi^{\star}$ of the SMF, rather than a change in the slope 
$\alpha$ or the characteristic stellar mass $M^{\star}_{\rm star}$, especially 
since redshift $z=3.0$. 
This is by itself a very interesting result, since it implies that the 
physical processes responsible in the building up of the stellar content 
of galaxies since $z=3.0$ do not change significantly the shape of the SMF 
(defined by $\alpha$ and $M^{\star}_{\rm star}$). The shape of the SMF is 
instead quite different at $z\sim3.5$, with a much steeper low-mass end 
slope. We note however that the SMF at $3.0 \leq z < 4.0$ is quite 
uncertain. We also find evidence for mass-dependent evolution of the SMF of 
galaxies with redshift. Specifically, galaxies below the characteristic 
stellar mass show stronger evolution with cosmic time with respect to the 
galaxies at the high-mass end, with the most massive galaxies 
($M_{\rm star}> 10^{11.5}$~M$_{\sun}$) showing a remarkable lack of evolution.

If both random and systematic errors affecting the observed SMFs are 
taken into account, the previous results are no longer robust, and 
significant progress is required to better constrain the high-redshift 
SMFs. The observed absolute number of the most massive galaxies is very small, 
only a handful in each of the targeted redshift interval. Surveys with much 
larger area are needed in order to significantly improve the statistics on 
these rare objects. Cosmic variance, i.e., field-to-field variations, remains 
a significant source of uncertainty, especially at the high-mass end and at 
$z<2$. Therefore, large surveys over multiple spatially disjoint fields are 
required to make significant progress in this respect. Last, but not least, 
most of the redshift information comes from photometric redshift estimates. 
At the very high-mass end, photometric redshift errors are a non-negligible 
source of errors, since the uncertainties will tend to scatter galaxies from 
the well-populated region around $M^{\star}_{\rm star}$ toward the sparsely 
populated high-mass end. Redshift errors are a significant source of 
uncertainties also at the highest considered redshift intervals.

Obtaining large numbers of spectroscopic redshifts for $K$-selected 
high-$z$ sources has proven difficult and extremely time consuming, 
even for very bright galaxies \citep{kriek08}. Note also that the 
galaxies belonging to the high-mass end 
($M_{\rm star}>10^{11}$~M$_{\sun}$) are generally very faint in the 
observed optical ($R\sim26$; \citealt{vandokkum06}), hence making 
optical spectroscopy completely unfeasible. Although the situation will 
be improved in the near future by the advent of multi-object NIR 
spectrographs, which will allow to construct large sample of high-$z$ 
$K$-selected galaxies with spectroscopic redshift measurements, these 
projects will only probe the brightest population, leaving the high-mass 
end of the SMF still prone to photometric redshift uncertainties affecting 
the rest of the galaxy population.

The on-going NEWFIRM Medium-Band Survey \citep{vandokkum09} will be able, 
upon completion, to improve on all of the above aspects. This deep and 
wide-field NOAO/Yale survey uses the newly commissioned NEWFIRM instrument 
(mounted on the 4~m Mayall telescope) with custom-made medium band-width 
filters over the wavelength range 1--1.8~$\mu$m to obtain well-sampled SEDs 
and high quality photometric redshifts for $K^{\rm tot}_{\rm S}<23.4$ galaxies. 
The full survey will provide accurate redshift measurements 
($\Delta z / (1+z) \approx 0.02$) for $\sim8000$ galaxies at $1.5<z<3.5$ over 
a total area of $\sim$0.5~square degree in the COSMOS and AEGIS fields, 
allowing to drastically reduce the impact of random uncertainties and cosmic 
variance in the measurements of the high-redshift SMFs.

We also note that we cannot constrain the low-mass end of the SMFs very 
well at $z>2$. If the SMF is much steeper at the low-mass end than 
indicated by our maximum-likelihood analysis, we could be missing a 
significant fraction of the integrated stellar mass (see \citealt{reddy08}). 
\citet{reddy08} suggest that up to $\sim$50\% of the total stellar mass at 
$1.9 \le z<3.4$ is in faint-UV galaxies with masses smaller than 
$\sim10^{10}$~M$_{\sun}$ (compared to $\sim$10\%-20\% from an extrapolation 
of our Schechter fits). However, \citet{reddy08} do not measure stellar 
masses directly, but convert UV luminosity to stellar mass. Ultra-deep 
near-IR imaging can directly determine whether there is a large population 
of very low-mass galaxies at high redshift which contribute significantly 
to the total stellar mass density. Deep surveys with WFC3 on HST, as well as 
deep ground-based surveys such as the UKIDSS Ultra-Deep Survey 
\citep{lawrence07} and the Ultra-VISTA survey\footnote{\url{http://www.eso.org/sci/observing/policies/PublicSurveys/sciencePublic\-Sur\-veys.html}}, will 
allow us to much better constrain the low-mass end of the SMFs.

Progress in observations have to be supported by significant progress 
in the theoretical arena. Systematic uncertainties due to different 
SED-modeling assumptions are a significant, if not dominant, contribution 
to the total error budget. Convergence on the stellar population synthesis 
models is of paramount importance not only to significantly decrease the 
systematic uncertainties on the derived SMF, but also to have a correct 
understanding of the properties of galaxies, such as age, star formation rate, 
etc. Understanding the IMF and its evolution as function of cosmic time is 
also extremely important, especially for studies at high-redshift. Significant 
work is however required theoretically to better understand what physical 
parameter is most responsible in shaping the IMF and its evolution with cosmic 
time.

The observed SMFs have been compared with the SMFs predicted from the 
semi-analytic models of \citet{monaco07}, \citet{wang08}, and 
\citet{somerville08}. The model-predicted SMFs are generally too steep 
with respect to the observed SMFs, resulting in significant over-prediction 
of the number densities of galaxies at the low mass end. While relatively 
good agreement is observed at the high-mass end at $z\lesssim2.5$, some 
models tend to under-predict the observed number densities of massive 
galaxies at $2.5 \lesssim z \lesssim 4.0$. The discrepancy at the 
high-mass end is susceptible to uncertainties in the models and the data, 
but the discrepancy at the low-mass end may be more difficult to explain. 
These results are robust, even when all random and systematic uncertainties 
are included, suggesting that current models do not yet provide a complete 
description of galaxy formation and evolution since $z=4.0$.

%============================================================================

\acknowledgments

We are grateful to C. Maraston, for providing stellar population 
synthesis models with bottom-light IMFs; S. Charlot, for providing 
the unpublished CB08 stellar population synthesis models; R. Somerville 
and F. Fontanot for providing the predictions of the semi-analytic models 
used in this work and for helpful clarifications. We also thank G. Brammer 
for constant support with EAZY. We thank the anonymous referee for his/her 
detailed and insightful comments, which helped improve the paper. We thank 
all the members of the MUSYC collaboration for their contribution to this 
research. MUSYC has greatly benefited from the support of Fundaci\'on Andes 
and the Yale Astronomy Department. DM is supported by NASA LTSA NNG04GE12G. 
The authors acknowledge support from NSF CARRER AST-0449678.

\clearpage

%============================================================================

\appendix

%============================================================================

\section{Data reduction and photometry of the {\it Spitzer}-IRAC data in MUSYC:} \label{app-iracdata}

Here we describe the data taken with the Infrared Array Camera (IRAC; 
\citealt{fazio04}) on board {\it Spitzer} over the deep NIR MUSYC 
fields, their reduction, and the creation of $K$-selected catalogs 
with IRAC photometry. The $K$-selected catalogs with IRAC photometry 
included is publicly available at \url{http://www.astro.yale.edu/musyc}.

%----------------------------------------------------------------------------

\subsection{{\it Spitzer}-IRAC data}

The IRAC data over the deep NIR MUSYC fields come from two different 
sources. Specifically, the IRAC data over the HDFS-2, the SDSS-1030, and 
the CW-1255 fields come from the Spitzer Space Telescope Cycle-3 program 
GO-30873 (P.I.: Labb\'e), while the IRAC data over the HDFS-1 field are 
part of the GTO-214 program (P.I.: Fazio). Table~\ref{tab-appiracdata1} 
summarizes the characteristics of the IRAC data over the deep MUSYC 
fields, such as the total exposure time, full-width half maximum (FWHM), 
limiting depth, positional accuracy, and Galactic extinction in each of 
the four IRAC bands. The total exposure times vary from 30 min for the 
HDFS-2 field, to 1 hr in the SDSS-1030 and CW-1255 fields across the entire 
$10^{\prime} \times 10^{\prime}$ field. The exposure time across the 
HDFS-1 is instead not homogeneous, being $\sim4.2$~hrs in the small area 
overlapping with the HDF-S Proper field, and 20~min everywhere else. All 
fields were covered in all four IRAC channel, namely the 3.6, 4.5, 5.8, 
and 8.0~$\mu$m bands. 

\begin{deluxetable}{lccccc}
\tablecaption{Characteristics of the IRAC observations
\label{tab-appiracdata1}}
\tablehead{\colhead{Filter} & \colhead{Exposure Time} & \colhead{FWHM} & \colhead{Total Limiting Magnitude\tablenotemark{a}} & \colhead{Positional Accuracy\tablenotemark{b}} & \colhead{Galactic Extinction}\\
($\mu$m) & (hr) & (arcsec) & (3~$\sigma$, AB mag) & (arcsec) & (mag)}
\startdata
SDSS-1030: & \multicolumn{5}{c}{} \\
3.6 & 1.0 & 1.4 & 24.53 & 0.07 & 0.005 \\
4.5 & 1.0 & 1.4 & 24.14 & 0.08 & 0.004 \\
5.8 & 1.0 & 1.7 & 22.35 & 0.07 & 0.004 \\
8.0 & 1.0 & 2.0 & 22.16 & 0.08 & 0.003 \\
\hline
CW-1255: & \multicolumn{5}{c}{} \\
3.6 & 1.0 & 1.5 & 24.74 & 0.05 & 0.003 \\
4.5 & 1.0 & 1.4 & 24.37 & 0.07 & 0.003 \\
5.8 & 1.0 & 1.8 & 22.52 & 0.05 & 0.002 \\
8.0 & 1.0 & 2.1 & 22.45 & 0.07 & 0.002 \\
\hline
HDFS-2: & \multicolumn{5}{c}{} \\
3.6 & 0.5 & 1.4 & 24.49 & 0.08 & 0.005 \\
4.5 & 0.5 & 1.4 & 24.00 & 0.09 & 0.005 \\
5.8 & 0.5 & 1.8 & 22.21 & 0.08 & 0.004 \\
8.0 & 0.5 & 2.2 & 22.08 & 0.09 & 0.004 \\
\hline
HDFS-1: & \multicolumn{5}{c}{} \\
3.6 & 0.3-4.2 & 1.7 & 24.41$-$25.06\tablenotemark{c} & 0.06 & 0.005 \\
4.5 & 0.3-4.2 & 1.7 & 24.12$-$24.83\tablenotemark{c} & 0.07 & 0.005 \\
5.8 & 0.3-4.2 & 2.2 & 22.39$-$23.36\tablenotemark{c} & 0.06 & 0.004 \\
8.0 & 0.3-4.2 & 2.1 & 22.35$-$23.24\tablenotemark{c} & 0.07 & 0.004 \\
\enddata
\tablenotetext{a}{The total limiting magnitudes 
were estimated using the empty aperture method, 
corrected for the flux missed outside the 
3$^{\prime \prime}$ aperture used for photometry. 
This correction is $\sim0.18$~mag.}
\tablenotetext{b}{The rms difference between 
bright star positions in IRAC and $K$-band image, 
after pointing refinement.}
\tablenotetext{c}{The first number corresponds 
to the shallower area, while the second depth 
corresponds to the deeper area overlapping with 
the HDF-S Proper field.}
\end{deluxetable}

%----------------------------------------------------------------------------

\subsection{Data reduction}

The reduction started with the basic calibrated data (BCD) as provided 
by the Spitzer Science Center pipeline. We applied a series of procedures 
to reject cosmic rays and remove artifacts such as column pulldown, 
muxbleed, maxstripe, and the ``first-frame effect'' \citep{hora04}. 
Then, the single background-subtracted frames were combined into a 
mosaic image large enough to hold all input frames (using WCS astrometry 
to align the images). In this step, bad pixels were also masked, and 
distortion corrections applied in WCS. This image, in combination with 
the $K$-band reference image, was then used to refine the pointing of 
the individual mosaics (6 mosaics, in a $2\times3$ grid). After pointing 
refinement, positional accuracy is $\sim0.05-0.09$~arcsec. We note that 
the source-fitting algorithm developed by I.~Labb\'e et al. (in 
preparation) that we have used to derive the IRAC photometry takes care 
of residual shifts. Finally, the individual pointing-refined frames were 
registered to and projected on the public $K$-band images 
(0.27$^{\prime \prime}$ pixel scale) of the four MUSYC fields 
\citep{quadri07},\footnote{All optical and NIR data and the $K$-selected 
catalogs from the deep NIR MUSYC survey are publicly available from 
\url{http://www.astro.yale.edu/musyc}.} and average-combined. Flux 
conservation has been forced throughout the reduction.

%----------------------------------------------------------------------------

\subsection{Seeing, zero-points, and limiting depths} \label{limdep}

FWHM of the IRAC images were derived by taking the median of the FWHM 
for a set of $\sim 15-30$ bright, isolated stars in the fields. The 
FWHM of each field and each IRAC band are listed in 
Table~\ref{tab-appiracdata1}, and amount to 1.4$-$1.7$^{\prime \prime}$, 
1.4$-$1.7$^{\prime \prime}$, 1.7$-$2.2$^{\prime \prime}$, and 
2.0$-$2.2$^{\prime \prime}$ for the 3.6, 4.5, 5.8, and 8.0~$\mu$m bands, 
respectively. 

Following \citet{quadri07}, the IRAC photometry in the $K$-selected 
catalogs are presented in units of flux, normalized so that the 
zero-points is 25 on the AB system. The use of flux, rather than 
magnitudes, avoids the problem of converting the measured flux 
uncertainties into magnitude uncertainties, the problem of asymmetric 
magnitude uncertainties for low S/N objects, and the loss of 
information for objects that have negative measured fluxes. Vega 
to AB magnitude conversions are 2.78, 3.26, 3.75, and 4.38 for the 
3.6, 4.5, 5.8, and 8.0~$\mu$m bands, respectively. We adopted the 
following transformation from Jy to Vega magnitudes:
\begin{equation}
m_{\rm Vega} = - 2.5 \log{(flux[Jy])} + B,
\end{equation}
with $B=6.12$, 5.64, 5.15, and 4.52 for the 3.6, 4.5, 5.8, and 
8.0~$\mu$m bands, respectively. The IRAC zero-point uncertainties 
are of the order of 2\% \citep{reach05}, which was included in the 
flux error budget.

The IRAC photometry has been corrected for Galactic extinction, using 
the extinctions listed in Table~\ref{tab-appiracdata1}. These values, 
taken from the Galactic Dust Extinction Service 
(\url{http://irsa.ipac.caltech.edu/applications/DUST}), were derived 
using the data and technique of \citet{schlegel98}.

The properties of the IRAC images were analyzed following the same 
approach as for the deep NIR MUSYC data \citep{quadri07}. Briefly, 
the technique uses aperture photometry distributed randomly over 
empty regions of the image to quantify the rms of background pixels 
within the considered aperture size. For a given aperture size, the 
distribution of empty aperture fluxes is well fitted by a Gaussian. 
The total limiting AB magnitudes (3~$\sigma$ for point sources, 
i.e., corrected for the flux missed outside the aperture used for 
photometry) are listed in Table~\ref{tab-appiracdata1}.

%----------------------------------------------------------------------------

\subsection{Photometry}

A source-fitting algorithm developed by I.~Labb\'e et al. (2008, in 
preparation), especially suited for heavily confused images for which 
a higher resolution prior (in this case the $K$-band image) is 
available, was used to extract the photometry from the IRAC images. 
A short description with illustration was also presented by 
\citet{wuyts07}. Since this program does not take into account 
large-scale background variations, these were removed a priori from the 
IRAC images. Briefly, the information on position and extent of 
sources bases on the higher resolution $K$-band segmentation map was 
used to model the lower resolution IRAC $3.6-8.0$~$\mu$m images. 
Each source was extracted separately from the $K$-band image and, under 
the assumption of negligible morphological $k$-corrections, convolved 
to the IRAC resolution using the local kernel coefficients. Convolution 
kernels were constructed using bright, isolated, unsaturated sources in 
the $K$ and the IRAC bands (derived by fitting a series of Gaussian-weighted 
Hermite functions to the Fourier transform of the sources, rejecting 
outlying or poorly-fitting kernels), and a smoothed 2D map of the kernel 
coefficients was stored. A fit to the IRAC image was then made for all 
sources simultaneously, where the fluxes of the objects were left as 
free parameters. Next, we subtracted the modeled light of neighboring 
objects and measured the flux on the cleaned IRAC map within a 
fixed $3^{\prime \prime}$ diameter aperture. Through a visual inspection 
of the IRAC residual image with all sources subtracted, we conclude that 
this method effectively removes contaminating sources (for an illustration 
of this technique, see also Figure~1 in \citealt{wuyts07}). In order to 
compute a consistent $K-IRAC$ color, we measured the source's flux 
$f_{\rm conv,~K}$ on a cleaned $K$-band image convolved to the IRAC 
resolution within the same aperture. We then scaled the photometry to 
the same color apertures that were used for the NIR photometry, allowing 
a straightforward computation of colors over $U-$to$-8$~$\mu$m 
wavelength baseline. For the IRAC photometry, this means that the 
catalog flux was computed as follows:
\begin{equation}
f_{\rm IRAC,col} = f_{\rm IRAC,3^{\prime \prime}} 
\frac{f_{\rm K,col}}{f_{\rm conv,K,3^{\prime \prime}}}.
\end{equation}

Note that the used source-fitting algorithm developed by I.~Labb\'e et 
al. (in prep.) takes into account the spatial extent of the sources 
on the reference $K$-band image, and it does not adopt the best-fit flux 
from the source fitting as final photometry, but rather measures the 
flux within an aperture on the cleaned image (followed by an aperture 
correction). This allows more robust photometry in cases where the 
object profile varies from the reference to the low-resolution image.
Uncertainties in the measured fluxes in the IRAC bands were derived 
as described in \citet{wuyts08}, accounting for the background rms 
and residual contamination of the subtracted neighbors (I.~ Labb\'e 
et al. 2008, in prep.).

%============================================================================

\section{Illustration of completeness estimation: the SDSS-1030 sample}
\label{app-compl}

We have used a different approach (described in \S~\ref{sec-compl}) 
to estimate the redshift-dependent completeness limit in stellar mass 
of our $K$-selected sample to be used to derive the SMFs of galaxies. 
In the following, we illustrate the estimation of the redshift-dependent 
completeness limit in stellar mass for the SDSS-1030 sample.

\begin{figure}
\epsscale{0.7}
\plotone{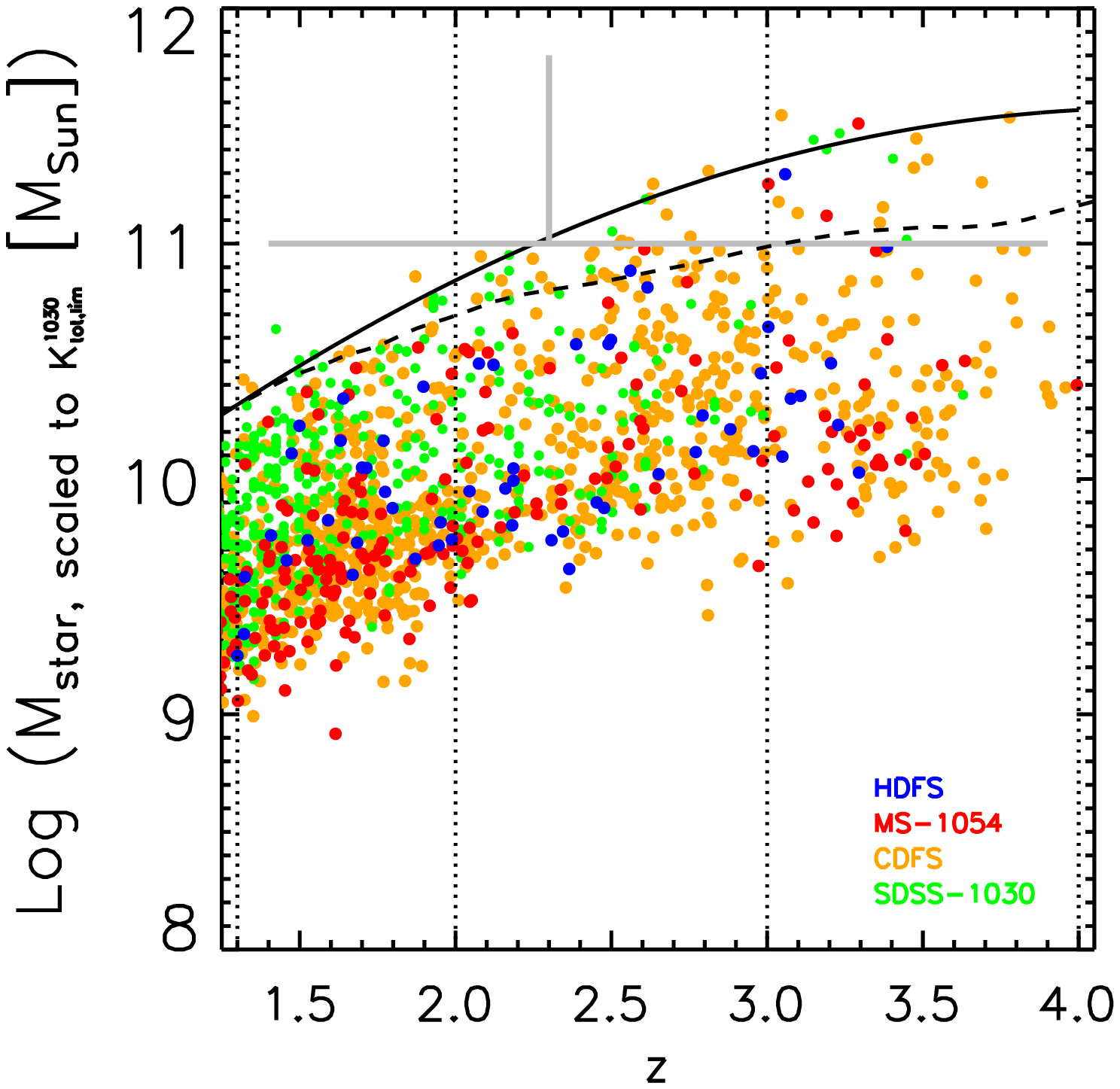}
\caption{Empirically-determined completeness in stellar mass 
as a function of redshift; black filled circles represent the 
stellar masses for the SDSS-1030 galaxies scaled down in flux to 
match the SDSS-1030 $K$-band 90\% completeness limit, and plotted 
as a function of redshift. The other circles show the stellar 
masses for the CDFS (orange), the MS-1054 (red), and the HDFS 
(blue) galaxies scaled in flux to match the SDSS-1030 $K$-band 90\% 
completeness limit. The solid curve represents the upper 
envelope of these points (encompassing 95\% of the points), and 
effectively defines, as a function of redshift, the limiting 
stellar mass corresponding to the observed flux limit of the 
SDSS-1030 sample. At $z\sim2.3$, the SDSS-1030 sample is approximately 
complete for stellar masses $M_{\rm star}>10^{11}$. We emphasize that 
the stellar masses plotted here are not the actual stellar masses, but 
the stellar masses scaled to the $K$-band limit of the SDSS-1030 field.
\label{fig-appcompl1}}
\end{figure}

First, we selected galaxies belonging to the available deeper samples, 
namely the HDFS, the MS-1054, and the CDFS samples. Second, we scale their 
fluxes and stellar masses to match the SDSS-1030 $K$-band 90\% completeness 
limit. This is illustrated in Figure~\ref{fig-appcompl1}. In this figure, 
the colored filled symbols represent objects from the deeper samples 
(HDFS in blue, MS-1054 in red, and CDFS in orange) scaled up in flux to the 
SDSS-1030 $K$-band 90\% completeness limit. These objects represent objects 
immediately at our detection limit. The upper envelope of these points 
in the $(M_{\rm star,scaled}-z)$ space represents the most massive 
galaxies that might escape detection/selection in our analysis. Therefore, 
this upper envelope, encompassing 95\% of the points, provides a 
redshift-dependent stellar mass completeness limit for the SDSS-1030 sample. 

This suggests that at $z \sim 2.3$, the SDSS-1030 sample is approximately 
complete for stellar masses $M_{\rm star}>10^{11}~M_{\sun}$. In the same 
panel, we also compare the empirically-derived completeness limit (solid 
curve) with the completeness derived assuming an SSP with no dust formed 
at $z=10$ and scaled to match the SDSS-1030 $K$-band 90\% completeness limit 
(dashed curve). It is obvious from this comparison that, for SDSS-1030, the 
SSP-derived completeness is similar to the empirically-derived completeness 
only for $z\lesssim2$, while at $z\gtrsim2$ the SSP-derived completeness 
implies a higher completeness in stellar mass than empirically derived. The 
difference between the empirically- and the SSP-derived completeness is 
a function of redshift, with the differences increasing going to higher 
redshift. This could be due, i.e, to the fact that dust extinction 
becomes progressively more important for galaxies in the high-mass end 
with increasing redshift.

%============================================================================
				   
\section{Comparison with previously published galaxy stellar mass functions:}
\label{app-comp}

Here we compare our results to previous studies on the SMFs of 
galaxies at $z>1$. These works include the work of \citet{drory05}, 
\citet{fontana06}, \citet{pozzetti07}, \citet{elsner08}, and 
\citet{perez08}. None of these works have included a comprehensive 
analysis of the uncertainties (random and systematic) on the 
derived SMFs. As already pointed out, our work represents the first 
analysis of the SMFs at high-redshift with a comprehensive analysis 
of the errors. The statistically significant disagreements among the 
different works mostly stem from the lack of a complete analysis of 
the errors of the SMFs from the literature . Once a complete analysis 
of the errors is performed, as done for the first time in our work, 
the disagreements between the different measurements of the SMFs are 
no longer statistically significant. 

Our $K$-selected composite sample is unique in that it combines very 
high-quality multi-waveband data from surveys with a large range of 
depths and areas in a self-consistent way.
%for its combination of depth and surveyed area, as well as very high 
%quality of the %multi-waveband data. 
The available large number of single large area 
fields allows for the empirically estimate of the contribution of 
cosmic variance to the error budget, which is the dominant source of 
random errors in the lowest targeted redshift range and at the high-mass 
end. The combination of deep and ultra-deep samples allows us to 
empirically derive the completeness in stellar mass of the $K$-selected 
sample, without relying on stellar models, such as a passively 
evolving single stellar population formed at very high redshift 
($z\sim10$) without any dust extinction (SSP-derived completeness), 
as done in most of the studies in the literature.

We therefore conclude that, with respect to the works in the 
literature, our SMFs are a significant improvement due to 1) the 
comprehensive analysis of the errors of the SMFs, both random 
and systematic (completely missing in the literature); 2) the large 
range in stellar masses probed by the composite $K$-selected sample, 
which allowed for better sampling of both the high- and the low-mass 
end; 3) the large surveyed area and the large number of independent 
fields, allowing for a decrease of the uncertainty due to cosmic 
variance (with respect to previous works) and to empirically quantify 
its contribution; 4) the availability of samples with different depths, 
allowing us to empirically derive, for the first time, the 
redshift-dependent completeness limits in stellar mass.

\subsubsection{Drory et al. (2005):}

\citet{drory05} derived the SMFs of galaxies from $z=0.25$ to $z=5$ 
from two different samples: an $I$-selected sample over 40~arcmin$^{2}$ 
of the FORS Deep Field (FDF; \citealt{heidt03}) with $UBgRIzJK$ coverage, 
consisting of 5557 galaxies down to $I\sim26.8$ (50\% completeness); and 
a $K$-selected sample over 50~arcmin$^{2}$ in the GOODS-CDFS field with 
UBVRIJHK coverage. No mid-IR IRAC photometry was used. The quoted photometric 
redshift accuracy in $\Delta z / (1+z_{\rm spec})$ is $\sim0.03$.
The stellar masses were derived by fitting the observed SEDs with a 
model grid of BC03 models. The star-formation history (SFH) was 
parameterized by a two-component model, with a main component with a 
smooth SFH modulated by a burst of star formation. The main component 
is parameterized by an exponentially declining SFH with an e-folding 
timescale $\tau \in [0.1,\infty]$~Gyr, and a metallicity of 
$-0.6<[Fe/H]<0.3$. The age was allowed to vary between 0.5~Gyr 
and the age of the universe at the object's redshift. This 
component was linearly combined with a burst modeled as a 100~Myr 
old CSF rate episode of solar metallicity, restricting the burst 
fraction $0<\beta<0.15$ in mass. A \citet{salpeter55} IMF truncated 
at 0.1 and 100~M$_{\sun}$ was adopted for both components. Finally, 
both components are allowed to exhibit a different and variable amount 
of extinction by dust. 
SMFs were derived with the $1/V_{\rm max}$ method, although it 
is unclear how the completeness is stellar mass as a function of 
redshift was derived. The Schechter function parameters were 
derived by fitting the $1/V_{\rm max}$ points with a Schechter 
function.

\begin{figure}
\epsscale{0.75}
\plotone{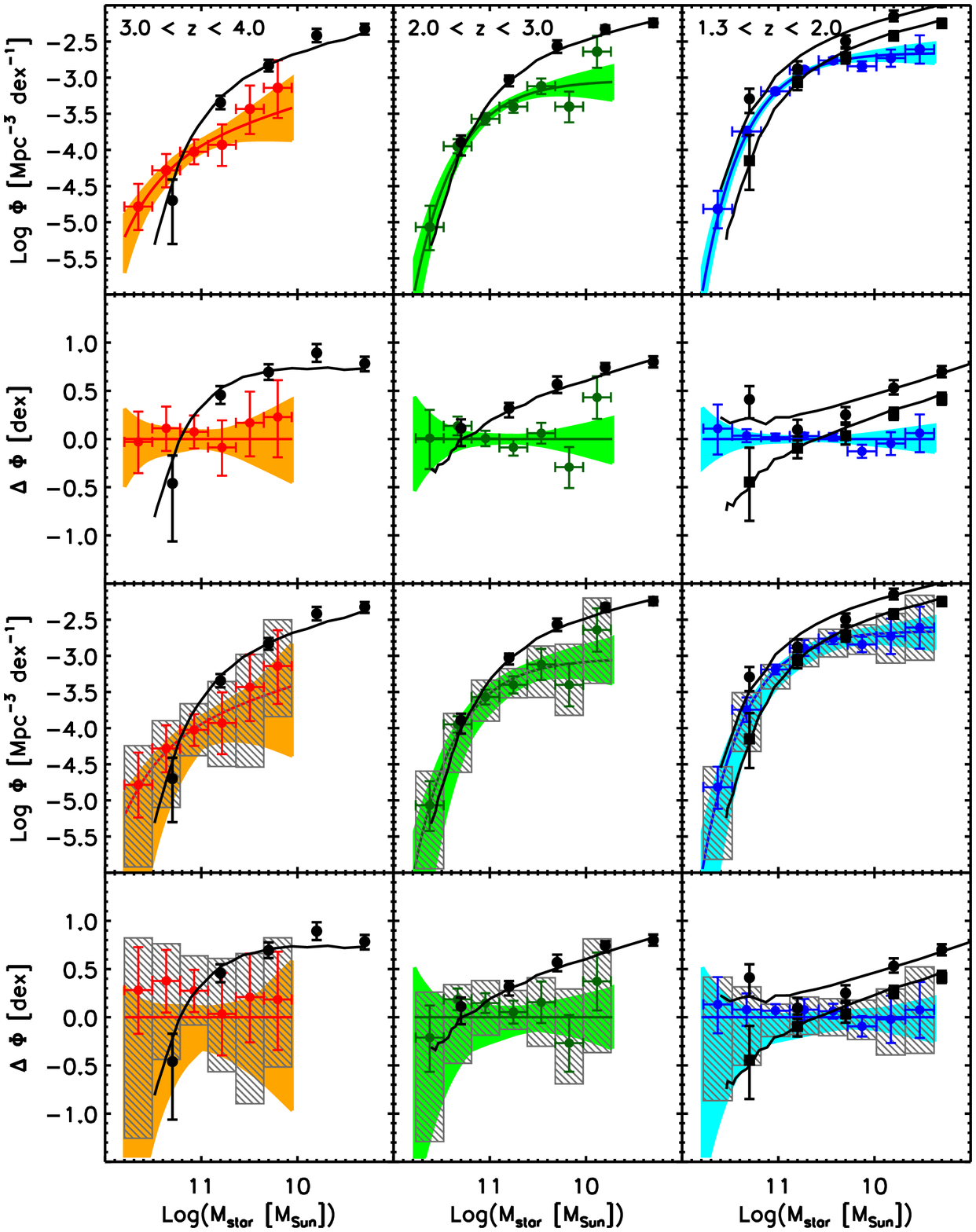}
\caption{Comparison between the SMFs from \citet{drory05} and from 
this work at $3.0 \leq z<4.0$ (left), $2.0 \leq z<3.0$ (middle), 
and $1.3 \leq z<2.0$ (right). Symbols as in Fig.~\ref{fig-all_comp}, 
except black filled symbols and black curves representing the 
SMFs derived in \citet{drory05}. The SMFs of \citet{drory05} 
have been scaled by -0.2~dex along the $x$-axis to take into 
account difference in the adopted IMF. Note that the SMFs 
from \citet{drory05} are actually derived in the following 
redshift intervals: $3.00<z<4.00$ (left panels), $2.25<z<3.00$ 
(middle panels), $1.25<z<1.75$ and $1.75<z<2.25$ (right panels; 
filled black circles and squares, respectively). 
\label{fig-drory}}
\end{figure}

Their surveyed area is $sim6.5$ times smaller than the surveyed area 
of our work, making their derived SMFs very much affected by cosmic 
variance. A direct comparison between the SMFs of \citet{drory05} and 
the SMFs measured in our work is shown in Figure~\ref{fig-drory}.
Note that the redshifts bins used in \citet{drory05} are not exactly 
the same as used in the present work.

The top panels of Figure~\ref{fig-drory} show the SMFs derived in 
our work without the inclusion of cosmic variance and systematic 
uncertainties in the plotted error bars; error bars include Poisson 
errors and photometric redshift random errors. The errors of the 
\citet{drory05} SMFs only include random uncertainties. From the top 
panels of Figure~\ref{fig-drory}, the SMFs from \citet{drory05} are 
consistent with those derived in our work only at the high-mass end, 
while their number densities are significantly higher at the low-mass 
end. The slopes of the SMFs at the low-mass end are steeper than those 
derived in our work, especially at $z\sim1.6$. In the highest redshift 
range, where Poisson statistics dominates the error budget, the two 
SMFs are consistent within 2~$\sigma$ in the overlapping stellar mass 
regime. 

The SMFs of \citet{drory05} become fully consistent with ours at the 
high-mass end once all sources of errors are included in the error 
budget (bottom panels of Figure~\ref{fig-drory}). However, the 
discrepancy at the low-mass end is still present and significant. 
The much higher densities derived by \citet{drory05} at the low-mass 
end (hence, the steeper slope $\alpha$) are very likely caused by 
the use of the $1/V_{\rm max}$ method in combination with an 
inappropriate redshift-dependent completeness in stellar mass. 
While it is unclear how the completeness in stellar mass has been 
derived in \citet{drory05}, we notice that, if an SSP-derived 
completeness were to be adopted in place of the empirically-derived 
completeness, steeper densities at the low-mass end would be derived, 
due to the more conservative completeness of the former with respect 
to the latter. We therefore believe that the SMFs of \citet{drory05} 
at the low-mass end are systematically too large, perhaps due to 
the use of incorrect redshift-dependent completeness limits in stellar 
mass.

\subsubsection{Fontana et al. (2006):}

\citet{fontana06} derived the SMFs of galaxies at $0.4<z<4.0$ from 
the GOODS-MUSIC sample \citep{grazian06}. Their final $K$-selected 
sample consists of $\sim$2931 galaxies (1762 with $H$-band coverage) 
complete down to $K_{\rm S}\approx23.5$, over an area of 
$\sim$143.2~arcmin$^{2}$ with $UB_{\rm 435}V_{\rm 606}i_{\rm 775}z_{\rm 850}JHK$ 
and IRAC bands coverage. Note that this dataset make use of the same public 
data of the FIREWORKS-CDFS dataset \citep{wuyts08} that we have used in our 
work, but without the inclusion of the WFI $BVRI$ band data. Consequently, the 
SEDs of sources in the FIREWORKS-CDFS catalog are better sampled. %Moreover, 
%while the FIREWORKS-CDFS dataset is purely $K$-selected, the GOODS-MUSIC 
%sample is to first order $z_{\rm 850}$-band selected, with an addition of 
%the remaining $K$-band sources that are undetected in the 
%$z_{\rm 850}$-band, making it less trivial to understand the completeness 
%of the sample. 
We also note that \citep{wuyts08} pointed out a systematic offset in the 
IRAC fluxes of the GOODS-MUSIC catalog with respect to the IRAC fluxes of 
the FIREWORKS-CDFS catalog, with the former averagely fainter than the 
latter by $\sim30$\%. The observed offset in the IRAC photometry was largely 
attributed to the use of an early version of the IRAC PSF by 
\citet{grazian06} and a bug in the normalization of the smoothing kernel 
for the IRAC data by \citet{grazian06} (see \citealt{wuyts08} for details). 

Stellar masses were derived by fitting the observed SEDs with a set 
of templates computed with the BC03 spectral synthesis models. A 
\citet{salpeter55} IMF was adopted with various metallicities 
(from $Z=0.02~Z_{\sun}$ to $Z=2.5~Z_{\sun}$), dust extinction 
($0<E(B-V)<1.1$) with a \citet{calzetti00} extinction curve, and 
e-folding timescales ($\tau \in [0.1,15]$~Gyr) in an exponentially 
declining SFH. 

SMFs were derived with the standard $1/V_{\rm max}$ 
formalism and the maximum-likelihood analysis assuming a Schechter 
function with Schechter parameters $\alpha$, $M^{\star}_{\rm star}$, 
and $\Phi^{\star}$ evolving with redshift (for a total of seven free 
parameters, three of which constrained by the local SMFs derived in 
\citealt{cole01}). A treatment has been included to correct for the 
incompleteness in mass at the faintest levels (see \citealt{fontana04} 
for details). Briefly, they start from the threshold computed from a 
passively evolving system (SSP-derived completeness limit), below 
which only a fraction of objects of given mass will be observed. Then, 
at any redshift, the observed distribution of $M/L$ was obtained for 
objects close to the magnitude limit of the sample. Using this 
distribution, the fraction of lost galaxies as function of redshift 
and mass was computed. The correction is finally applied to the volume 
element $V_{\rm max}$ of any galaxy in the $1/V_{\rm max}$ binned SMFs 
as well as in the number of detected galaxies entering the 
maximum-likelihood analysis. We note that this method assumes that 
the sample is complete for objects with stellar masses larger 
than the SSP-derived limit. This assumption is not necessarily valid 
at all redshifts, and depends on the specific depth of the sample 
(see the right panel of Figure~\ref{fig-seccompl1}). Moreover, their 
correction for incompleteness also assumes that the distribution of 
$M/L$ ratio is independent of $L$. As shown in the left panel of 
Figure~\ref{fig-seccompl1}, 
this assumption might not be valid at all
redshift, depending on the specific depth of the sample. In fact, 
at faint luminosities, the distribution of $M/L$ is visibly different 
from that at the bright end. 
While it is hard to predict the differences on the low-mass end 
of the SMFs derived with their method with respect to the method 
used in our work (empirically-derived completeness limit), we notice 
that their correction has been applied only for sources for which 
the correction factor is smaller that 0.5 (usually affecting only 
the last $1/V_{\rm max}$ point). Moreover, as shown in the right panel of 
Figure~\ref{fig-seccompl1}, the SSP-derived completeness limit of 
the CDFS sample is quite similar to the empirically-derived completeness 
limit. Therefore, we do not expect large differences in the derived 
SMFs at the low-mass end, except for the lowest stellar mass bins. 

\begin{figure}
\epsscale{0.75}
\plotone{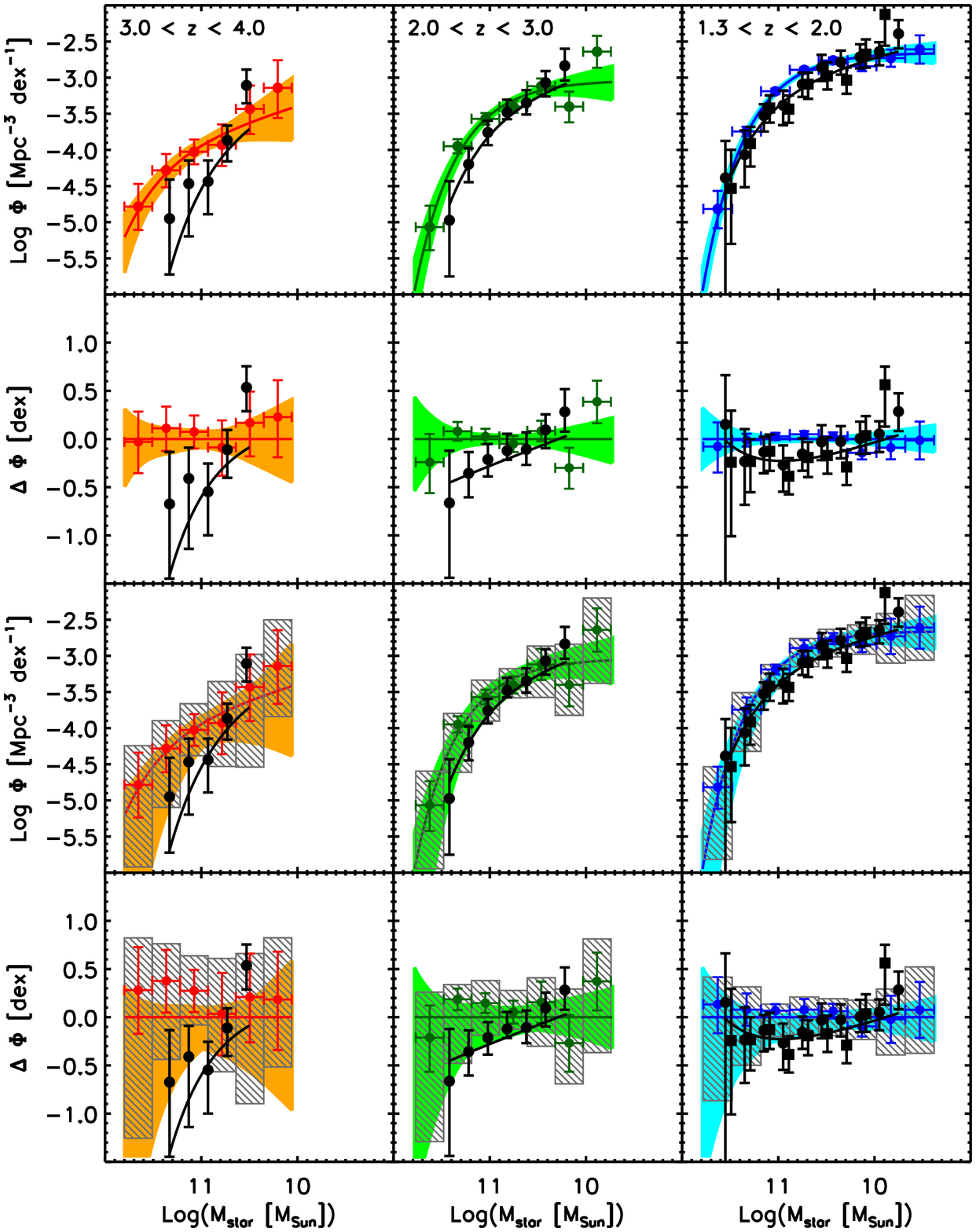}
\caption{Comparison between the SMFs from \citet{fontana06} 
and from this work at $3.0 \leq z<4.0$ (left), $2.0 \leq z<3.0$ (middle), 
and $1.3 \leq z<2.0$ (right). Symbols as in Fig.~\ref{fig-drory}.
The SMFs of \citet{fontana06} have been scaled by -0.2~dex 
along the $x$-axis to take into account difference in the 
adopted IMF. Note that the SMFs from \citet{fontana06} are 
actually derived in the following redshift intervals: 
$3.0<z<4.0$ (left panels), $2.0<z<3.0$ (middle panels), 
$1.3<z<1.6$ and $1.6<z<2.0$ (right panels; filled black circles 
and squares, respectively).
\label{fig-fontana}}
\end{figure}

A direct comparison between the SMFs of \citet{fontana06} and the 
SMFs measured in this work is shown in Figure~\ref{fig-fontana}. Again, 
top panels do not show the contribution of cosmic variance and systematic 
uncertainties in our SMFs, while they are included in the bottom panels 
of Figure~\ref{fig-fontana}. Error bars in the SMFs of \citet{fontana06} 
include Poisson errors and photometric redshift uncertainties, but no 
cosmic variance nor systematic uncertainties due to different 
SED-modeling assumptions. Note that the area surveyed in the GOODS-MUSIC 
sample is a factor of $\sim4$ smaller than the surveyed area in our work, 
and it consists of a single pointing, making their SMFs significantly 
affected by cosmic variance.

First of, as shown in the top panels of Fig.~\ref{fig-fontana}, 
the SMFs estimated from the $1/V_{\rm max}$ method are broadly 
consistent within the errors. There appears to be a systematic 
offset, such that the SMFs of \citet{fontana06} are systematically 
shifted to lower masses. Although it is not simple to fully 
understand the origin of this difference, it could partly be due 
to the found systematic offset in the IRAC photometry. The SMFs 
derived using the maximum-likelihood analysis show a larger degree 
of disagreement, especially in the high-mass end at $z\sim3.5$. 
We stress once again that cosmic variance is a dominant source of 
uncertainty in the SMFs of \citet{fontana06} since they are derived 
from a single, relatively small field. The found differences at the 
bright end can well be accounted for by field-to-field variations.

From the bottom panels of Figure~\ref{fig-fontana}, it is obvious 
that the SMFs of \citet{fontana06} are fully consistent with ours 
once all the source of uncertainties are taken into account, 
including the systematic effects due to different  SED-modeling 
assumptions. 

\subsubsection{Pozzetti et al. (2007):}

\citet{pozzetti07} derived the SMFs of galaxies from $z=0.05$ to 
$z=2.5$ from the $K$-selected sample of the VIMOS-VLT Deep Survey 
(VVDS; \citealt{lefevre05}) 02h field. Their $K$-selected sample 
consists of a shallow and a deeper component. The shallow component 
consists of 6720 galaxies at $0<z<2.5$ down to $K=22.34$ (90\% 
complete) over 442~arcmin$^{2}$, while the deeper component is made 
of 3440 galaxies down to $K=22.84$ (90\% complete) over 
172~arcmin$^{2}$. About 15\% of the galaxies have secure spectroscopic 
identification. The waveband coverage consists of $UBVRIugrizJK$, 
but no IRAC coverage (note that only 170~arcmin$^{2}$ have deep $J$ 
and $K$ coverage; \citealt{iovino05}).
Therefore, their surveyed area is $\sim$3 times smaller than the area 
surveyed in our work down to the same $K$-band limit of $K=22.8$.

The stellar masses were derived by fitting the observed SEDs with a 
model grid of BC03 models. The star-formation history (SFH) was 
parameterized with an exponentially declining SFH with an e-folding 
timescale $\tau \in [0.1,\infty]$~Gyr, and solar metallicity. The age 
was allowed to vary between 0.1~Gyr and the age of the universe at the 
object's redshift. A \citet{chabrier03} IMF was adopted, and extinction 
modeled using the \citet{calzetti00} curve with $A_{\rm V} \in [0,2.4]$.

SMFs were derived with the standard $1/V_{\rm max}$ formalism and 
the maximum-likelihood analysis assuming a \citet{schechter76} function. 
While \citet{pozzetti07} stress the important of using only galaxies with 
stellar masses above the stellar mass limit where all the SEDs are 
potentially observable (very restrictive limit), they ended up using as 
a lower limit of the mass range the minimum mass above which late-type 
SEDs are potentially observable. Therefore, we caution about potential 
biases in the estimate of the low-mass end of their SMFs. 

\begin{figure}
\epsscale{0.6}
\plotone{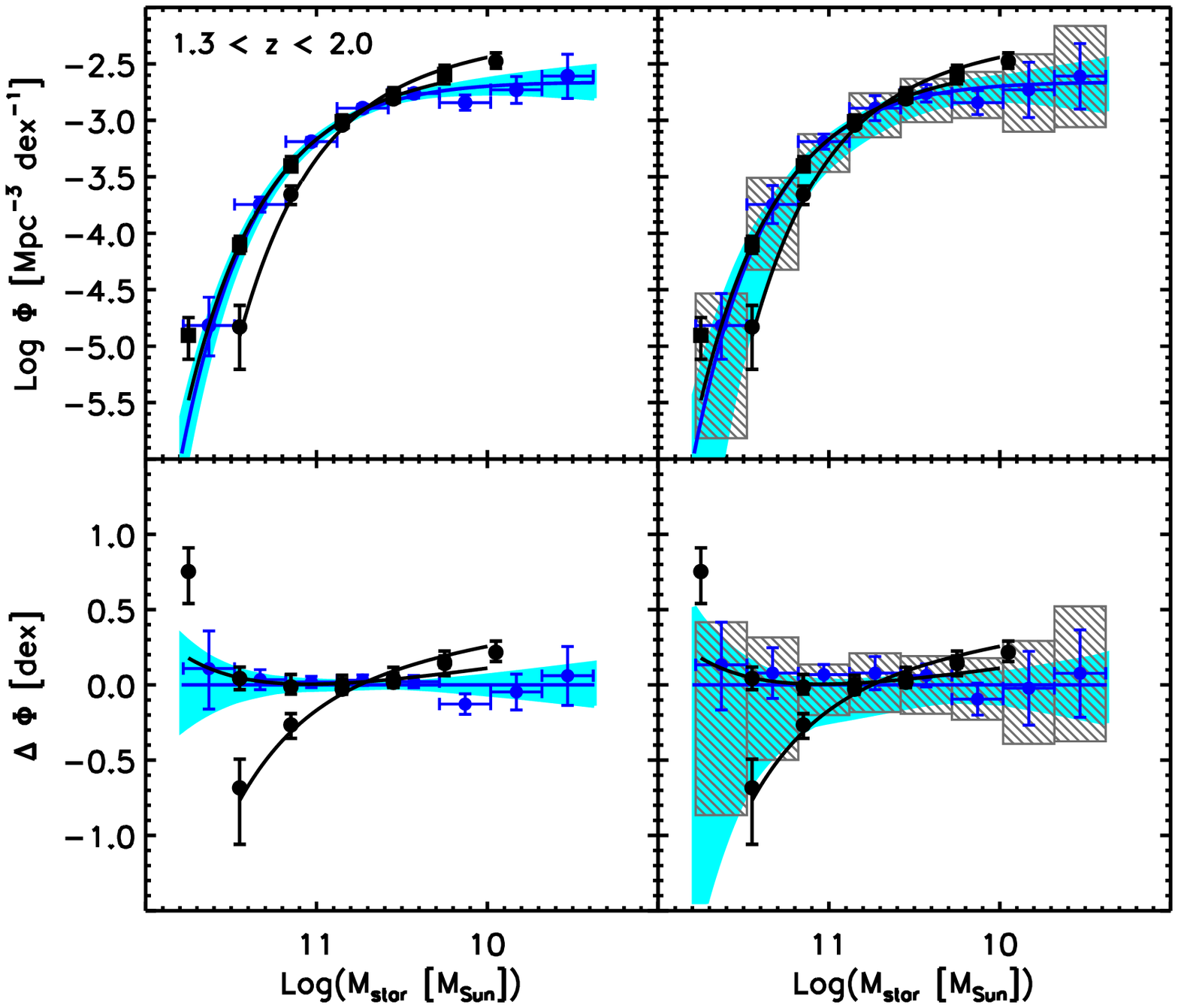}
\caption{Comparison between the SMFs from \citet{pozzetti07} 
and from this work at $1.3 \leq z<2.0$. Symbols as in 
Fig.~\ref{fig-drory}. Note that the SMFs from 
\citet{pozzetti07} are actually derived at $1.2<z<1.6$ 
(filled black squares) and $1.6<z<2.5$ (filled black circles). 
\label{fig-pozzetti}}
\end{figure}

A direct comparison between the SMFs of \citet{pozzetti07} and the SMFs 
measured in our work is shown in Figure~\ref{fig-pozzetti}. Note that the 
redshift bins used in \citet{pozzetti07} are not exactly the same as 
used in the present work. The SMFs derived in our work are shown in 
the left panel without the contribution of cosmic variance and the systematic 
uncertainties, and in the right panel with all errors included.

As shown in the left panel of Figure~\ref{fig-pozzetti}, the SMF from 
\citet{pozzetti07} at $1.6<z<2.5$ is perfectly consistent with ours. 
On the contrary, their SMF at $1.2<z<1.6$ is significantly different 
at the high-mass end. If the two SMFs at $1.2<z<1.6$ and $1.6<z<2.5$ 
are averaged, the obtained SMF is in very good agreement with the 
SMF derived in our work, except at the very low-mass end. We note 
that the low-mass end becomes consistent within the errors when the 
systematic uncertainties are taken into account in the total error 
budget, highlighting once again the importance of a comprehensive 
analysis of the errors. Overall, the agreement between the SMF of 
\citet{pozzetti07} and ours is very good.

\subsubsection{Elsner et al. (2008):}

\citet{elsner08} have derived SMFs of galaxies from $z=5.0$ to 
$z=0.25$ from the GOODS-MUSIC $z+K$-selected catalog of \citet{grazian06}, 
the same used in \citet{fontana06}. The catalog used in \citet{elsner08} 
is to first order $z_{\rm 850}$-band selected sample, with the addition of 
the remaining $K$-band sources that are undetected in the $z_{\rm 850}$-band, 
making it less trivial to understand the completeness of the sample, 
especially in stellar mass. This sample comprises $\sim$14800 galaxies down 
to $z_{\rm lim} \approx 26.0$ (90\% completeness level) over a total area of 
143.2~arcmin$^{2}$. 

The stellar masses were derived by fitting the observed SEDs with a 
model grid of BC03 models. The star-formation history (SFH) was 
parameterized with an exponentially declining SFH with an e-folding 
timescale $\tau \in [0.5,20]$~Gyr, and solar metallicity. The age 
was allowed to vary between 0.2~Gyr and the age of the universe at the 
object's redshift. A \citet{salpeter55} IMF truncated at 0.1 and 100 
M$_{\sun}$ was adopted. Dust extinction was modeled using the 
\citet{calzetti00} curve with $A_{\rm V} \in [0,1.5]$. In addition 
to this main component, a starburst was superimposed which was allowed 
to contributed at most 20\% to the $z$-band luminosity in the rest-frame. 
This component was modeled as a 50~Myr old episode of constant star 
formation with an independent extinction up to $A_{\rm V}=2.0$~mag.

SMFs were derived with the $1/V_{\rm max}$ method after correcting 
the data points for incompleteness due to the flux-limited sample. 
The completeness limit in stellar mass was estimated by scaling the 
$z$-band completeness limit by the calculated 95\% quantile in 
$M/L_{\rm z}$ as a function of redshift, e.g. the limit below which 
95\% of the $M/L_{\rm z}$ ratios of the sample are located. 
The Schechter function parameters were then derived by fitting the 
$1/V_{\rm max}$ points with a Schechter function after fixing the 
low-mass end slope at its error-weighted mean value. 

\begin{figure}
\epsscale{0.75}
\plotone{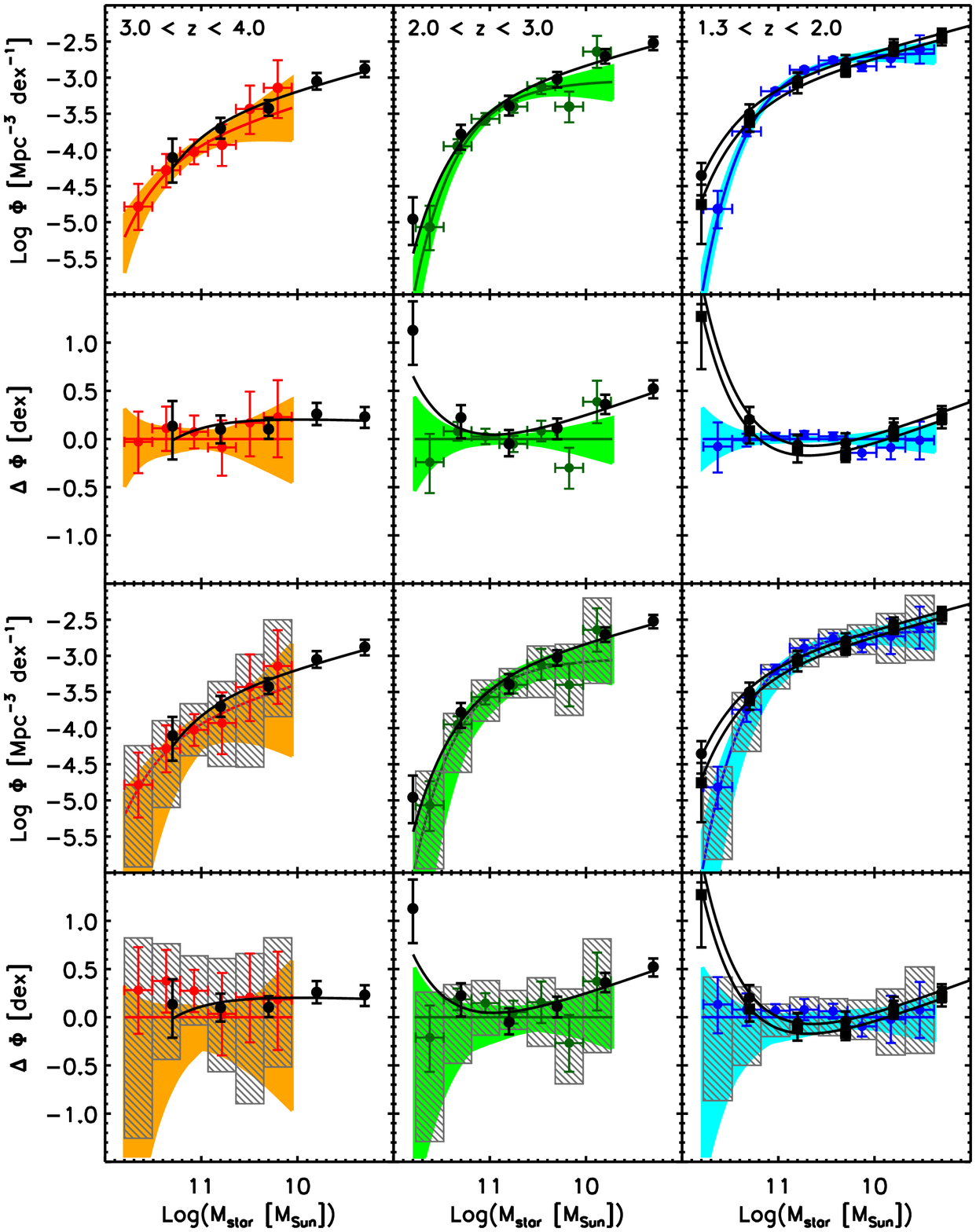}
\caption{Comparison between the SMFs from \citet{elsner08} 
and from this work at $3.0 \leq z<4.0$ (left), $2.0 \leq z<3.0$ (middle), 
and $1.3 \leq z<2.0$ (right). Symbols as in Fig.~\ref{fig-drory}. 
The SMFs of \citet{elsner08} have been scaled by -0.2~dex 
along the $x$-axis to take into account difference in the 
adopted IMF. Note that the SMFs from \citet{elsner08} are 
actually derived at $3.01 \leq z < 4.01$ (left panels), 
$2.25 \leq z < 3.01$ (middle panels), $1.25 \leq z < 1.75$ 
and $1.75 \leq z < 2.25$ (right panels; filled black circles 
and squares, respectively).
\label{fig-elsner}}
\end{figure}

A direct comparison between the SMFs of \citet{elsner08} and the SMFs 
measured in our work is shown in Figure~\ref{fig-elsner}. Note that 
the redshifts bins used in \citet{elsner08} are not exactly the same 
as used in the present work.

The SMFs estimated from \citet{elsner08} are in general good agreement 
with the SMFs measured in our work, especially at $z\sim3.5$ and 
$z\sim2.5$. At the lower targeted redshift interval ($1.3 \leq z < 2.0$), 
the SMF of \citet{elsner08} shows a higher number density for the most 
massive galaxies with respect to our measurements. We note however that 
the sample of \citet{elsner08} is constructed from the single, relatively 
small GOODS field ($\sim$4 times smaller than the area surveyed in our 
work). Therefore, the SMFs derived by \citet{elsner08} are significantly 
affected by field-to-field variations, especially at low redshift and at 
the high-mass end. Note that the error bars in the SMFs of \citet{elsner08} 
do not include the error due to cosmic variance, which we have shown being 
the dominant contribution to the total random error budget at $z \lesssim 2$. 
We therefore conclude that the SMFs of \citet{elsner08} are fully consistent 
with our measurements, and that the disagreements at the high-mass end can 
be fully accounted by field-to-field variations.

We finally note the large significant discrepancies between the SMFs 
derived from \citet{elsner08} and from \citet{fontana06}. The used 
catalog is exactly the same, i.e., the GOODS-MUSIC catalog. The 
differences are in the way the stellar mass completeness limits are 
estimated (affecting the low-mass end of the SMF) and in the assumptions 
of the SED-modeling. Specifically, the metallicity in \citet{elsner08} is 
fixed to solar, while it is left as a free parameter in \citet{fontana06}. 
Also, larger $A_{\rm V}$ are allowed in \citet{fontana06}, as well as no 
secondary starburst component added, contrary to what done in 
\citet{elsner08}. The SMFs from \citet{fontana06} are systematically 
lower at similar stellar mass, particularly at $z\sim3.5$ and at the 
high-mass end. While the differences are statistically significant, as 
already pointed out by \citet{elsner08}, they can be due to the 
different choices of SED-modeling assumptions, stressing the importance 
of a comprehensive analysis of the errors, both random and systematics.

\subsubsection{P\'erez-Gonz\'alez et al. (2008):}

\citet{perez08} derived the SMFs of galaxies from $z=0$ to $z=4$ 
from a 3.6~$\mu$m and 4.5~$\mu$m  {\it Spitzer}-IRAC selected 
sample. Their sample consists of $\sim$19400 sources down to the 
75\% completeness limit ($\sim$23.3~mag at 3.6~$\mu$m) over three 
fields, namely the HDF-N, the CDF-S, and the Lockman Hole fields, 
for a total surveyed area of 664~arcmin$^{2}$, a factor of $\sim$1.14 
larger than the total area surveyed by the sample used in the present 
work ($\sim$583~arcmin$^{2}$). Their 90\% completeness levels are in the 
range 22.0-22.4~mag at 3.6~$\mu$m and 4.5~$\mu$m, about a magnitude 
shallower than the 75\% completeness levels.

The stellar masses were derived by fitting the observed SEDs with a 
grid of models created with the PEGASE code \citep{fioc97}. The 
star-formation history (SFH) was parameterized with an exponentially 
declining SFH with an e-folding timescale $\tau \in [0.001,100]$~Gyr, 
with allowed ages from 1~Myr to the age of the universe at the object's 
redshift. Seven discrete values of the metallicity were used, from 
$Z=0.005$~Z$_{\sun}$ to $Z=5.0$~Z$_{\sun}$. A \citet{salpeter55} IMF 
truncated at 0.1 and 100 M$_{\sun}$ was adopted. Dust extinction was 
modeled using the \citet{calzetti00} curve with $A_{\rm V} \in [0,5]$. 
While other SED-modeling assumptions were used to test how the stellar 
masses changed by changing IMF, extinction curve, stellar population 
synthesis model, and star formation history, the systematic effects of 
these changes on the derived SMFs were not explicitly quantified nor 
discussed in \citet{perez08}.

SMFs were derived by integrating the bivariate luminosity-stellar 
mass function (the estimation of which was performed with a stepwise 
maximum likelihood technique) over all luminosities. The resulting 
SMFs were then fitted with a \citet{schechter76} function. The low-mass 
end of the SMF at $z>1.6$ was constrained by combining their results 
with other estimates of the SMFs from the literature. The 
redshift-dependent completeness limit were derived assuming to be 
complete for stellar masses larger than the stellar mass corresponding to 
a passively evolving stellar population formed at $z\sim \infty$ with 
no extinction and having a 3.6~$\mu$m flux equal to the 75\% completeness 
level of the IRAC sample. Only galaxies with stellar masses larger than 
this completeness level were used in their analysis, and no completeness 
correction was carried out to recover the SMF at smaller masses.

\begin{figure}
\epsscale{0.75}
\plotone{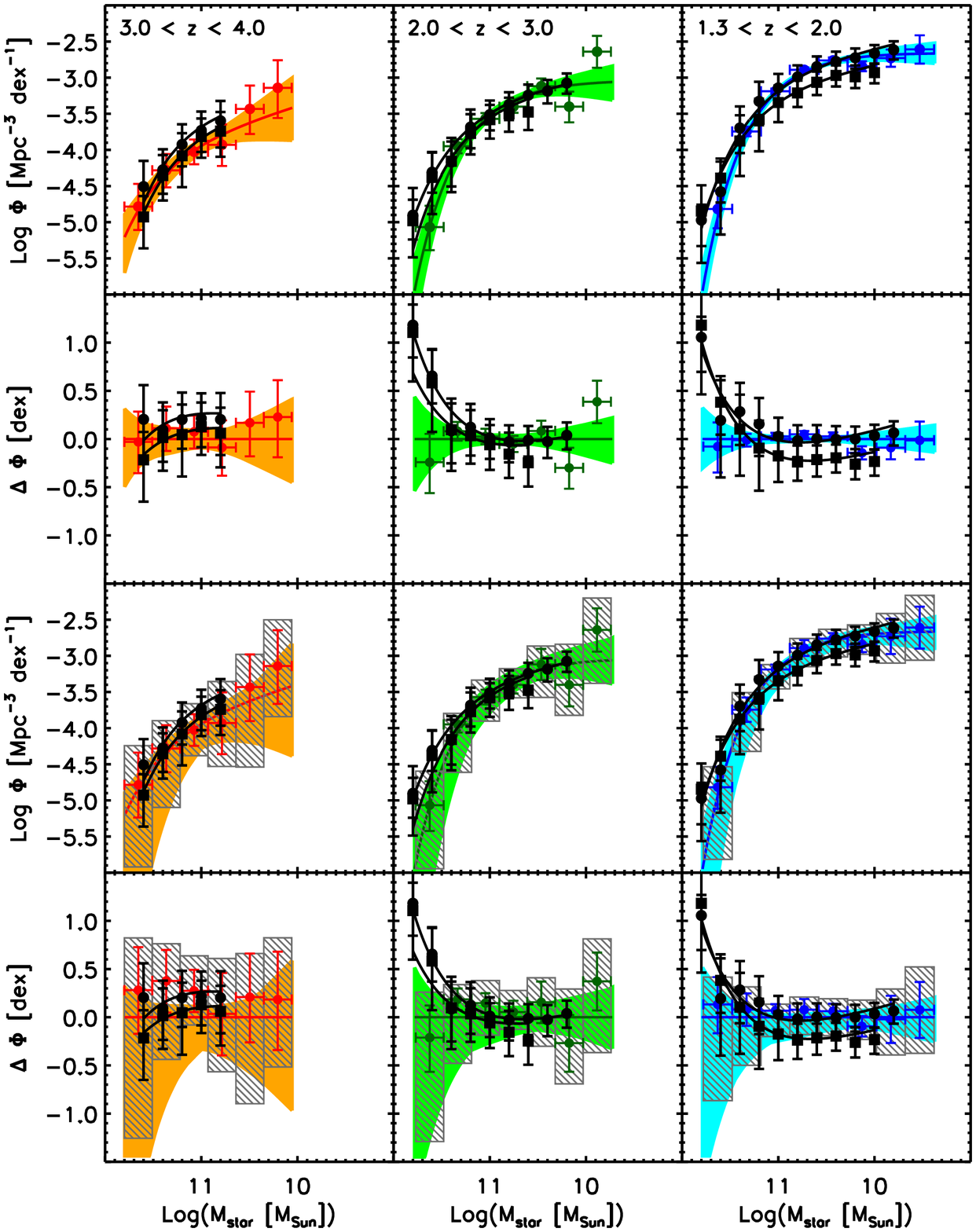}
\caption{Comparison between the SMFs from \citet{perez08} and 
from this work at $3.0 \leq z<4.0$ (left panels), $2.0 \leq z<3.0$ (middle 
panels), and $1.3 \leq z<2.0$ (right panels). Symbols as in 
Fig.~\ref{fig-drory}. The SMFs of \citet{perez08} have been 
scaled by -0.2~dex along the $x$-axis to take into account 
difference in the adopted IMF. Note that the SMFs from 
\citet{perez08} are actually derived at $3.0<z \leq 3.5$ and 
$3.5<z \leq 4.0$ (left panels; filled black circles and squares, 
respectively); $2.0<z \leq 2.5$ and $2.5<z \leq 3.0$ (middle 
panels; filled black circles and squares, respectively); 
$1.3<z \leq 1.6$ and $1.6<z \leq 2.0$ (right panels; filled black 
circles and squares, respectively). \label{fig-perez}}
\end{figure}

A direct comparison between the SMFs of \citet{perez08} and the SMFs 
measured in our work is shown in Fig.~\ref{fig-perez}. Note that the 
redshifts bins used in \citet{perez08} are smaller than the redshift 
intervals used in our work, i.e., our redshift intervals are further 
split in two in \citet{perez08}.

As shown in Fig.~\ref{fig-perez}, the SMFs from \citet{perez08} 
are generally in good agreement with the SMFs derived in our 
work. at all redshifts, especially at the low-mass end. Only at 
the very high-mass end in the redshift ranges $2.0 \leq z<3.0$ and 
$1.3 \leq z<2.0$, the number densities derived from \citet{perez08} 
look slightly larger than those derived in our work. These 
differences are only barely significant, and are definitely 
not statistically significant at all once cosmic variance and 
systematic uncertainties are included in the total error budget. 
We therefore conclude that the SMFs derived from \citet{perez08} 
are in good agreement with the SMFs derived in our work, with the 
latter better sampling the low-mass end of the SMFs by 0.3--0.6 dex 
in stellar mass.\\

%============================================================================

\clearpage

\end{document}